\documentclass[fleqn,usenatbib]{mnras}

\usepackage{newtxtext,newtxmath}

\usepackage[T1]{fontenc}

\DeclareRobustCommand{\VAN}[3]{#2}
\let\VANthebibliography\thebibliography
\def\thebibliography{\DeclareRobustCommand{\VAN}[3]{##3}\VANthebibliography}

\usepackage{graphicx}	
\usepackage{amsmath}	
\usepackage{CJKutf8}
\usepackage{subcaption}
\usepackage[dvipsnames]{xcolor}
\usepackage{orcidlink}

\graphicspath{{fig/}}

\newcommand{\aref}[1]{\hyperref[#1]{Appendix \ref*{#1}}}

\newcommand{\coone}{\mbox{CO(1-0)}}
\newcommand{\cotwo}{\mbox{CO(2-1)}}

\title[WISDOM XXVII.\ GMCs of NGC~1387]{WISDOM Project  -- XXVII.\ Giant molecular clouds of the lenticular galaxy NGC~1387: similarities with spiral galaxy clouds}

\author[F.-H.\ Liang et al.]{
\begin{CJK*}{UTF8}{gbsn}Fu-Heng Liang (梁赋珩)\end{CJK*},$^{\orcidlink{0000-0003-2496-1247}1,2,3}$\thanks{Email: \href{mailto:fuheng.liang@physics.ox.ac.uk}{fuheng.liang@physics.ox.ac.uk}, \href{mailto:ericfuhengliang@gmail.com}{ericfuhengliang@gmail.com}}
Martin Bureau,$^{\orcidlink{0000-0003-4980-1012}1}$
Lijie Liu,$^{4,5,6}$ 
Pandora Dominiak,$^{1}$ 
Woorak Choi,$^{\orcidlink{0000-0001-5033-7208}7,8}$ 
\newauthor{
Timothy A.\ Davis,$^{9}$ 
Jacob Elford,$^{9}$ 
Jindra Gensior,$^{\orcidlink{0000-0001-6119-9883}10}$ 
Anan Lu,$^{11}$ 
Ilaria Ruffa,$^{9,12}$  
Sel\c{c}uk Topal,$^{\orcidlink{0000-0003-2132-5632}1,13}$ 
}
\newauthor{
Thomas G.\ Williams$^{1,14,15}$ 
and Hengyue Zhang$^{1}$
}
\\
$^{1}$Sub-department of Astrophysics, Department of Physics, University of Oxford, Keble Road, Oxford OX1~3RH, UK\\
$^{2}$European Southern Observatory, Karl-Schwarzschild-Stra\ss e 2, Garching 85748, Germany\\
$^{3}$Astronomisches Rechen-Institut, Zentrum f\"{u}r Astronomie der Universit\"{a}t Heidelberg, M\"{o}nchhofstra\ss e 12-14, Heidelberg 69120, Germany\\
$^{4}$Cosmic Dawn Center (DAWN), Copenhagen, Denmark\\
$^{5}$Niels Bohr Institute, University of Copenhagen, Lyngbyvej 2, Copenhagen \O\ 2100, Denmark \\
$^{6}$DTU-Space, Technical University of Denmark, Elektrovej 327, Kgs.\ Lyngby 2800, Denmark \\
$^{7}$Department of Astronomy, Yonsei University, 50 Yonsei-ro, Seodaemun-gu, Seoul 03722, Republic of Korea\\
$^{8}$Department of Physics and Astronomy, McMaster University, 1280 Main Street West, Hamilton, ON L8S~4M1, Canada\\
$^{9}$Cardiff Hub for Astrophysics Research \& Technology, School of Physics \& Astronomy, Cardiff University, Queens Buildings, The Parade, Cardiff CF24~3AA, UK\\
$^{10}$Institute for Astronomy, University of Edinburgh, Royal Observatory, Blackford Hill, Edinburgh EH9~3HJ, UK \\
$^{11}$Trottier Space Institute and Department of Physics, McGill University, 3600 University Street, Montreal, QC H3A~2T8, Canada\\
$^{12}$INAF, Arcetri Astrophysical Observatory, Largo Enrico Fermi 5, I-50125 Florence, Italy\\
$^{13}$Department of Physics, Van Yüzüncü Yıl University, Van 65080, T\"{u}rkiye\\
$^{14}$Max Planck Institut f\"{u}r Astronomie, K\"{o}nigstuhl 17, Heidelberg 69117, Germany \\
$^{15}$UK ALMA Regional Centre Node, Jodrell Bank Centre for Astrophysics, Department of Physics and Astronomy, The University of Manchester, Oxford Road, \\Manchester M13 9PL, UK \\
}

\date{Accepted 2026 January 28. Received 2025 December 30; in original form 2025 June 26}

\pubyear{2026}

\begin{document}
\label{firstpage}
\pagerange{\pageref{firstpage}--\pageref{lastpage}}
\maketitle

\begin{abstract}
Molecular gas is crucial to understanding
star formation and galaxy evolution, but the giant molecular clouds (GMCs) of early-type galaxies (ETGs) have rarely been studied.
Here, we present analyses of the spatially resolved GMCs of the lenticular galaxy NGC~1387, exploiting high spatial resolution ($0\farcs15$ or $14$~pc) $^{12}${\cotwo} line observations from the Atacama Large Millimeter/submillimeter Array.
We identify $1285$ individual GMCs and measure the fundamental properties (radius, velocity dispersion and molecular gas mass) of each with a modified version of the {\tt CPROPStoo} package.
Unusually for an ETG, the GMCs of NGC~1387 follow scaling relations very similar to those of the Milky Way disc and Local Group galaxy clouds, and most are virialised.
GMCs with large masses and radii and/or small galactocentric distances have their angular momenta aligned with the large-scale galactic rotation, while other GMCs do not.
{These results show that ETGs have more diversified GMC properties than previously thought. We discuss potential reasons for such diversity, and viewing-angle dependency is a plausible candidate.}
\end{abstract}



\section{Introduction}

Giant molecular clouds (GMCs) are the fuel for and the sites of star formation, and are thus key to galaxy evolution.
A thorough understanding of GMC properties in all types of galaxies is essential to effectively tackling questions regarding star formation, quenching, stellar feedback, turbulence, etc.\ {\citep[e.g.][]{2010MNRAS.406.2065H, 2011ApJ...729..133M, 2023A&A...672A.153S}}.
Despite the progress so far, most previous studies of molecular gas either have been confined to the Local Group or lack the resolution to spatially and spectrally resolve individual GMCs.
The former studies suffer from a limited diversity of galaxy types (exclusively late-type galaxies, LTGs, and dwarfs), while the latter studies lead to difficulties interpreting empirical relations derived from unresolved GMC populations. 
With the advent of the Atacama Large Millimeter/submillimeter Array (ALMA), resolved observations of GMCs beyond the Local Group have become feasible for a large number of galaxies. 

Early studies of the Milky Way (MW) GMCs established the so-called Larson {scaling} relations among GMC properties such as radius, velocity dispersion and molecular gas mass (or equivalently molecular gas mass surface density; \citealt{1981MNRAS.194..809L, 1987ApJ...319..730S}).
Subsequent studies within Local Group and other nearby galaxies reported similar {scaling} relations despite minor methodological differences \citep[e.g.][]{2007prpl.conf...81B, 2008ApJ...686..948B, 2013ApJ...772..107D, 2014ApJ...784....3C}.
Beyond the Local Group, \citet{2018ApJ...860..172S,2020ApJ...901L...8S} also reported similar {scaling relations among some of the same physical properties} for molecular gas probed at a spatial scale of $\approx100$~pc (without identifying individual clouds), using a large sample of LTGs as part of the Physics at High Angular resolution in Nearby GalaxieS (PHANGS) survey.

Studies of the MW centre (i.e.\ the central molecular zone, CMZ) and other galaxy centres in the Local Group revealed scaling relations different from those of the MW disc clouds, such as a steeper and higher (i.e.\ larger zero-point) size -- line width relation \citep[e.g.][]{2001ApJ...562..348O, 2005ApJ...623..826R, 2017MNRAS.468.1769F,2020ApJ...901L...8S}.
Beyond the Local Group, the first GMC catalogue of an early-type galaxy (ETG), NGC~4526 (distance $16.4$~Mpc), was created from Combined Array for Research in Millimeter-wave Astronomy observations by \citet{2015ApJ...803...16U}.
The properties of the NGC~4526 GMCs also deviate from those of MW disc clouds, with much larger line widths and luminosities (i.e.\ molecular gas masses) at a given radius and no size -- line width correlation.
\citet{2021MNRAS.505.4048L} studied another ETG with ALMA, NGC~4429 (distance $16.5$~Mpc), and the properties of its GMCs again deviate from those of the MW disc clouds, {with elevated GMC line widths at a given radius and elevated virial parameters}.
{Yet another ETG, NGC~524 (distance $18.7$~Mpc), was studied by \citet{2024MNRAS.531.3888L} at a physical resolution of $37$~pc, enough to marginally resolve large GMCs.
The very few identified GMCs also show elevated line widths compared to the MW disc cloud relation.
They concluded that galactic shear tears apart most of the GMCs.} 
These {three} galaxies are the only ETGs with GMC populations studied in a spatially (and spectrally) resolved manner so far, apart from NGC~5128 with its highly disturbed and externally accreted gas (\citealt{2021MNRAS.504.6198M}; see also \autoref{sec:mw_compare}).
It is thus natural to ask whether the GMC properties of all ETGs systematically differ from those implied by the Larson relations.

As part of the mm-Wave Interferometric Survey of Dark Object Masses (WISDOM) project \citep{2022MNRAS.512.1522D}, we study here another ETG with ALMA, NGC~1387.
{Given the small number of ETGs with resolved GMC studies as of today, any additional one can improve our understanding of GMC properties in ETGs. Moreover, the near-face-on orientation (inclination of $13\fdg6$; see \autoref{sec:data}) makes it special against previously studied ETGs. Its location near the Fornax Cluster centre indicates a probable internal origin of the molecular gas (see \autoref{sec:mw_compare}). Therefore, it is an ideal laboratory to test the influence of strong shear on the properties of GMCs when viewed near face-on.
}

We use the $^{12}${\cotwo} emission line (hereafter {\cotwo}) as a tracer of the molecular gas at a spatial resolution of $0\farcs15$ or $14$~pc.
In \autoref{sec:data}, we provide details of the target and our data and create a GMC catalogue using a modified {\tt CPROPStoo} package.
In \autoref{sec:property}, 
we follow previous studies {to measure the basic properties of the GMCs (size, velocity dispersion, molecular gas mass, etc.), check their consistency with the Larson relations of the MW, and contrast the GMC gaseous and virial masses.}
In \autoref{sec:kine}, we present a kinematic analysis constraining the origin of the internal rotation of the GMCs. {It is again the first time for such an analysis to be carried out for GMCs of a (near) face-on ETG.}
In \autoref{sec:toomre}, we calculate the radial profile of the Toomre instability parameter and discuss cloud-cloud collisions as potential regulators of GMC properties.
In \autoref{sec:gradient}, we discuss the steep galactocentric radial gradients of GMC properties observed, which appear unique to NGC~1387. 
In \autoref{sec:physics_discussion}, we discuss the {potential physical origin of the} similarities between the NGC~1387 and MW disc GMCs.
We summarise our results in \autoref{sec:sum}.


\section{DATA AND GMC IDENTIFICATION}
\label{sec:data}

\subsection{Target overview}
\label{sec:target}

NGC~1387 is a nearby galaxy at a distance of $19.3\pm0.8$~Mpc, as measured from surface brightness fluctuations \citep{2009ApJ...694..556B}.
At this distance, $1\arcsec$ corresponds to $\approx94$~pc.
In this work, we adopt a systemic velocity\footnote{We use barycentric velocities derived using the radio convention throughout this paper.} of $1286.4$~km~s$^{-1}$, as calculated in \autoref{sec:mom} and consistent within the galaxy redshift of $0.004345$ \citep{2019A&A...627A.136I}.
NGC~1387 has a morphological type of SB0 \citep{1981rsac.book.....S,1989AJ.....98..367F, 2007dvag.book.....B}. It has a prominent bulge with an effective (half-light) radius $R_\mathrm{e}=4\farcs4$ ($0.4$~kpc) at $K_\mathrm{s}$ band, a weak bar, a nuclear ring with a radius of $6\arcsec$ ($0.6$~kpc), a stellar disc extending from $\approx30\arcsec$ ($3$~kpc) to $\approx90\arcsec$ ($8$~kpc) from the centre and no spiral arm \citep[][]{2006AJ....132.2634L}.
Its total stellar mass is $4.70\times10^{10}$~M$_\odot$, derived using a colour-dependent $i$-band mass-to-light ratio of $1.31~(\mathrm{M/L})_{i,\odot}$ \citep{2019A&A...623A...1I}.
The effective radius of its entire stellar component is $50\farcs1$ ($4.7$~kpc) in the $B$ band \citep{1989AJ.....98..367F, 2018A&A...616A.121S} and $35\farcs5$ ($3.3$~kpc) in the $r$ band \citep{2019A&A...627A.136I}.
The stellar velocity dispersion reaches a maximum of $\approx200$~km~s$^{-1}$ at the galaxy centre,
while the luminosity-weighted average within the $r$-band effective radius is $143$~km~s$^{-1}$ \citep{2019A&A...627A.136I}.
NGC~1387 is a fast rotator \citep{2019A&A...627A.136I}.

\begin{figure*}
    \includegraphics[width=0.9\textwidth]{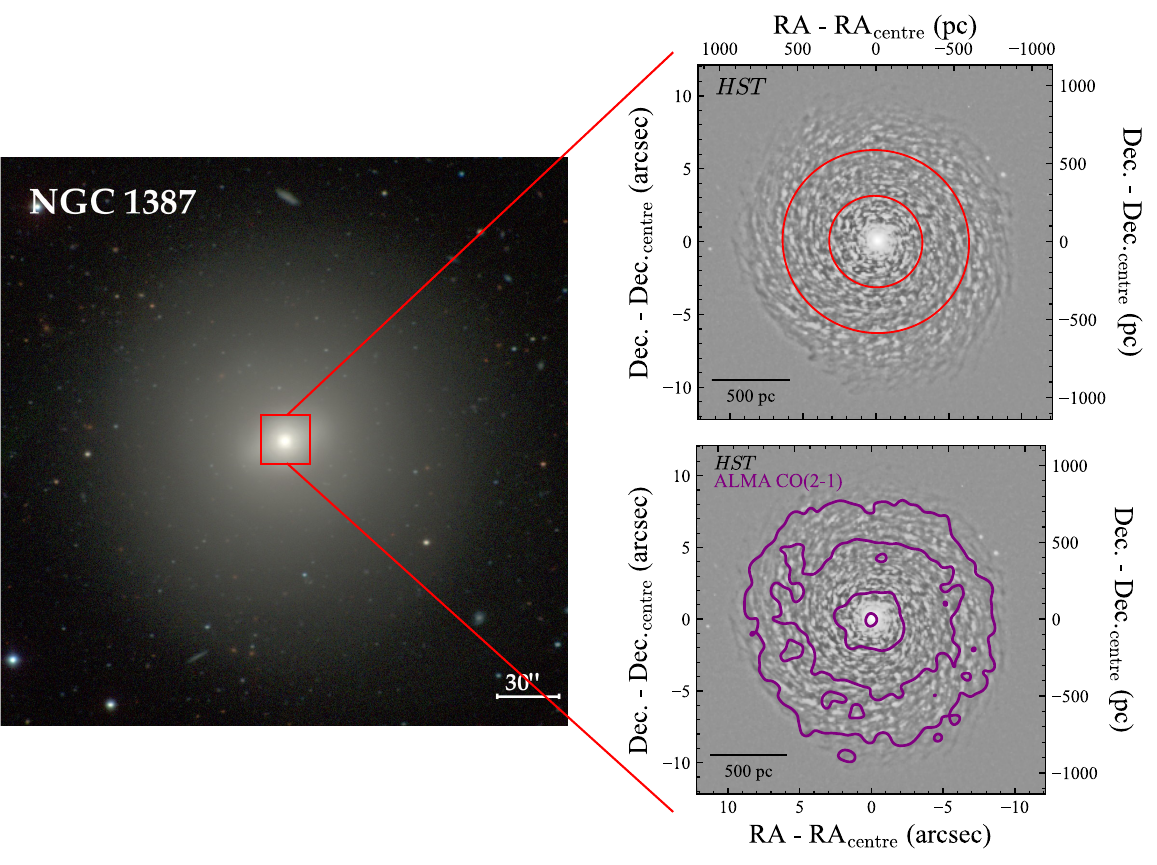}
    \caption{Overview of NGC~1387 and its molecular gas content. \textbf{Left:} optical image from the CGS survey.
    \textbf{Top-right:} unsharp-masked {\it HST} ACS/WFC F475W image of the central region only, highlighting dust features.
    The red ellipses overlaid indicate the boundaries of the three regions defined in \autoref{sec:mom} and discussed in the text.
    \textbf{Bottom-right:} the same unsharp-masked {\it HST} ACS/WFC F475W image, with the contours of the (smoothed) {\cotwo} total intensity map overlaid in purple (see \autoref{sec:co}).
    A scale bar is provided in each panel.}
    \label{fig:NGC1387_overview}
\end{figure*}

\autoref{fig:NGC1387_overview} shows an overview of NGC~1387 and its molecular gas content.
An optical image ($B$-, $V$- and $I$-band composite) from the Carnegie-Irvine Galaxy Survey\footnote{\url{https://cgs.obs.carnegiescience.edu/CGS/Home.html}} (CGS; \citealt{2011ApJS..197...21H}) is shown in the left panel, while the right panels show unsharp-masked images from the {\it Hubble Space Telescope} ({\it HST}) Advanced Camera for Surveys (ACS) Wide Field Channel (WFC) F475W filter (PI: Andres Jordan) created with \texttt{scikit-image}\footnote{\url{https://scikit-image.org}} \citep{2014PeerJ...2..453V}, with (bottom right) and without (top right) the {\cotwo} intensity contours overlaid (see Sections~\ref{sec:co} and \ref{sec:mom}).
Those images {confirm the findings of \citet{2019MNRAS.483.2251Z}}, that the molecular gas is in a regular disc perfectly co-spatial with a flocculent dust disc embedded within the bulge.
Both discs are co-spatial with a kinematically decoupled stellar core \citep{2019A&A...627A.136I}, have sharp edges and occupy a roughly circular region of $10\arcsec$ ($0.9$~kpc) radius. 
This radius is $\approx20\%$ of the $B$-band effective radius, typical of Virgo Cluster ETGs \citep{2013MNRAS.429..534D}.
The red ellipses in the top-right panel divide the galaxy into three regions based on the molecular gas mass surface density profile; these regions are used later for our GMC analyses (see \autoref{sec:mom}).

NGC~1387 (also known as FCC~184) belongs to the Fornax Cluster of galaxies and is {close to} the brightest cluster galaxy NGC~1399 {($17\arcmin$ away on the sky or $97~$kpc in projection)}.
Previous studies have found evidence of a gravitational interaction between the two galaxies in the stellar component (e.g.\ a stellar bridge and an overdensity of globular clusters; \citealt{2016ApJ...819L..31D, 2016ApJ...820...42I, 2018MNRAS.477.1880S}).
The cluster environment can significantly affect the cold gas properties of galaxies.
For example, recent observations revealed molecular gas discs smaller (once normalised by the size of the stellar component) than those of field galaxies, this for both LTGs \citep{2022ApJ...940..176V} and ETGs \citep{2013MNRAS.429..534D}.
This is interpreted as a loss of angular momentum and subsequent inward migration of molecular gas due to {interaction with the intracluster medium} \citep{2009ApJ...694..789T}, which can also cause steep or truncated radial profiles of molecular gas mass surface density \citep{2017MNRAS.467.4282M, 2022ApJ...933...10Z} and affect star formation \citep[e.g.][]{2016AJ....151...78P,2016MNRAS.456.4384M, 2018ApJ...866L..25V, 2022ApJ...940..176V, 2022ApJ...936..133C}.

The cold molecular gas of NGC~1387 was first studied by \citet{2019MNRAS.483.2251Z} using $^{12}${\coone} (hereafter {\coone}) ALMA observations at an angular resolution of $2\farcs8$.
This revealed a smooth, symmetric and regularly rotating molecular gas disc and several $3$-mm continuum sources due to nuclear activity and dust. 
\citet{2019MNRAS.483.2251Z} report an integrated {\coone} flux of $83.3\pm8.3$~Jy~km~s$^{-1}$ and we establish that the {\cotwo} disc has the same spatial extent as the {\coone} disc (see \autoref{sec:mom} and \autoref{fig:maps}).
We thus use {\coone} as the benchmark for the total molecular gas of NGC~1387. 
Throughout this paper, we adopt a common constant CO-to-molecular gas conversion factor $\alpha_\mathrm{CO(1-0)}=4.3$~$\mathrm{M_\odot~(K~km~s^{-1}~pc^2)^{-1}}$ (equivalent to a CO-to-H$_2$ conversion factor $X_\mathrm{CO(1-0)}=2\times10^{20}$~$\mathrm{cm^{-2}~(K~km~s^{-1})^{-1}}$; \citealt{2013ARA&A..51..207B}). 
{We note that this widely adopted $\alpha_\mathrm{CO(1-0)}$ and the commonly used uncertainty of 30\% are not well constrained in the environment of lenticular galaxy centres {(see \autoref{sec:alphaCO} for more details)}. We do not propagate this uncertainty into other gas properties of NGC~1387 because it is more likely to be a systematic error than a random error.}
Using equation~(3) of \citet{2013ARA&A..51..207B}, the total molecular gas mass is $(3.2\pm0.3)\times10^8$~M$_\odot$ or $0.7\%$ of the total stellar mass, similar to ETGs both inside and outside the Virgo Cluster \citep{2011MNRAS.414..940Y}.
The (inclination-corrected) mean molecular gas mass surface density of the molecular gas disc is $127 \pm 13$~$\mathrm{M_\odot~pc^{-2}}$. 

No atomic hydrogen (\ion{H}{i}) was detected in both the \ion{H}{i} Parkes All Sky Survey and using the Australia Telescope Compact Array, the latter providing the more stringent upper limit of $3\times10^7$~$\mathrm{M_\odot}$ \citep{2021A&A...648A..31L}. 
Optical spectra and emission-line ratio analyses indicate that the stars and the molecular gas of NGC~1387 are co-rotating, and that the centre of the galaxy is a low-ionisation nuclear emission-line region (\citealt{2019A&A...627A.136I, 2020MNRAS.496.2155Z}).
The star-formation rate (SFR) derived from the H$\alpha$ emission line luminosity is reported to be in the range of $0.008$ (considering only pure star-forming regions) to $0.082$~$\mathrm{M_\odot~yr^{-1}}$ (also including composite regions; \citealt{2019A&A...627A.136I}). 
At X-ray wavelengths (energy range $0.3$ -- $8$~keV), \citet{2011ApJS..192...10L} reported a point source associated with NGC~1387 with a luminosity of $3.22\times10^{39}$~$\mathrm{erg~s^{-1}}$.
The total $1.4$-GHz radio power is $1.8\times10^{20}$~W~Hz$^{-1}$ \citep{2011ApJ...731L..41B}. 

{Key parameters of NGC~1387 are summarised in \autoref{tab:N1387_para}.}

\begin{table}
    \centering
    \caption{{Key parameters of NGC~1387.}}
    \label{tab:N1387_para}
    \begin{tabular}{c|l|c}
    \hline \hline
    Row  & Quantity & Value  \\ \hline
    (1)  & RA  & $03^\mathrm{h}36^\mathrm{m}57\fs0333(1)$ \\
    (2)  & Dec. & $-35\degr30^\prime23\farcs6790(9)$ \\
    (3)  & Distance (Mpc) & $19.3 \pm 0.8$ \\
    (4)  & Linear scale of $1\arcsec$ (pc) & $94 \pm 4$ \\
    (5)  & Redshift & $0.004345(2)$ \\
    (6)  & Systemic velocity (km~s$^{-1}$) & $1286.4\pm0.4$ \\
    (7)  & Morphological type & SB0 \\
    (8)  & Inclination (degree) & $13\fdg6^{+0\fdg8}_{-0\fdg9}$\\
    (9)  & Position angle (degree) & $243\fdg9\pm0\fdg3$\\
    (10)  & $\log(M_*/{\rm M}_\odot)$ & $10.7 \pm 0.1$ \\
    (11)  & $S_{\rm CO(1-0)}$ (Jy~km~s$^{-1}$) & $83.3 \pm 8.3$ \\
    (12) & $\alpha_\mathrm{CO(1-0)} (\mathrm{M_\odot~(K~km~s^{-1}~pc^2)^{-1}})$ & $4.3$ \\
    (13) & $\log(M_{\rm gas, tot}/{\rm M}_\odot)$ & $8.51 \pm 0.04$ \\
    (14) & $<\Sigma_{\rm mol}> ({\rm M_\odot~pc^{-2}})$ & $127 \pm 13$ \\
    (15) & $M_{\rm \ion{H}{i} } ({\rm M}_\odot)$ & $<3\times10^7$ \\
    \hline \hline
    \end{tabular}
    \parbox{\columnwidth}{{Rows: (1) -- (2) J2000 equatorial coordinates determined from the central continuum source (see \autoref{sec:co}), with $1\sigma$ uncertainties of the last digits in brackets.
    (3) -- (4) Distance from \citet{2009ApJ...694..556B} and corresponding linear scale.
    (5) Redshift from \citet{2019A&A...627A.136I}, with uncertainty of the last digit in brackets.
    (6) Systemic velocity in the barycentric frame derived using the radio convention (see \autoref{sec:mom}).
    (7) Hubble morphological type reported in multiple sources  \citep[e.g.][]{1981rsac.book.....S, 1989AJ.....98..367F, 2007dvag.book.....B}.
    (8) -- (9) Inclination and position angle determined in \autoref{sec:hst} and \citet{2024MNRAS_dominiak}. 
    (10) Total stellar mass from \citet{2019A&A...623A...1I}.
    (11) Integrated {\coone} flux from \citet{2019MNRAS.483.2251Z}.
    (12) CO-to-molecular gas conversion factor adopted in this work \citep{2013ARA&A..51..207B}.
    (13) Total molecular gas mass derived using the quantities in rows 11 and 12 and equation~(3) of \citet{2013ARA&A..51..207B}.
    (14) Inclination-corrected mean molecular gas mass surface density of the molecular gas disc assuming a disc radius of $0.9$~kpc.
    (15) Upper limit ($3\sigma$) of the total atomic hydrogen mass \citep{2021A&A...648A..31L}.
    }
    }
\end{table}

\subsection{ALMA data}
\label{sec:co}

Observations were carried out with the ALMA $12$-m array under project 2016.1.00437.S (PI: Timothy Davis) on 4 Nov.\ 2016, 21 Dec.\ 2016 and 7 Sept.\ 2017, for a total on-source time of $20.2$~min.
Observations were also obtained with the $7$-m Atacama Compact Array (ACA) under project 2016.2.00053.S (PI: Lijie Liu) on 28 Jul.\ 2017, for a total on-source time of $17.1$~min.
These observations provide shorter baselines, therefore improving flux recovery.
The baselines range from $9$ to $7552$~m after combining all observations, with a maximum recoverable scale of $28\arcsec$ (2.7~kpc). 
For the $12$-m array, the ALMA correlator (band~6) was configured with one spectral window centred on the systemic velocity of the galaxy ({\cotwo} rest frequency of $230.538$~GHz), with a bandwidth of $1.875$~GHz ($\approx2450$~km~s$^{-1}$) and channels of $488$~kHz ($\approx0.6$~km~s$^{-1}$).
The remaining three spectral windows were used to observe the continuum, if any, each with a bandwidth of $2$~GHz and channels of $15.625$~MHz.
For the ACA, one spectral window was also centred on {\cotwo} but with a bandwidth of $2$~GHz ($\approx2610$~km~s$^{-1}$) and channels of $977$~kHz ($\approx1.3$~km~s$^{-1}$).
The remaining three windows for continuum detection again each had a bandwidth of $2$~GHz and channels of $15.625$~MHz.

Using the {\tt Common Astronomy Software Applications} ({\tt CASA}; \citealt{2022PASP..134k4501C}), we re-ran the standard calibration pipeline (version 4.7).
After masking out channels with emission, the continuum (assumed constant with frequency\footnote{{By imaging the upper sideband and the lower sideband of the ALMA data separately, we measure the spectral index to be $-0.03\pm1.09$, consistent with zero (although with large uncertainties). The assumption of a constant is thus valid in terms of the wavelength coverage and the quality of these ALMA data.}}) was fitted in the $uv$ domain and subtracted from each integration separately.
We then combined all tracks and used Briggs weighting \citep{1995AAS...18711202B} with a robustness parameter of $1.0$ and no taper nor clipping to image the data.
This was chosen to achieve a physical resolution better than the characteristic size of GMCs while maintaining adequate sensitivity.
The final data cube has a synthesised beam with major and minor axes of $\theta_\mathrm{maj}\times\theta_\mathrm{min}=0\farcs17\times0\farcs14$ (or $16\times13$~pc$^2$) full-widths at half-maximum (FWHM), with a position angle of $89\fdg6$ (measured from north through east).
We adopt a $0\farcs04$ pixel size to sufficiently sample the synthesised beam. 
We also bin the raw data spectrally to obtain a final $2$~km~s$^{-1}$ channel width, sufficient to spectrally resolve most GMCs. 
Interactive cleaning was carried out with the Hogbom algorithm \citep{1974A&AS...15..417H} implemented in {\tt CASA} to a depth of $1.5$~mJy~beam$^{-1}$ (slightly larger than the root-mean-square, RMS, noise level; see below).
In each channel, an emission mask used for cleaning (denoted `mask$_{\rm clean}$')
was generated by the automatic masking algorithm {\tt AUTO-MULTITHRESH} \citep{Kepley:2020}.
The resulting cube (hereafter the `uncorrected' cube) has a uniform RMS noise $\sigma_\mathrm{RMS}=1.11$~mJy~beam$^{-1}$ ($1.12$~K in brightness temperature units) per $2$~km~s$^{-1}$ channel. Primary beam (PB) correction was then applied, producing our final `corrected' cube. The fields of view of both cubes are $\approx30\arcsec$, corresponding to the diameter of the area with a PB response $\ge0.2$ at the observed frequency. 
{Key information of the ALMA observations and the CO(2-1) data cube properties are summarised in \autoref{tab:alma_co}, while \citet{2024MNRAS_dominiak} provide more information on individual observation tracks.}

\begin{table}
    \centering
    \caption{{ALMA CO(2-1) data cube properties.}}
    \label{tab:alma_co}
    \begin{tabular}{l|c}
    \hline \hline
    Synthesised beam size  &  $0\farcs17\times0\farcs14$ ($16\times13$~pc$^2$) \\
    Pixel scale (arcsec pix$^{-1}$) & 0.04 \\
    Channel width (km~s$^{-1}$) &  2 \\
    $T_{\rm CO(2-1)} / I_{\nu,\rm CO(2-1)}$ (K~(Jy~beam$^{-1}$)$^{-1}$) & 1015 \\
    RMS noise (K~channel$^{-1}$) & 1.12 \\ 
    CO(2-1) integrated flux (Jy~km~s$^{-1}$) & $255\pm3$ \\
    CO line ratio $T_{\rm CO(2-1)} / T_{\rm CO(1-0)}$ (K/K) & $0.77\pm0.11$ \\
    \hline \hline
    \end{tabular}
    \parbox{\columnwidth}{{Note: $T$ is the brightness temperature and $I_{\nu}$ the specific intensity.}
    }
\end{table}

The continuum emission was also imaged (and PB corrected). 
A point source is detected at the centre of the galaxy, most likely a low-luminosity active galactic nucleus (AGN). 
{A Gaussian fit with the {\tt CASA} task {\tt imfit} yields} an integrated flux of $1.19\pm0.04$~mJy {(\citealt{2024MNRAS.528..319E} reported $1.00\pm0.05$~mJy using a different measurement method)} and a position of $\mathrm{RA}=03^\mathrm{h}36^\mathrm{m}57{\fs}0333 \pm 0{\fs}0001$, $\mathrm{Dec.}=-35\degr30\arcmin23\farcs6790 \pm 0\farcs0009$ (J2000). 
We adopt this position as the centre of the galaxy.
There is also an extended continuum source $10\farcs3$ ($970$~pc) north-west of the galaxy centre. 
{The continuum image and some properties of this second continuum source are presented in \aref{sec:app_continuum}.}

\subsection{{\it HST} image and stellar mass distribution}
\label{sec:hst}

We model the stellar mass distribution of NGC~1387 using a multi-Gaussian expansion (MGE; \citealt{Emsellem:1994}) constructed using the {\tt MGE\_FIT\_SECTORS} package\footnote{\url{https://www-astro.physics.ox.ac.uk/~cappellari/software/\#mge}} of \citet{Cappellari:2002}.
We use an {\it HST} Wide Field Camera~3 F160W-filter
image (PI: Benjamin Boizelle) and a constant stellar mass-to-light ratio of $0.98$~(M/L)$_\mathrm{F160W,\odot}$, derived by \citet{2024MNRAS_dominiak} by fitting the CO gas kinematics with the publicly available package {\tt Kinematic Molecular Simulation} ({\tt KinMS}\footnote{\url{https://github.com/TimothyADavis/KinMS}}; \citealt{2013MNRAS.429..534D}).
The best-fitting inclination is $13\fdg6 ^{+0\fdg8}_{-0\fdg9}$ and the best-fitting position angle $243\fdg9\pm0\fdg3$. 
The resulting three-dimensional (3D) stellar-mass model and one-dimensional circular velocity curve ($v_\mathrm{circ}(R_\mathrm{gal})$, where $R_\mathrm{gal}$ is the galactocentric distance) will be used in the kinematic study of the GMCs (\autoref{sec:kine}) and the Toomre parameter calculation (\autoref{sec:toomre}).

\subsection{Moment maps and region definition}
\label{sec:mom}

\begin{figure*}
    \centering
        \begin{subfigure}[c]{\textwidth}
            \centering
                \includegraphics[width=\textwidth]{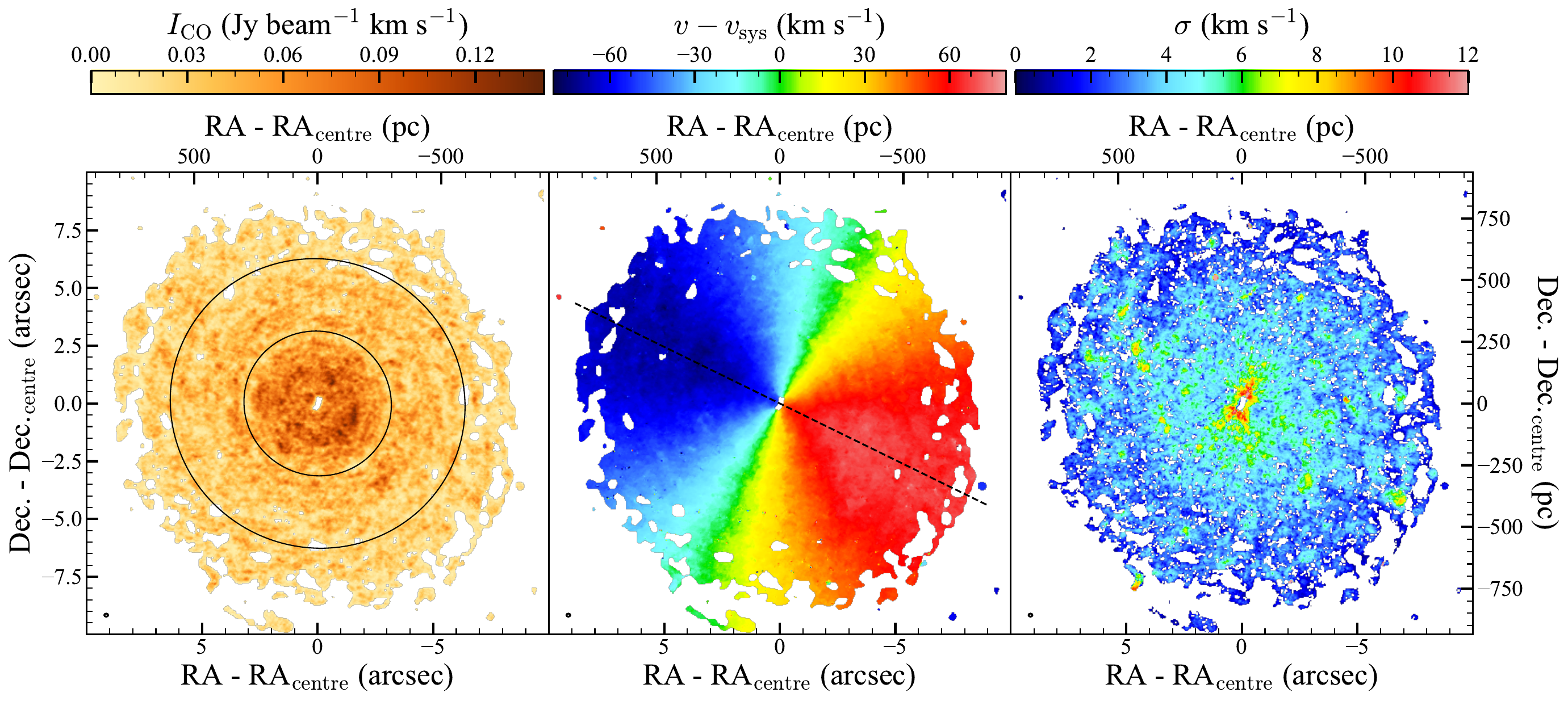} 
        \end{subfigure}
        \begin{subfigure}[c]{0.42\textwidth}
            \centering
                \includegraphics[width=\textwidth]{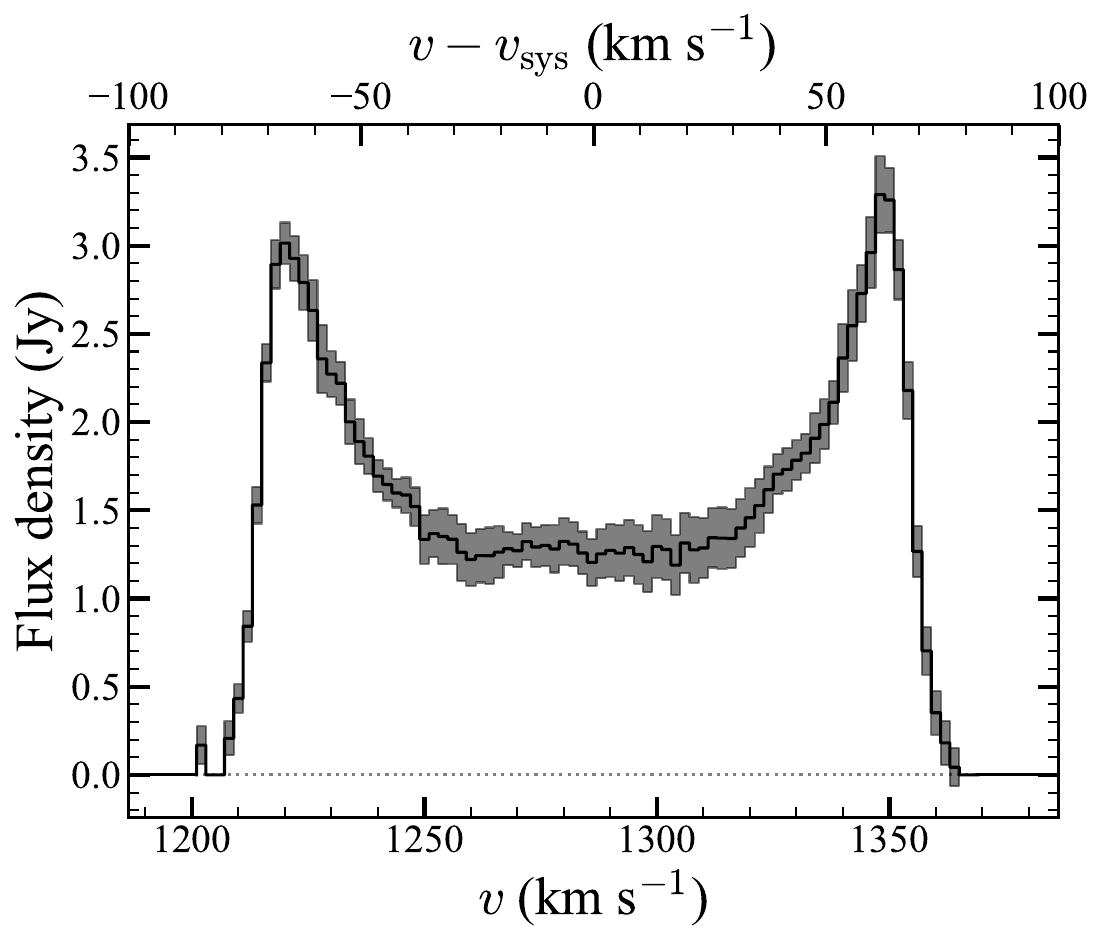} 
        \end{subfigure}
        \hspace{0.5cm}
        \begin{subfigure}[c]{0.45\textwidth}
            \centering
                \includegraphics[width=\textwidth]{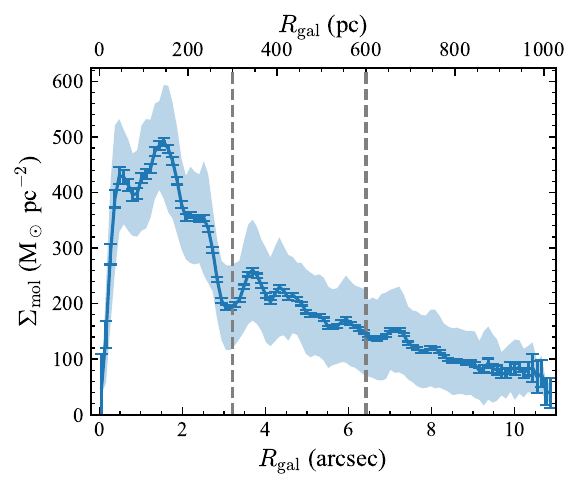} 
        \end{subfigure}
    \caption{Molecular gas distribution of NGC~1387. {\bf Top:} {\cotwo} zeroth (integrated-intensity), first (intensity-weighted mean line-of-sight, LoS, velocity) and second (intensity-weighted LoS velocity dispersion) moment maps.
    The black ellipses in the left panel indicate the boundaries of the three regions defined in \autoref{sec:mom}.
    The dashed line in the middle panel shows the kinematic major axis.
    The synthesised beam is shown in the bottom-left corner of each panel as a black ellipse.
    {\bf Bottom-left:} spatially integrated {\cotwo} spectrum, corrected to an infinite cleaning depth channel by channel (see \autoref{sec:mom}).
    Uncertainties ($1$~$\sigma$) are shown as grey shading.
    {\bf Bottom-right:} azimuthally averaged inclination-corrected molecular gas mass surface density radial profile.
    Error bars indicate the $1$~$\sigma$ uncertainty on the mean of each radial bin, while the blue shaded region shows the scatter (standard deviation) within each radial bin.
    The grey dashed vertical lines again indicate the boundaries of the three regions defined in \autoref{sec:mom}.
    }
    \label{fig:maps}
\end{figure*}

We use a moment-masking technique inspired by \citet{2011arXiv1101.1499D} and {adapted from} the python package \texttt{pymakeplots}\footnote{The public version is available here: \url{https://pypi.org/project/pymakeplots/}. {Our adapted version is made available at: \url{https://github.com/ericliang45/NGC1387-GMC/blob/main/python_output/pymakeplots.py}.}}
to generate the moment maps (see \autoref{fig:maps}). 
We start by defining the line-free channels to be those channels outside of the $1200$ -- $1370$~km~s$^{-1}$ velocity range, and the line-free spaxels to be those outside a $20\arcsec\times20\arcsec$ box centred on the galaxy centre, based on visual inspection of the data cube.
We then smooth the uncorrected cube (i.e.\ the cube before PB correction) with a 3D Gaussian kernel. The kernel's FWHM in the two spatial dimensions are both twice that of the synthesised beam major axis and the standard deviation in the spectral dimension is two channels.
Because of smoothing, real emission should remain significant while noise should decrease.
We then create a mask (denoted `mask$_{\rm RMS}$') that includes all the pixels of this smoothed uncorrected cube with an amplitude greater than five times the RMS noise measured in the line-free channels of this same cube. We then only select the pixels from the (unsmoothed) corrected cube within this mask to create the moment maps. Maps of the moment uncertainties are created using the python package {\tt Uncertainties}\footnote{\url{https://pythonhosted.org/uncertainties/}}
and standard error propagation.

In the top-left panel of \autoref{fig:maps}, the zeroth-moment (integrated-intensity) map shows that the large-scale molecular gas distribution has a high filling factor and the molecular gas disc is fragmented into a very large number of small clumps.
This morphology is reminiscent of the flocculent spiral arms seen at optical wavelengths (\autoref{fig:NGC1387_overview}).
There is also a small central hole ($\approx0\farcs2$ in {deprojected} radius).
In the top-middle panel of \autoref{fig:maps}, the first-moment (intensity-weighted mean line-of-sight, LoS, velocity) map reveals an extremely regularly rotating molecular gas disc, with no {evident} non-circular motions.
In the top-right panel of \autoref{fig:maps}, the second-moment (intensity-weighted LoS velocity dispersion) map shows that the molecular gas velocity dispersion gradually decreases away from the galaxy centre.
{The median of the velocity dispersion is $3.2\pm0.5$~km~s$^{-1}$.}
The enhancement along the minor axis and some of the central peak are however most likely due to beam smearing.
{Despite} the central hole, the morphology and kinematics of the molecular gas disc make it {feasible} to measure the mass of a putative central supermassive black hole, as done by \citet{2024MNRAS_dominiak}, where the position-velocity diagram of the molecular gas disc is also presented. 

To accurately construct a synthesised integrated spectrum and recover the total flux, we apply the formula in Appendix~A.2 of \citet{1995AJ....110.2037J}.
This corrects for the finite cleaning depth of our data, which causes a bias in the total flux of each channel due to the difference between the dirty synthesised beam and the (Gaussian) clean synthesised beam.
Using a data cube cleaned to a depth of $3.0$~mJy~beam$^{-1}$ in addition to our adopted data cube, we apply the aforementioned formula to each channel to derive the total flux at infinite cleaning depth. 
Using the python routine {\tt scipy.ndimage.binary\_dilation},\footnote{\url{https://docs.scipy.org/doc/scipy/reference/generated/scipy.ndimage.binary_dilation.html}} we also gradually enlarge mask$_{\rm RMS}$ to encompass more faint diffuse emission, until the derived total flux stops increasing. This happens at an enlargement of $\approx40$ pixels ($\approx11$~beams or $\approx150$~pc).
Due to the masking scheme, the uncertainty of the integrated flux of each channel is calculated in a rather complicated manner: for each line channel, we take the associated two-dimensional (2D) mask, independently sum every line-free channel of the corrected cube over that mask, calculate the standard deviation of those sums, and adopt that standard deviation as the uncertainty of the integrated flux of that line channel.

The resulting synthesised integrated spectrum (corrected to an infinite cleaning depth channel by channel) is shown in the bottom-left panel of \autoref{fig:maps}.
It reveals the classic `double-horn' shape of a rotating disc. 
The total spatially integrated {\cotwo} flux (also corrected to an infinite cleaning depth) is $255\pm3$~Jy~km~s$^{-1}$ (assuming standard error propagation for a sum). The absolute flux calibration of ALMA is however only accurate to $\approx 10\%$, translating to an additional systematic error of $\approx10\%$ on every channel and thus on the integrated flux.
Using the {\coone} integrated flux of \citet{2019MNRAS.483.2251Z}, we derive a {\cotwo}/{\coone} line ratio of $0.77\pm0.11$ in brightness temperature units (the error has included the systematic uncertainty of flux calibration). 
This is consistent with the average ratio of nearby massive disc galaxies within $1$~$\sigma$ ($0.50$ -- $0.83$; \citealt{2022ApJ...927..149L}).
{This line ratio yields an effective $\alpha_\mathrm{CO(2-1)}=5.6$~$\mathrm{M_\odot~(K~km~s^{-1}~pc^2)^{-1}}$, which can be used to directly convert CO(2-1) flux to molecular gas mass.}
The flux-weighted mean LoS velocity calculated from this spectrum, $1286.4\pm0.4$~$\mathrm{km~s^{-1}}$, is adopted as the systemic velocity of NGC~1387.

The azimuthally averaged inclination-corrected molecular gas mass surface density radial profile $\Sigma_\mathrm{mol}(R_\mathrm{gal})$ is shown in the bottom-right panel of \autoref{fig:maps}.
The error bars indicate the $1$~$\sigma$ uncertainty on the mean in each radial bin (propagated from the zeroth-moment map), while the blue-shaded region shows the scatter (standard deviation) within each radial bin.
The central hole is reflected by the sharp drop in the central few radial bins.

Molecular gas and GMCs in different regions of the same galaxy can exhibit significantly different properties.
In NGC~1387, we thus divide the molecular gas disc into three regions: an inner region ($R_\mathrm{gal}<300$~pc), an intermediate region ($300\le R_\mathrm{gal}<600$~pc) and an outer region ($R_\mathrm{gal}\ge600$~pc).
These boundaries are based on a visual inspection of the zeroth-moment map and reflect significant changes {(especially for the inner boundary)} of the molecular gas mass surface density radial profile.
Although the exact location of the second boundary at $600$~pc is somewhat arbitrary, it does not affect the results discussed in the following sections.

\subsection{GMC identification}
\label{sec:cprops}

With the corrected cube as input, we adopt a modified version of the \texttt{CPROPStoo} package \citep{2006PASP..118..590R, 2021MNRAS.505.4048L} for GMC identification and property measurements.
{{\tt CPROPStoo} first estimates a 2D noise map ($\sigma_{\rm RMS,local}$) of the cube, mainly reflecting the ALMA primary beam.}
It then uses a mask to identify regions of real emission, for which we adopt a slightly manually enlarged version of mask$_{\rm clean}$ encompassing faint diffuse emission. 
Within this mask, \texttt{CPROPStoo} identifies pixels above $2$~$\sigma_\mathrm{RMS,local}$, accretes neighbouring pixels above $1.5$~$\sigma_\mathrm{RMS,local}$ and requires those regions to have an area larger than $32$ pixels (roughly twice the synthesised beam area) to define `islands' considered to be real emission.
These parameters were selected to ensure that almost all diffuse emission is retained while avoiding noise.
As the channel width of $2$~$\mathrm{km~s^{-1}}$ is comparable to the smallest line width of the molecular gas in the disc (see \autoref{fig:maps}), we do not require any minimum number of channels for island identification.

Within these islands, \texttt{CPROPStoo} identifies local maxima and their uniquely associated spaxels (i.e.\ pixels brighter than the merging contour level with adjacent local maxima) as the seeds of GMCs.
It then compares these maxima and associated spaxels with a minimum area, a minimum spectral width, a minimum contrast $\Delta T_\mathrm{min}$ between the highest and the lowest intensity and a minimum convexity $S_\mathrm{min}$ (enforced to prevent individual GMCs from having too much sub-structure; \citealt{lin2003hybrid}).
A more detailed description of the methodology can be found in \citet{2021MNRAS.505.4048L}.
To avoid a significant bias toward small structures when a small minimum area is used, the local maximum search cycles through a range of minimum areas from $160$ ($\approx10$ synthesised beam areas) to $16$ ($\approx1$ synthesised beam area) spaxels, with a step size of $16$ spaxels.
The remaining spaxels in the islands (i.e.\ those not uniquely associated with a single local maximum) are assigned to clouds using the `friend-of-friend' {\tt CLUMPFIND} algorithm \citep{2006PASP..118..590R,2021MNRAS.505.4048L}.
In the smallest minimum area search cycle (i.e.\ $16$ spaxels), the aforementioned cloud criteria are checked after the remaining spaxel assignment, while during previous cycles, the criteria are checked using only uniquely associated spaxels.
We set the minimum spectral width to be $2$ channels, the minimum contrast to be $2$~K ($\approx1.8$~$\sigma_\mathrm{RMS}$) and the minimum convexity to be $0.50$.

These parameters were chosen via trial and error.
After each trial, we examined the GMCs identified to verify that most had reasonable structures and most over-densities were indeed identified as GMCs.
Different GMC identification results (with different $\Delta T_\mathrm{min}$ and $S_\mathrm{min}$) are presented in \aref{sec:robustness}, confirming that all reasonable sets of parameters yield similar GMC properties and do not significantly alter the results discussed in the following sections.

\begin{figure*}
    \includegraphics[width=0.8\textwidth]{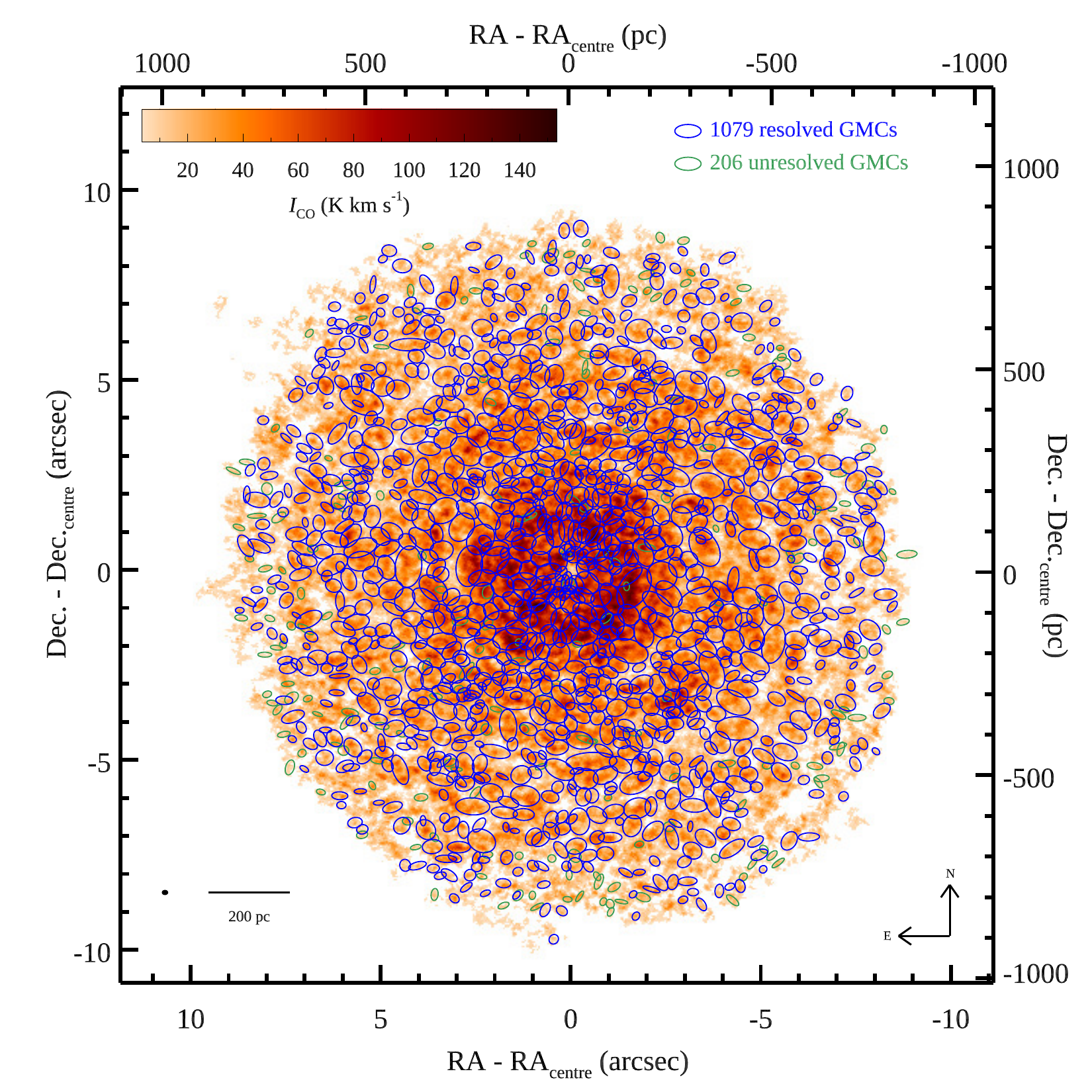}
    \caption{NGC~1387 GMCs identified by \texttt{CPROPStoo}, schematically overlaid on the zeroth-moment map as ellipses.
    There are $1079$ spatially and spectrally resolved GMCs (blue ellipses) and $206$ unresolved GMCs (green ellipses). 
    The synthesised beam is shown in the bottom-left corner as a black filled ellipse.
    A scale bar is also shown in the bottom-left corner.}
    \label{fig:cprops_identification}
\end{figure*}

The final GMC catalogue is visualised in \autoref{fig:cprops_identification}, where the $1079$ spatially and spectrally resolved GMCs and the $206$ unresolved GMCs are schematically overlaid on the zeroth-moment map.
For a GMC to be considered resolved, both its deconvolved radius and its deconvolved line width (see \autoref{sec:property}) must be larger than the instrumental resolution (half of the synthesised beam size and half of the channel width, respectively).

\section{GMC properties}
\label{sec:property}

We measure the properties of each GMC in the same manner as \citet{2021MNRAS.505.4048L} and references therein \citep[e.g.][]{2006PASP..118..590R}.
Briefly, the GMC radius $R_\mathrm{c}$ is the geometric mean of the deconvolved intensity-weighted second moment in each direction ($\sigma_\mathrm{x}$, $\sigma_\mathrm{y}$), scaled up by {the empirically calibrated} factor $\eta=1.9$ \citep{1987ApJ...319..730S,2006PASP..118..590R}.
The GMC line width $\sigma_\mathrm{obs,los}$ is measured from the stacked spectrum of each GMC and is deconvolved to account for the finite channel width.
The GMC molecular gas mass $M_\mathrm{gas}$ is calculated from the total {\cotwo} luminosity ($L_\mathrm{CO(2-1)}$) of each cloud assuming the aforementioned global {\cotwo}/{\coone} line ratio, a standard CO-to-molecule conversion factor $\alpha_\mathrm{CO(1-0)}=4.3$~$\mathrm{M_\odot~(K~km~s^{-1}~pc^2)^{-1}}$ (yielding an effective $\alpha_\mathrm{CO(2-1)}=5.6$~$\mathrm{M_\odot~(K~km~s^{-1}~pc^2)^{-1}}$) and our adopted distance to NGC~1387.
{The uncertainty of the $\alpha_\mathrm{CO(1-0)}$ factor is further discussed in \autoref{sec:alphaCO}.}
The molecular gas mass surface density of each GMC is simply $\Sigma_\mathrm{gas}\equiv M_\mathrm{gas}/\pi R_\mathrm{c}^2$.
It is not corrected for the cloud's inclination angle as a spherical geometry is assumed {for each individual cloud}.
An extrapolation is applied when measuring $\sigma_\mathrm{x}$, $\sigma_\mathrm{y}$, $\sigma_\mathrm{obs,los}$ and $L_\mathrm{CO(2-1)}$, whereby each quantity is measured as a function of the intensity threshold and is extrapolated to zero intensity (equivalent to having infinite sensitivity).
The uncertainty of each quantity is derived from $1000$ bootstrap realisations of the pixels of each GMC.

The GMCs identified and their measured properties are listed in \autoref{tab:gmc}.
For each GMC, additional properties listed include the (intensity-weighted) central position, (intensity-weighted barycentric) systemic velocity ($V_\mathrm{bc}$), 
brightness temperature of the brightest pixel ($T_\mathrm{max}$), internal velocity gradient ($\omega_\mathrm{obs}$), projected position angle of the angular momentum vector ($\phi_\mathrm{rot}$) and (deprojected) galactocentric distance ($R_\mathrm{gal}$), where the kinematic parameters $\omega_\mathrm{obs}$ and $\phi_\mathrm{rot}$ will be discussed in \autoref{sec:kine}.
{The $1$~$\sigma$ uncertainties of all quantities are also listed in \autoref{tab:gmc} and used for the rest of this paper.}
The uncertainty of the galaxy distance $D$ is not propagated into the tabulated uncertainties of the measured properties, as an error on the distance translates to a systematic (rather than random) scaling of some quantities (no effect on the others), i.e.\ $R_\mathrm{c}\propto D$, $L_\mathrm{CO(2-1)}\propto D^2$, $M_\mathrm{gas}\propto D^2$, $\omega_\mathrm{obs}\propto D^{-1}$ and $R_\mathrm{gal}\propto D$.

\begin{table*}
    \begin{center}
    \caption{NGC~1387 GMC catalogue.}
    \label{tab:gmc}
    \resizebox{\textwidth}{!}{
    \begin{tabular}{cccccccccccc}
        \hline
        \hline
        ID & RA & Dec. & $V_\mathrm{bc}$ & $R_\mathrm{c}$ & $\sigma_\mathrm{obs,los}$ & $L_\mathrm{CO(2-1)}$ & $M_\mathrm{gas}$ & $T_\mathrm{max}$ & $\omega_\mathrm{obs}$ & $\phi_\mathrm{rot}$ & $R_\mathrm{gal}$\\
         & (h:m:s) & ($\degr:\arcmin:\arcsec$) & (km~s$^{-1}$) & (pc) & (km~s$^{-1}$) & ($10^4$~K~km~s$^{-1}$~pc$^2$) & ($10^5$~M$_\odot$) & (K) & (km~s$^{-1}$~pc$^{-1}$) & (degree) & (pc)\\
        \hline
        $\phantom{000}1$ & $3$:$36$:$57.252$ & $-35$:$30$:$21.60$ & $\phantom{}$$1208.5\phantom{}$ & $\phantom{}$$14\phantom{} \pm \phantom{}26\phantom{}$ & $\phantom{}$$1\phantom{.0} \pm \phantom{}2\phantom{.0}$ & $\phantom{}$$1\phantom{.0} \pm \phantom{}2\phantom{.0}$ & $\phantom{}$$0.6\phantom{} \pm \phantom{}0.9\phantom{}$ & $\phantom{}$$5.5\phantom{}$ & $\phantom{}$$0.13\phantom{0} \pm \phantom{}0.02\phantom{0}$ & $\phantom{-0}$$21\phantom{} \pm \phantom{}16\phantom{}$ & $\phantom{}$$316\phantom{}$ \\
        $\phantom{000}2$ & $3$:$36$:$57.242$ & $-35$:$30$:$21.29$ & $\phantom{}$$1208.7\phantom{}$ & $\phantom{}$$12\phantom{} \pm \phantom{}12\phantom{}$ & $\phantom{}$$3\phantom{.0} \pm \phantom{}3\phantom{.0}$ & $\phantom{}$$1.1\phantom{} \pm \phantom{}0.5\phantom{}$ & $\phantom{}$$0.6\phantom{} \pm \phantom{}0.3\phantom{}$ & $\phantom{}$$5.4\phantom{}$ & $\phantom{}$$0.08\phantom{0} \pm \phantom{}0.02\phantom{0}$ & $\phantom{-}$$134\phantom{} \pm \phantom{}14\phantom{}$ & $\phantom{}$$328\phantom{}$ \\
        $\phantom{000}3$ & $3$:$36$:$57.519$ & $-35$:$30$:$21.70$ & $\phantom{}$$1211.5\phantom{}$ & $\phantom{}$$11\phantom{} \pm \phantom{}15\phantom{}$ &                            -                            & $\phantom{}$$1.4\phantom{} \pm \phantom{}0.3\phantom{}$ & $\phantom{}$$0.8\phantom{} \pm \phantom{}0.2\phantom{}$ & $\phantom{}$$8.4\phantom{}$ &                              -                              &                            -                            & $\phantom{}$$585\phantom{}$ \\
        $\phantom{000}4$ & $3$:$36$:$57.500$ & $-35$:$30$:$21.11$ & $\phantom{}$$1212.0\phantom{}$ & $\phantom{}$$10\phantom{} \pm \phantom{}42\phantom{}$ & $\phantom{}$$3\phantom{.0} \pm \phantom{}3\phantom{.0}$ & $\phantom{}$$1.1\phantom{} \pm \phantom{}0.7\phantom{}$ & $\phantom{}$$0.6\phantom{} \pm \phantom{}0.4\phantom{}$ & $\phantom{}$$4.6\phantom{}$ & $\phantom{}$$0.10\phantom{0} \pm \phantom{}0.02\phantom{0}$ & $\phantom{-0}$$57\phantom{} \pm \phantom{}15\phantom{}$ & $\phantom{}$$584\phantom{}$ \\
        $\phantom{000}5$ & $3$:$36$:$57.663$ & $-35$:$30$:$22.30$ & $\phantom{}$$1213.1\phantom{}$ & $\phantom{}$$12\phantom{} \pm \phantom{}31\phantom{}$ &                            -                            & $\phantom{}$$2\phantom{.0} \pm \phantom{}3\phantom{.0}$ & $\phantom{}$$1.2\phantom{} \pm \phantom{}1.5\phantom{}$ & $\phantom{}$$5.2\phantom{}$ &                              -                              &                            -                            & $\phantom{}$$732\phantom{}$ \\
        $\phantom{000}6$ & $3$:$36$:$57.374$ & $-35$:$30$:$21.95$ & $\phantom{}$$1214.7\phantom{}$ & $\phantom{0}$$7\phantom{} \pm \phantom{}16\phantom{}$ & $\phantom{}$$4\phantom{.0} \pm \phantom{}2\phantom{.0}$ & $\phantom{}$$2.1\phantom{} \pm \phantom{}0.6\phantom{}$ & $\phantom{}$$1.2\phantom{} \pm \phantom{}0.3\phantom{}$ & $\phantom{}$$6.6\phantom{}$ & $\phantom{}$$0.075\phantom{} \pm \phantom{}0.008\phantom{}$ & $\phantom{-0}$$14\phantom{} \pm \phantom{0}9\phantom{}$ & $\phantom{}$$422\phantom{}$ \\
        $\phantom{000}7$ & $3$:$36$:$57.444$ & $-35$:$30$:$21.71$ & $\phantom{}$$1213.9\phantom{}$ &                           -                           & $\phantom{}$$1.0\phantom{} \pm \phantom{}1.3\phantom{}$ & $\phantom{}$$0.6\phantom{} \pm \phantom{}0.7\phantom{}$ & $\phantom{}$$0.3\phantom{} \pm \phantom{}0.4\phantom{}$ & $\phantom{}$$4.4\phantom{}$ &                              -                              &                            -                            & $\phantom{}$$503\phantom{}$ \\
        $\phantom{000}8$ & $3$:$36$:$57.474$ & $-35$:$30$:$21.20$ & $\phantom{}$$1215.1\phantom{}$ & $\phantom{}$$16\phantom{} \pm \phantom{}13\phantom{}$ & $\phantom{}$$3\phantom{.0} \pm \phantom{}2\phantom{.0}$ & $\phantom{}$$3\phantom{.0} \pm \phantom{}2\phantom{.0}$ & $\phantom{}$$1.9\phantom{} \pm \phantom{}1.0\phantom{}$ & $\phantom{}$$8.7\phantom{}$ & $\phantom{}$$0.125\phantom{} \pm \phantom{}0.008\phantom{}$ & $\phantom{0}$$-49\phantom{} \pm \phantom{0}5\phantom{}$ & $\phantom{}$$554\phantom{}$ \\
        $\phantom{000}9$ & $3$:$36$:$57.428$ & $-35$:$30$:$21.09$ & $\phantom{}$$1214.6\phantom{}$ & $\phantom{}$$14\phantom{} \pm \phantom{0}5\phantom{}$ & $\phantom{}$$5\phantom{.0} \pm \phantom{}2\phantom{.0}$ & $\phantom{}$$3.1\phantom{} \pm \phantom{}0.6\phantom{}$ & $\phantom{}$$1.7\phantom{} \pm \phantom{}0.4\phantom{}$ & $\phantom{}$$7.8\phantom{}$ & $\phantom{}$$0.107\phantom{} \pm \phantom{}0.007\phantom{}$ & $\phantom{-0}$$12\phantom{} \pm \phantom{0}4\phantom{}$ & $\phantom{}$$511\phantom{}$ \\
        $\phantom{00}10$ & $3$:$36$:$57.487$ & $-35$:$30$:$18.62$ & $\phantom{}$$1213.5\phantom{}$ & $\phantom{}$$10\phantom{} \pm \phantom{}20\phantom{}$ & $\phantom{}$$1.8\phantom{} \pm \phantom{}1.4\phantom{}$ & $\phantom{}$$1.1\phantom{} \pm \phantom{}1.2\phantom{}$ & $\phantom{}$$0.6\phantom{} \pm \phantom{}0.7\phantom{}$ & $\phantom{}$$6.6\phantom{}$ & $\phantom{}$$0.06\phantom{0} \pm \phantom{}0.02\phantom{0}$ & $\phantom{-}$$142\phantom{} \pm \phantom{}24\phantom{}$ & $\phantom{}$$703\phantom{}$ \\
        ... & ... & ... & ... & ... & ... & ... & ... & ... & ... & ... & ...\\
        $\phantom{}1285$ & $3$:$36$:$56.827$ & $-35$:$30$:$27.01$ & $\phantom{}$$1361.7\phantom{}$ &                           -                           & $\phantom{}$$2\phantom{.0} \pm \phantom{}2\phantom{.0}$ & $\phantom{}$$2\phantom{.0} \pm \phantom{}2\phantom{.0}$ & $\phantom{}$$1.0\phantom{} \pm \phantom{}0.9\phantom{}$ & $\phantom{}$$5.4\phantom{}$ &                              -                              &                            -                            & $\phantom{}$$394\phantom{}$ \\
        \hline
        \hline
    \end{tabular}
    }
    \end{center}
    \parbox{\textwidth}{Notes:
    Measurements of $M_\mathrm{gas}$ assume a fixed {\cotwo}/{\coone} line ratio of $0.77\pm0.08$ (in brightness temperature units) and a standard CO-to-molecule conversion factor $\alpha_\mathrm{CO(1-0)}=4.3$~$\mathrm{M_\odot~(K~km~s^{-1}~pc^2)^{-1}}$.
    {The uncertainties of all quantities are quoted at the $1$~$\sigma$ level.}
    The uncertainty of the galaxy distance $D$ is not propagated into the tabulated uncertainties of the measured properties, as an error on the distance translates to a systematic (rather than random) scaling of some quantities (no effect on the others), i.e.\ $R_\mathrm{c}\propto D$, $L_\mathrm{CO(2-1)}\propto D^2$, $M_\mathrm{gas}\propto D^2$, $\omega_\mathrm{obs}\propto D^{-1}$ and $R_\mathrm{gal}\propto D$.
    \autoref{tab:gmc} is available in its entirety in machine-readable form in the electronic edition.
    }
\end{table*}

\subsection{GMC property distributions}
\label{sec:GMC_properties}

\begin{figure*}
    \centering
        \begin{subfigure}[t]{0.33\textwidth}
            \centering
                \includegraphics[width=\linewidth]{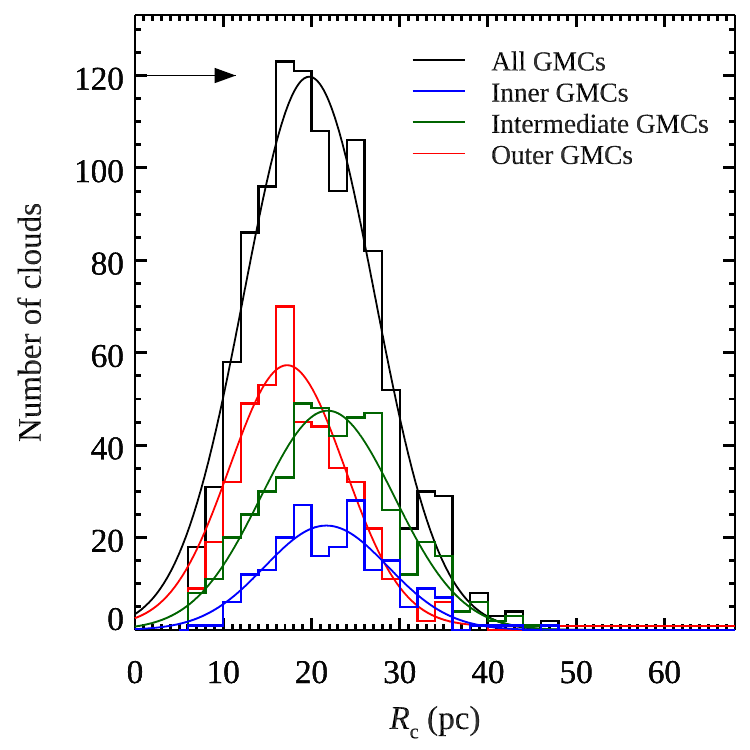} 
        \end{subfigure}
        \begin{subfigure}[t]{0.33\textwidth}
            \centering
                \includegraphics[width=\linewidth]{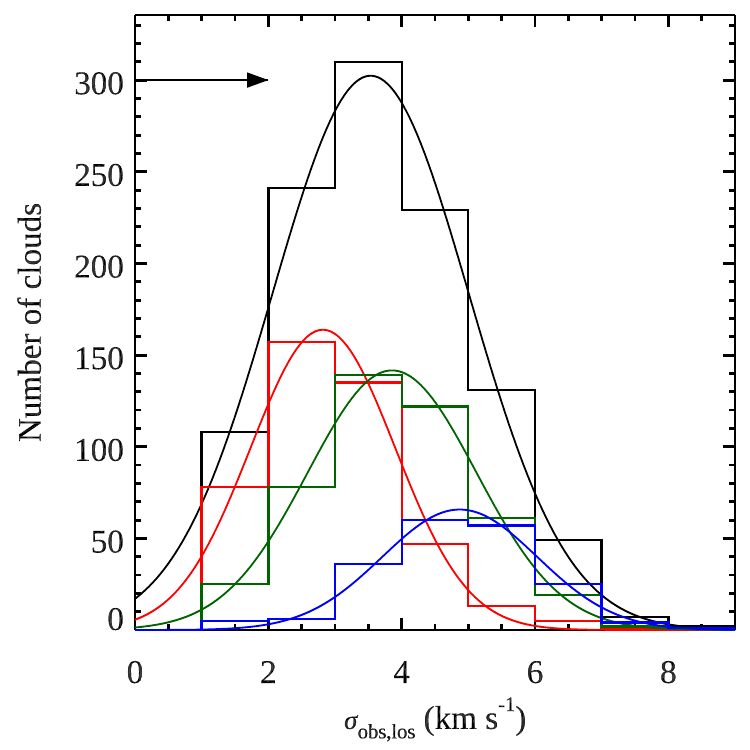} 
        \end{subfigure}
        \begin{subfigure}[t]{0.33\textwidth}
            \centering
                \includegraphics[width=\linewidth]{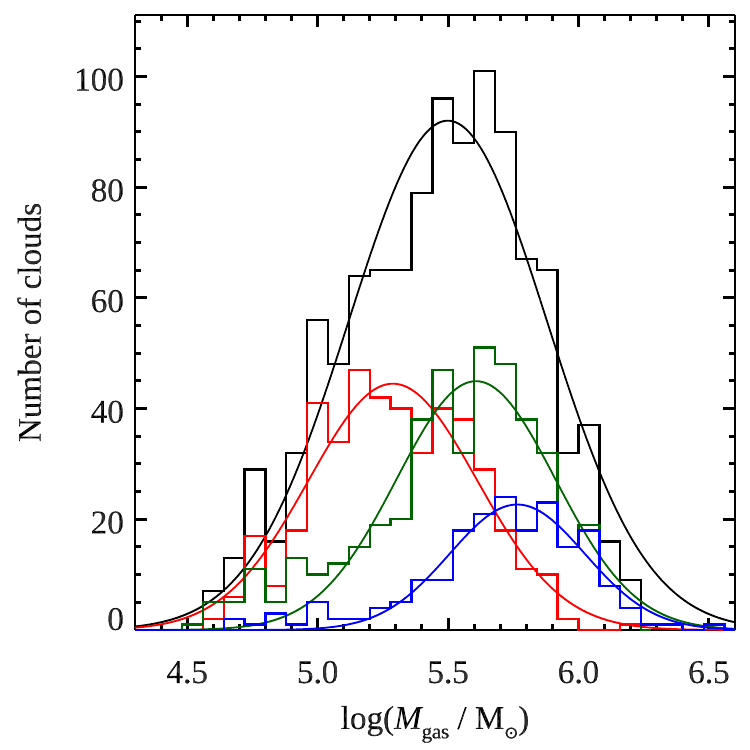} 
        \end{subfigure}
        \begin{subfigure}[t]{0.33\textwidth}
            \centering
                \includegraphics[width=\linewidth]{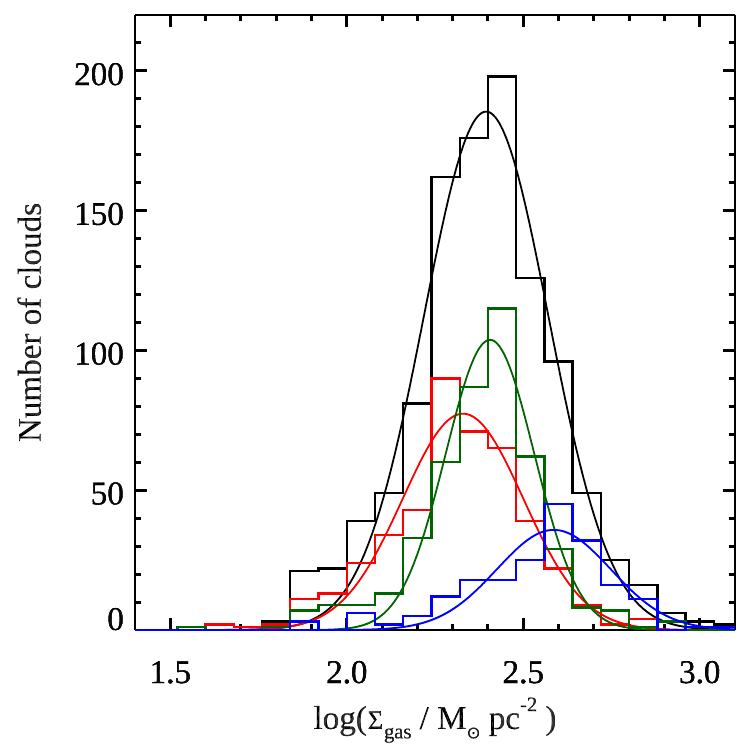} 
        \end{subfigure}
        \begin{subfigure}[t]{0.33\textwidth}
            \centering
                \includegraphics[width=\linewidth]{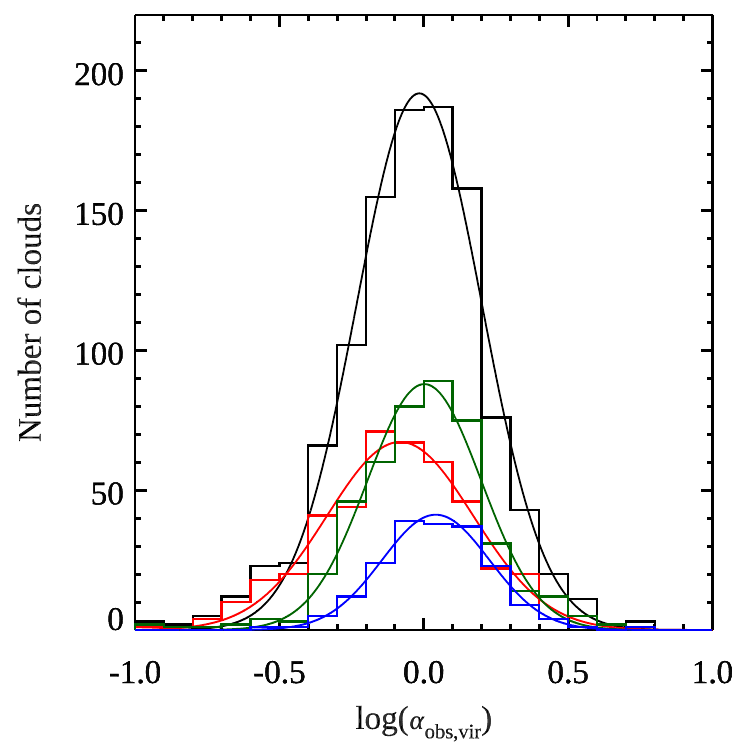} 
        \end{subfigure}
        \caption{Distributions of the fundamental GMC properties $R_\mathrm{c}$ (top-left), $\sigma_\mathrm{obs,los}$ (top-middle), $M_\mathrm{gas}$ (top-right), $\Sigma_\mathrm{gas}$ (bottom-left) and $\alpha_{\rm obs,vir}$ (bottom-right) of the NGC~1387 GMCs.
        Only (spatially and spectrally) resolved GMCs are included, and the different regions are shown in different colours (black for the entire sample).
        Gaussian fits are overlaid in matching colours.
        The spatial and spectral resolutions are indicated by arrows in the panels of $R_\mathrm{c}$ and $\sigma_\mathrm{obs,los}$, respectively. For $R_\mathrm{c}$, the arrow length is $\eta\sqrt{\sigma_\mathrm{maj}\sigma_\mathrm{min}}=11.4$~pc, where the standard deviation of the synthesised beam major (minor) axis is $\sigma_\mathrm{maj\,(min)}=\theta_\mathrm{maj\,(min)}/2.35$.
        For $\sigma_\mathrm{obs,los}$, the arrow length is the channel width of $2$~km~s$^{-1}$.}
        \label{fig:properties}
\end{figure*}

The most fundamental properties of GMCs are $R_\mathrm{c}$, $\sigma_\mathrm{obs,los}$ and $M_\mathrm{gas}$ (or $\Sigma_\mathrm{gas}$).
\autoref{fig:properties} shows their distributions, for resolved (both spatially and spectrally) GMCs only and each region separately (colour coded). 
Gaussian fits are overlaid in matching colours.
The means (standard deviations) of the fits are $\langle R_\mathrm{c}\rangle=20$~pc ($7$~pc), $\langle\sigma_\mathrm{obs,los}\rangle=3.5$~km~s$^{-1}$ ($1.5$~km~s$^{-1}$), $\langle\log(M_\mathrm{gas}/\mathrm{M_\odot})\rangle=5.5$ ($0.4$) and $\langle\log(\Sigma_\mathrm{gas}/\mathrm{M_{\odot}~pc^{-2}})\rangle=2.4$ ($0.2$).

There are some variations across the three regions.
The strongest trend is that all four quantities decrease with radius, i.e.\ on average each quantity is the largest in the inner region and the smallest in the outer region (except for $R_\mathrm{c}$ in the inner and intermediate regions).
These galactocentric radial gradients will be quantified and discussed in more detail in \autoref{sec:gradient}.
The mean $R_\mathrm{c}$ ($20$~pc) is only slightly larger than the synthesised beam size of $14$~pc, so there might still be a tendency to identify GMCs of approximately such size, despite the range of scales probed (see \autoref{sec:cprops}).

{
We compare the resolved GMCs of NGC~1387 with those of the previously studied ETGs, NGC~4429 \citep{2021MNRAS.505.4048L} and NGC~4526 \citep{2015ApJ...803...16U} below and in other relevant sections of this paper (while those of NGC~524, \citealt{2024MNRAS.531.3888L}, and NGC~5128, \citealt{2021MNRAS.504.6198M}, are not directly comparable due to the limited spatial resolution and the external gas origin, respectively).
{These comparisons are tabulated in \autoref{tab:prop_compare} and the main findings are discussed below.}
The $\langle R_\mathrm{c}\rangle$ of all three galaxies are well within one standard deviation of each other.
The $\langle\sigma_\mathrm{obs,los}\rangle$ of the NGC~1387 GMCs is one standard deviation smaller than that of the NGC~4429 GMCs. The spectral resolution of the NGC~4526 data ($10~$km~s$^{-1}$) is much poorer than that of the other two datasets ($2~$km~s$^{-1}$) and the measurement method of the GMC line widths was also different, making those $\sigma_\mathrm{obs,los}$ unsuitable for a direct comparison.
The $\langle M_\mathrm{gas}\rangle$ of the NGC~1387 GMCs is larger than that of the NGC~4429 GMCs by almost one standard deviation, but it is smaller than that of the NGC~4526 GMCs by $1.4$ standard deviation ($0.6$~dex).
Given these, as expected the $\langle\Sigma_\mathrm{gas}\rangle$ of the NGC~1387 GMCs is larger by one standard deviation than that of the NGC~4429 GMCs, but it is much lower (by four standard deviations or $0.8$~dex) than that of the NGC~4526 GMCs.
The much larger $\langle M_\mathrm{gas}\rangle$ and $\langle\Sigma_\mathrm{gas}\rangle$ of the NGC~4526 GMCs may be (partially) due to LoS confusion given its poorer spectral resolution.}

The wide distribution of $\Sigma_\mathrm{gas}$ both within individual galaxies and across galaxies 
is different from the {historically} presumed constant {(or mass-independent)} molecular gas mass surface density of MW and Local Group GMCs \citep{1987ApJ...319..730S,1993prpl.conf..125B}. 
This has been shown by many previous studies {of both LTGs \citep[e.g.][]{2005ApJ...623..826R,2009ApJ...699.1092H,2014ApJ...784....3C,2017ApJ...834...57M} and ETGs \citep{2023MNRAS.525.4270W}.}
The {overall} range of molecular gas mass surface densities is $2\lesssim\log(\Sigma_\mathrm{gas}/\mathrm{M_\odot~pc~^{-2}})\lesssim3$. 
It is within the reported range of the molecular gas mass surface densities of barred LTG centres\footnote{{The entire molecular gas disc of NGC~1387 would sit inside the `galaxy centre' according to their definition, not to be confused with the `inner region' of GMCs defined in this paper.}} {(defined by near-infrared morphology, \citealt{2021A&A...656A.133Q}}) by \citet{2020ApJ...901L...8S} in the left panel of their fig.~2. Both ranges are higher (i.e.\ denser) than the ranges of unbarred LTGs {and of gas outside galaxy centres (defined as above)}.
This is consistent with NGC~1387 being weakly barred (despite the vastly different spatial resolutions; $120$~pc in \citealt{2020ApJ...901L...8S}).
{The previously studied ETGs, NGC~4526 and NGC~4429, also agree with this trend by having bars \citep{2013MNRAS.432.1796A,2022ApJ...933...90Y} and denser GMCs \citep{2015ApJ...803...16U,2021MNRAS.505.4048L} than unbarred LTGs and outside galaxy centres.}

{A more detailed comparison of GMC $\Sigma_\mathrm{gas}$ across galaxies can potentially advance the understanding of self-shielding, magneto-gravitational instability, and other physical mechanisms \citep{2015ARA&A..53..583H}. Nonetheless, it requires a more accurate calibration of the $\alpha_\mathrm{CO(2-1)}$ factor (see \autoref{sec:alphaCO}) and the correction of the finite-cleaning effect on flux for individual GMCs (see \autoref{sec:mom} and \citealt{1995AJ....110.2037J}), as well as proper treatment of the different resolutions, sensitivities, and GMC boundary definitions across different studies. These are beyond the scope of this paper, but are worth exploration in the future.}

\begin{table*}
    \caption{ {GMC properties of ETGs.}}
    \label{tab:prop_compare}
    \begin{tabular}{lccccccl}
    \hline \hline
    Galaxy & Synthesised beam size & $\langle R_\mathrm{c} \rangle$  & Channel width & $\langle \sigma_\mathrm{obs,los} \rangle$  & $\langle\log(M_\mathrm{gas}$ / M$_\odot)\rangle$ & $\langle\log(\Sigma_\mathrm{gas}$ /   M$_\odot$~pc$^{-2})\rangle$ & Reference \\ 
     & (pc) & (pc) & (km~s$^{-1}$) & (km~s$^{-1}$) & & & \\ \hline
    NGC~1387 & 14 & 20 (7) & \phantom{0}2 & 3.5 (1.5) & 5.5 (0.4) & 2.4 (0.2) & This work \\
    NGC~4429 & 12 & 16 (6) & \phantom{0}2 & 5.2 (2.4) & 5.2 (0.3) & 2.2 (0.2) & \citet{2021MNRAS.505.4048L} \\
    NGC~4526 & 18 & 18 (8) & 10 & - & 6.1 (0.2) & 3.2 (0.3) & \citet{2015ApJ...803...16U} \\
    \hline \hline
    \end{tabular}
    \parbox{\textwidth}{Notes: The $\langle R_\mathrm{c} \rangle$, $\langle \sigma_\mathrm{obs,los} \rangle$, $\langle\log(M_\mathrm{gas}$ / M$_\odot)\rangle$, and $\langle\log(\Sigma_\mathrm{gas}$ /   M$_\odot$~pc$^{-2})\rangle$ are the means of the respective GMC properties.
    The numbers in parentheses are the standard deviations of the property distributions (rather than the uncertainties of the means) of the GMC populations in each galaxy. The spatial and spectral resolutions are provided to aid the comparisons. 
    See \autoref{sec:GMC_properties} for details.}
\end{table*}

\subsection{Mass spectra}
\label{sec:mass}

\begin{figure}
    \includegraphics[width=\columnwidth]{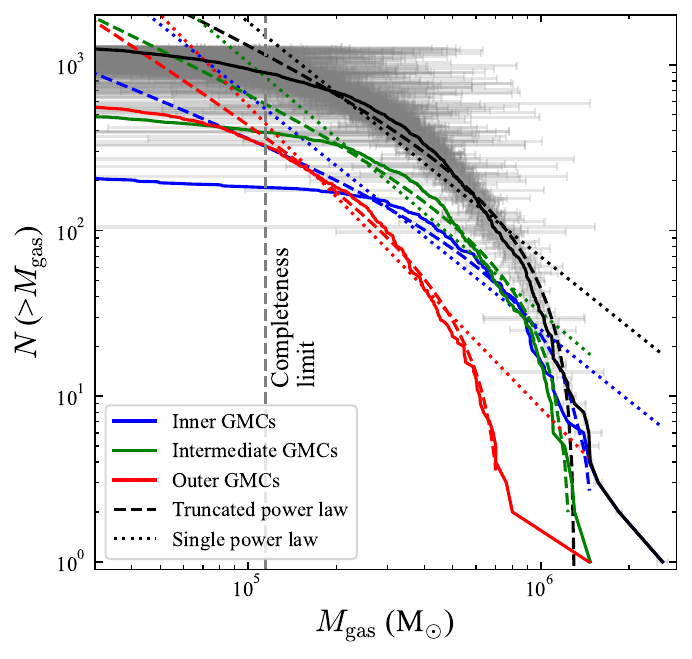}
    \caption{Molecular gas mass spectra (i.e.\ cumulative mass distribution functions) of the NGC~1387 GMCs. Both resolved and unresolved GMCs are included, and the different regions are shown in different colours (black for the entire sample).
    Best-fitting single power laws (dotted lines) and truncated power laws (dashed lines) are shown in matching colours.
    The nine most massive GMCs were excluded from the fit for the whole sample and the three most massive GMCs for each individual region, to avoid sensitivity to these few most massive bins.
    Grey horizontal error bars show the uncertainty of each GMC mass at the location where it contributes to the distribution.
    The grey vertical dashed line indicates our mass completeness limit (see \autoref{sec:mass}).}
    \label{fig:mass}
\end{figure}

The distribution of GMC molecular gas masses can help reveal GMC formation and evolution mechanisms \citep[e.g.][]{2005PASP..117.1403R, 2014ApJ...784....3C, 2016ApJ...821..125F, 2017MNRAS.468.1769F}.
\autoref{fig:mass} shows the mass spectra (i.e.\ the cumulative mass distribution functions) of the NGC~1387 GMCs, where both resolved and unresolved GMCs are included, again for each region separately (colour coded).
Horizontal error bars show the uncertainty of each GMC mass at the location where it contributes to the distribution.
Although these are not strictly the uncertainties of the cumulative mass distribution functions, they provide a useful illustration of those uncertainties.
We estimate the completeness limit of our data by adopting the common estimator $M_\mathrm{min}+10\delta_\mathrm{M}$, where the first term is the minimum resolved GMC mass in our catalogue, $3.5\times10^4$~M$_{\odot}$, and $\delta_\mathrm{M}$ is the uncertainty induced by noise ($\delta_\mathrm{M}=1.48\,MAD\times2\Delta v \times \alpha_\mathrm{CO(2-1)}\times A_\mathrm{beam}=9.9\times10^3$~M$_{\odot}$, where $MAD=1.32$~K is the median absolute deviation of the whole corrected cube, the factor of $1.48$ converts $MAD$ to the commonly quoted standard deviation, $A_\mathrm{beam}=225$~pc$^2$ is the area of the (clean) synthesised beam and $2\Delta v=4$~km~s$^{-1}$ is the minimum spectral width of the GMCs).
Our mass completeness limit is therefore estimated to be $1.3\times10^5$~M$_{\odot}$.

For each mass spectrum, we fit (in logarithmic space) the part above the mass completeness limit with both a single power law (dotted lines in \autoref{fig:mass}) and a truncated power law (dashed lines in \autoref{fig:mass}).
The two functions have the following forms, respectively:
\begin{align}
    \log\left(N(>M_\mathrm{gas})\right) & =\log\left(\frac{M_\mathrm{gas}}{M_0}\right)^{\gamma+1}\,\,,\\
    \log\left(N(>M_\mathrm{gas})\right) & =\log N_0+\log\left(\left(\frac{M_\mathrm{gas}}{M_0}\right)^{\gamma+1}-1\right)\,\,,
\end{align}
where $N(>M_\mathrm{gas})$ is the number of GMCs with a mass higher than $M_\mathrm{gas}$, $\gamma$ is the power-law slope, $M_0$ is the normalisation of the single power law but the characteristic cut-off mass of the truncated power law and $N_0$ serves as the normalisation of the truncated power law.
The power-law index ($\gamma+1$) is normally negative.
Therefore, in the truncated power law, $M_\mathrm{gas}$ must be smaller than $M_0$ for the formula to be meaningful.
A truncation occurs when $M_\mathrm{gas}$ approaches $M_0$ and $\log\left(N(>M_\mathrm{gas})\right)$ decreases more quickly than a single power law.
As can be seen in \autoref{fig:mass}, the advantage of the truncated power law is that it can better reproduce the spectra at the high-mass ends, while the disadvantage is that the best-fitting slope is more sensitive to the few highest-mass bins than a single power-law fit.
In fact, due to this sensitivity, and to reasonably match the totality of the spectra, before fitting we remove the nine most massive GMCs from the whole sample and the three most massive GMCs from each individual region (for both functions).

The best fitting parameters are listed in \autoref{tab:mass}.
As demonstrated by \autoref{fig:mass}, the truncated power laws always match the mass spectra better than the single power laws, especially at the high-mass ends.
These truncations possibly indicate a cessation of active mass accumulation at the high-mass ends, but they could also be due to a selection bias towards GMCs of radii similar to the synthesised beam size.
Regarding the different regions, although the four most massive GMCs are located in the inner region (blue curves), the intermediate region has roughly the same number of GMCs with $M_\mathrm{gas}>10^6$~M$_\odot$ as the inner region.
The inner-region GMCs have the shallowest power-law slopes while the outer-region GMCs have the steepest, irrespective of the function used.
As expected, the slopes of the truncated power laws are always shallower than those of the single power laws.
The best-fitting truncated power-law slope of the whole GMC sample is $-1.820 \pm 0.013$.
{As summarised in \autoref{tab:slope_compare},} this is between the slopes of inner MW GMCs ($-1.6\pm0.1$, where `inner' includes those GMCs closer to the MW centre than the Sun) and of outer MW GMCs ($-2.2\pm0.1$, where `outer' includes those GMCs farther away from the MW centre than the Sun; \citealt{2016ApJ...822...52R}).
In fact, the slope of the NGC~1387 GMCs has a similar radial trend, steepening from $\approx-1.7$ in the inner region to $\approx-2.1$ in the outer region. 
The best-fitting slope of the whole GMC sample of NGC~1387 is slightly steeper than that of the overall mass spectrum of Local Group GMCs ($\approx-1.7$; \citealt{2007prpl.conf...81B}), it is shallower than that of the ETG NGC~4429 ($-2.18\pm0.21$; \citealt{2021MNRAS.505.4048L}), and it is much shallower than that of another ETG NGC~4526 ($-2.39\pm0.03$; \citealt{2015ApJ...803...16U}).

The cut-off mass of the truncated power-law fit of the whole sample is $(1.533\pm0.004)\times10^{6}$~M$_\odot$, similar to that of the outer MW GMCs, $(1.5\pm0.5)\times10^6$~M$_\odot$ \citep{2016ApJ...822...52R}, but much smaller than that of the inner MW GMCs, $(1.0\pm0.2)\times10^7$~M$_\odot$ \citep{2016ApJ...822...52R}, {as summarised in \autoref{tab:cutoff_compare}}. 
This cut-off mass of the NGC~1387 GMCs is between those of the two previously studied ETGs, NGC~4429 with $(8.8\pm1.3)\times10^5$~M$_\odot$ \citep{2021MNRAS.505.4048L} and NGC~4526 with $(4.12\pm0.08)\times10^6$~M$_\odot$ \citep{2015ApJ...803...16U}.

{
The differences of the GMC mass spectra among the three ETGs are significant. The physical driver behind GMC mass spectra is intriguing but not fully resolved yet. \citet{2016IAUS..315...30H} reported a correlation among a sample of nearby galaxies, where flatter slopes are found in galaxies with higher surface mass densities (of stars or total cold gas or cold molecular gas). This can be explained by enhanced gas accumulation and gravitational instability in denser regions \citep{2005PASP..117.1403R,2014ApJ...784....3C}.
The comparison between NGC~1387 and NGC~4429 supports this trend. NGC~1387 has a flatter slope, a higher central stellar mass surface density \citep[][]{2022MNRAS.512.1522D}, and a higher molecular gas surface density \citep{2021MNRAS.505.4048L} than NGC~4429.
The trend also holds when different regions of NGC~1387 are compared --- the slope gets flatter and the surface density (of both stars and gas) gets higher towards the galaxy centre.
The comparisons of the cut-off mass between NGC~1387 and NGC~4429, and across different regions in NGC~1387 provide a consistent picture. A higher cut-off mass always correspond to a flatter slope, both indicating the existence of more high-mass GMCs.
On the other hand, NGC~4526 is more complicated. It exhibits the steepest slope yet the highest cut-off mass among the three ETGs. It is thus not straightforward to state whether NGC~4526 is rich or poor in high-mass GMCs. Regarding the slope -- density trend discussed above, NGC~4526 has the highest molecular gas surface density\footnote{{The {\it stellar} surface density of NGC~4526 is not available from the same source as NGC~1387 and NGC~4429, and the varied measurements from other papers lead to varied comparative results.}} \citep{2015ApJ...803...16U} yet the steepest slope among the three ETGs, opposite to the trend. The LoS confusion with NGC~4526 data is perhaps related to the complexity, which is due to the high inclination angle and the poorer spectral resolution.
Overall, this comparison calls for GMC studies of more ETGs to build larger statistics and further investigate the physical driver of GMC mass spectra.
}

{It is worth noting that the measured GMC mass spectrum parameters can be biased by observational limitations and effects such as blending (related to spatial resolution and spectral resolution), sensitivity, and GMC decomposition techniques \citep{2005PASP..117.1403R,2010ApJ...719..561R,2011ApJS..197...16W}. Although the typical spatial resolution ($\lesssim30$~pc) behind all the results in \autoref{tab:slope_compare} are adequate to resolve individual GMCs, the separation between neighbouring GMCs and the exact boundary of each GMC are less certain in some galaxies, due to a low contrast of gas surface density or a high inclination angle. 
The inhomogeneous sensitivity across surveys, the completeness limit estimation, the adopted GMC decomposition technique, and the $\alpha_{\rm CO(2-1)}$ factor can all affect the best-fit parameters as well \citep{2008ApJS..178...56F,2011ApJS..197...16W}. Overall, these effects should be minor for the synoptic comparison presented in this paper, but a more detailed comparison would need better treatment of these effects in the future.}

\begin{table*}
    \centering
    \caption{Best-fitting parameters of the NGC~1387 GMC molecular gas mass spectra.}
    \label{tab:mass}
    \begin{tabular}{lccccc}
        \hline
        \hline
        & \multicolumn{2}{c}{Single power law} & \multicolumn{3}{c}{Truncated power law}\\
        & $\gamma$ & $M_0$ & $\gamma$ & $M_0$ & $N_0$\\ 
        & & ($10^6$~M$_\odot$) & & ($10^6$~M$_\odot$) &\\
        \hline
        All GMCs & $-2.42\pm0.02$ & $23.3\pm1.2$ & $-1.820\pm0.013$ & $1.533\pm0.004$ & $180\pm\phantom{}5$\\
        Inner-region GMCs & $-2.42\pm0.05$ & $\phantom{}11.4\pm1.2$ & $-1.67\phantom{0}\pm0.05\phantom{0}$ & $1.82\phantom{0}\pm0.02\phantom{0}$ & $\phantom{0}69\pm8$\\
        Intermediate-region GMCs & $-2.51\pm0.03$ & $\phantom{}$$11.6\pm0.8$ & $-1.82\phantom{0}\pm0.02\phantom{0}$ & $1.487\pm0.006$ & $\phantom{0}95\pm\phantom{}5$\\
        Outer-region GMCs & $-2.83\pm0.03$ & $\phantom{0}$$3.8\pm0.2$ & $-2.13\phantom{0}\pm0.02\phantom{0}$ & $0.873\pm0.004$ & $\phantom{0}50\pm\phantom{}2$\\ 
        \hline
        \hline
    \end{tabular}
    \parbox{\textwidth}{Notes: The fitting parameters are described in \autoref{sec:mass}.
    Due to the sensitivity of the truncated power-law function to the few highest-mass bins, and to reasonably match the totality of the spectra, the nine most massive GMCs of the whole GMC sample were excluded from the fit and the three most massive GMCs of each individual region (for both functions).
    Uncertainties are quoted at the $1$~$\sigma$ level.}
\end{table*}

\begin{table}
    \centering
    {
    \caption{{GMC mass spectrum slopes of galaxies.}}
    \label{tab:slope_compare}
        \begin{tabular}{lcl}
        \hline \hline
        Galaxy (region)       & Slope $\gamma$             & Reference                   \\ \hline
        MW (inner)            & $-1.6 \phantom{00} \pm 0.1 \phantom{00}$     & \citet{2016ApJ...822...52R} \\
        NGC~1387 (inner)      & $-1.67 \phantom{0}\pm 0.05\phantom{0}$   & This work                   \\
        Local Group           & $-1.7\phantom{00 \pm 0.001}$             & \citet{2007prpl.conf...81B} \\
        NGC~1387 (overall)    & $-1.820 \pm 0.013$ & This work                   \\
        NGC~1387 (inter.)     & $-1.82\phantom{0} \pm 0.02\phantom{0}$   & This work                   \\
        NGC~1387 (outer)      & $-2.13 \phantom{0}\pm 0.02\phantom{0}$   & This work                   \\
        NGC~4429    & $-2.18 \phantom{0}\pm 0.21\phantom{0}$   & \citet{2021MNRAS.505.4048L} \\
        MW (outer)            & $-2.2 \phantom{00}\pm 0.1\phantom{00}$     & \citet{2016ApJ...822...52R} \\
        NGC~4526    & $-2.39\phantom{0}\pm0.03\phantom{0}$     & \citet{2015ApJ...803...16U} \\ \hline \hline
        \end{tabular}   
        \parbox{\columnwidth}{\vspace{0.1cm} Notes: For NGC~1387 GMCs, the truncated power law is adopted for comparison.
        The regions of NGC~1387 are defined in \autoref{sec:mom}. `MW (inner)' includes GMCs closer to the MW centre than the Sun, while `MW (outer)' includes GMCs farther away from the MW centre than the Sun \citep{2016ApJ...822...52R}.
        The table is ordered by decreasing mass spectrum slopes.}
}
\end{table}

\begin{table}
    \centering
    {
    \caption{{GMC mass spectrum cut-off mass of galaxies.}}
    \label{tab:cutoff_compare}
    \begin{tabular}{lcl}
    \hline \hline
    Galaxy (region)    & Cut-off mass $M_0$ (10$^6$~M$_\odot$) & Reference                   \\ \hline
    MW (inner)         & $10\phantom{.000}\pm2\phantom{.000}$              & \citet{2016ApJ...822...52R} \\
    NGC~4526           & $\phantom{0}4.12\phantom{0}\pm\phantom{}0.08\phantom{0}$    & \citet{2015ApJ...803...16U} \\
    NGC~1387 (inner)   & $\phantom{0}1.82\phantom{0}\pm\phantom{}0.02\phantom{0}$      & This work                   \\
    NGC~1387 (overall) & $\phantom{0}1.533\pm\phantom{}0.004$                                  & This work                   \\
    MW (outer)         & $\phantom{0}1.5\phantom{00} \pm \phantom{}0.5\phantom{00}$        & \citet{2016ApJ...822...52R} \\
    NGC~1387 (inter.)  & $\phantom{0}1.487\pm\phantom{}0.006$                                  & This work                   \\
    NGC~4429           & $\phantom{0}0.88\phantom{0}\pm\phantom{}0.13\phantom{0}$                                     & \citet{2021MNRAS.505.4048L} \\ 
    NGC~1387 (outer)   & $\phantom{0}0.873\pm\phantom{}0.004$                                   & This work                   \\
    \hline \hline
    \end{tabular}
    \parbox{\columnwidth}{\vspace{0.1cm} Notes: As \autoref{tab:slope_compare} but for the mass spectrum cut-off mass. {\citet{2007prpl.conf...81B} adopted single power law fitting, and thus did not have a cut-off mass.}}
    }
\end{table}

\subsection{Larson relations}
\label{sec:larson}

\begin{figure*}
    \includegraphics[width=\textwidth]{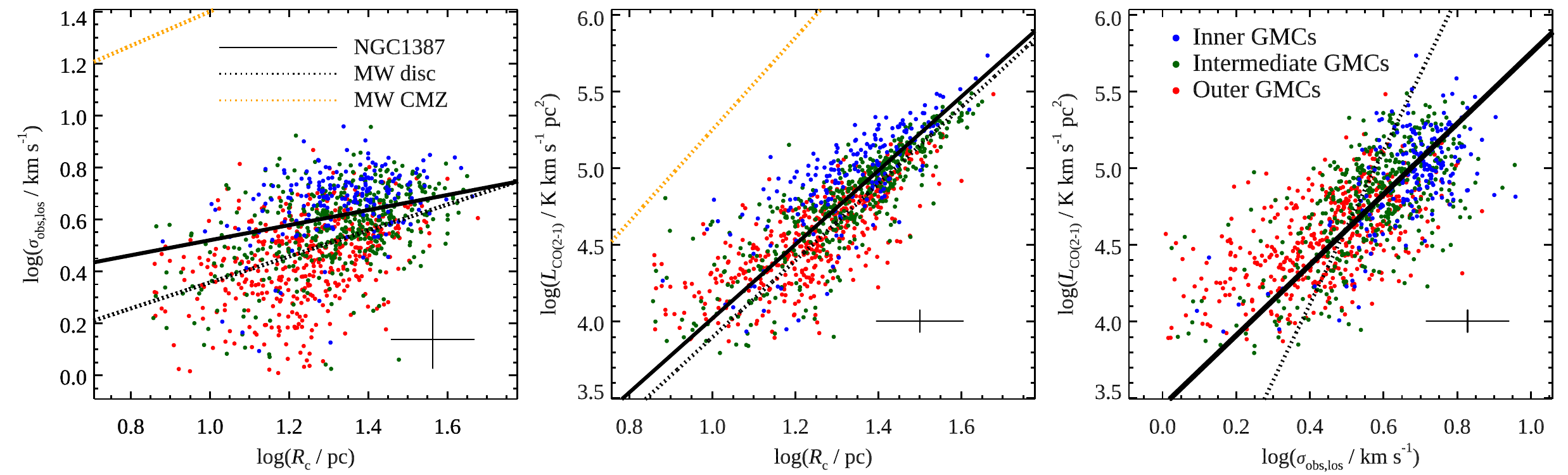}
    \caption{Larson relations of the resolved GMCs of NGC~1387.
    {\bf Left:} size -- line width relation.
    {\bf Middle:} size -- luminosity relation.
    {\bf Right:} line width -- luminosity relation.
    The GMCs of the different regions are shown in different colours.
    Solid lines show the best-fitting power-law relations of all the GMCs, while the black dotted lines show the Larson relations of the MW disc GMCs and the orange dotted lines those of the MW CMZ GMCs.
    The median error bar is shown in the bottom-right corner of each panel.}
    \label{fig:larson}
\end{figure*}

Nearly all previous GMC studies have used the Larson relations \citep{1981MNRAS.194..809L} as the common yardstick against which to gauge GMC properties.
\autoref{fig:larson} shows the Larson relations of the resolved GMCs of NGC~1387.
GMCs from different regions of NGC~1387 are shown in different colours. 
The left panel shows the size -- line width relation, thought to be a signature of the turbulence within GMCs.
The middle panel shows the size -- luminosity relation (equivalent to a size -- molecular gas mass relation, as we are assuming a constant mass conversion factor), providing insight into the internal structures of GMCs.
The right panel shows the line width -- luminosity relation (again equivalent to a line width -- molecular gas mass relation), the GMC analogue of the Faber-Jackson relation of galaxies \citep{1987ApJ...319..730S}.
These three relations together constitute the GMC `fundamental plane', but only two out of the three relations are independent as they are mathematically related.

The GMCs from different regions of NGC~1387 occupy different areas of parameter space in \autoref{fig:larson}, consistent with the regional variations of GMC properties (illustrated in \autoref{fig:properties}).
In the size -- line width relation, the inner-region GMCs (blue data points) occupy primarily the top-right portion of the distribution, at larger velocity dispersions than the intermediate-region GMCs (green data points), {but with a similar total range of sizes.}
The outer-region GMCs (red data points) have a distribution that is much more extended than that of the inner-region and intermediate-region GMCs, and is centred at smaller sizes and velocity dispersions.
The trends are similar for the other two Larson relations.
For each relation, we fit a power law to both the entire resolved GMC population and that in each region.
The fits are carried out in logarithmic space with the Interactive Data Language routine {\tt LINMIX\_ERR}.\footnote{\url{https://www.nv5geospatialsoftware.com/docs/linmix_err.html}}
The best-fitting relations of the entire population are shown as the black solid lines in \autoref{fig:larson}, where they can be compared to those of the MW disc GMCs (black dotted lines; \citealt{1987ApJ...319..730S, 2009ApJ...699.1092H}) and MW CMZ GMCs (orange dotted lines; \citealt{2017A&A...603A..89K, 2020ApJ...897...89L}).
The best-fitting parameters are listed in \autoref{tab:larson}.

\begin{table*}
    \centering
    \caption{NGC~1387 Larson relation best-fitting parameters.}
    \label{tab:larson}
    \resizebox{\textwidth}{!}{%
    \begin{tabular}{lccc|ccc|ccc}
        \hline
        \hline
        & \multicolumn{3}{c|}{Size -- line width relation}                                                     & \multicolumn{3}{c}{Size -- luminosity relation}                                                    & \multicolumn{3}{|c}{Line width -- luminosity relation}                                                 \\
        & Slope & Normalisation & $r_{\rm s}$ & Slope & Normalisation & $r_{\rm s}$ & Slope & Normalisation & $r_{\rm s}$ \\ \hline
        All GMCs          & $0.29\pm0.06$ & $\phantom{-}0.23\pm0.08$   & $0.47$                                                              & $2.41\pm0.07$ & $1.61\pm0.10$   & $0.82$ 
        & $2.30\pm0.11$   & $3.45\pm0.07$   & $0.70$  \\
        Inner-region GMCs        & $0.09 \pm 0.07$ & $\phantom{-}0.57 \pm 0.10$   & $0.31$                                                              & $2.13 \pm 0.11$ & $2.1\phantom{0} \pm 0.2\phantom{0}$   & $0.78$ 
        & $2.3\phantom{0} \pm 0.3\phantom{0}$ & $3.4\phantom{0} \pm 0.2\phantom{0}$   & $0.50$    \\
        Intermediate-region GMCs & $0.35 \pm 0.08$ & $\phantom{-}0.13 \pm 0.11$   & $0.43$
        & $2.25 \pm 0.08$ & $1.80 \pm 0.12$   & $0.87$       
        & $2.6\phantom{0} \pm 0.2\phantom{0}$   & $3.27 \pm 0.14$   & $0.60$  \\
        Outer-region GMCs        & $0.5\phantom{0} \pm 0.2\phantom{0}$ & $-0.1\phantom{0} \pm 0.2\phantom{0}$   & $0.39$
        & $2.38 \pm 0.15$ & $1.6\phantom{0} \pm 0.2\phantom{0}$   & $0.77$ 
        & $3.0\phantom{0} \pm 0.5\phantom{0}$   & $3.1\phantom{0} \pm 0.3\phantom{0}$  & $0.57$    \\ 
        \hline
        \hline
    \end{tabular}}\\
    \parbox{\textwidth}{Notes: Slopes and normalisations are defined in the logarithm space of all properties involved, as shown in \autoref{fig:larson}. Uncertainties are quoted at the $1$~$\sigma$ level. The $r_{\rm s}$ is the Spearman correlation coefficient. {The $p$-values of all the correlations are $\le1\times10^{-5}$.}}
\end{table*}

For the whole GMC population, the Spearman correlation coefficients $r_{\rm s}$ of the resolved GMCs of NGC~1387 are weak ($0.47$) for the size -- line width relation, strong ($0.82$) for the size -- luminosity relation and moderate ($0.70$) for the line width -- luminosity relation.
The correlation coefficients of the GMCs of the different regions considered separately are almost always weaker, suggesting that the aforementioned systematic differences between the different regions contribute to creating the correlations when they are considered together.
In other words, the galactocentric distance gradients of the three basic GMC parameters ($R_\mathrm{c}$, $\sigma_\mathrm{obs,los}$ and $L_\mathrm{CO(2-1)}$) are significant.
This will be shown more clearly in \autoref{sec:gradient}.
{The $p$-values of all the correlations across all (sub-)samples are $\le1\times10^{-5}$, implying that all the correlations have $4\sigma$ or higher statistical significance.}
{The tightness of the size -- line width relation varies in the literature, from the extremely tight early versions for the MW GMCs by \citet{1981MNRAS.194..809L} in his fig.~1 and by \citet{1987ApJ...319..730S} in their fig.~1, to the non-existence of a correlation ($r_{\rm s} < 0.2$) in some LTGs \citep{2012A&A...542A.108G,2014ApJ...784....3C} and ETGs \citep{2015ApJ...803...16U,2021MNRAS.505.4048L}. The physical explanation for such a discrepancy is still an open question.}

In all panels of \autoref{fig:larson}, the resolved GMCs of NGC~1387 follow the trend of the MW disc GMCs quite well, in terms of both slope and normalisation.
In turn, this implies that they are unlike those of {NGC~4526 and NGC~4429}, which typically have velocity dispersions higher than those of MW disc GMCs at any given size \citep{2015ApJ...803...16U,2021MNRAS.505.4048L} and luminosity \citep{2015ApJ...803...16U}.\footnote{{\citet{2015ApJ...803...16U} only reported velocity gradient-subtracted velocity dispersions, and thus the $\sigma_\mathrm{obs,los}$ (if measured as in this work) of NGC~4526 GMCs would be even higher than the velocity dispersions actually reported in that paper.}}
Despite the fact that all of the NGC~1387 GMCs are located within its bulge, the properties of these GMCs deviate significantly from those of the MW CMZ GMCs.
The size -- line width relation of the NGC~1387 GMCs even has a shallower slope ($0.29\pm {0.06}$) than that of the MW disc ($0.5\pm0.1$) or Local Group ($0.6\pm0.1$) GMCs.
This is rather unique and will be discussed in light of a GMC collision model in \autoref{sec:toomre}.
When considering the regions separately, the Larson relation best-fitting parameters can vary significantly. Specifically, the inner-region GMCs have an extremely flat size -- line width relation with a slope of {$0.09 \pm 0.07$}, again contrary to the MW in which the CMZ GMCs have a much steeper size -- line width relation ($0.66\pm0.18$).
The variation of the normalisation again mostly reflects the galactocentric distance gradients.

\subsection{Virial parameter}
\label{sec:virial_parameter}

We estimate the virial parameters $\alpha_\mathrm{obs,vir}$ of the resolved GMCs of NGC~1387. The virial parameter is defined as the ratio of the virial mass (i.e.\ the dynamical mass assuming virial equilibrium $M_\mathrm{obs,vir}$; \citealt{1988ApJ...333..821M}) and the molecular gas mass of a GMC, yielding
\begin{equation}
    \alpha_\mathrm{obs,vir}\equiv\frac{M_\mathrm{obs,vir}}{M_\mathrm{gas}}=\frac{\sigma_\mathrm{obs,los}^2R_\mathrm{c}/b_\mathrm{s} {\rm G}}{M_\mathrm{gas}}\,\,,
\end{equation}
where G is the gravitational constant and the geometric parameter $b_\mathrm{s}$ is taken to be $1/5$, corresponding to a homogeneous spherical mass volume density distribution (and allowing to compare with most previous works on GMCs).

\begin{figure}
    \includegraphics[width=\columnwidth]{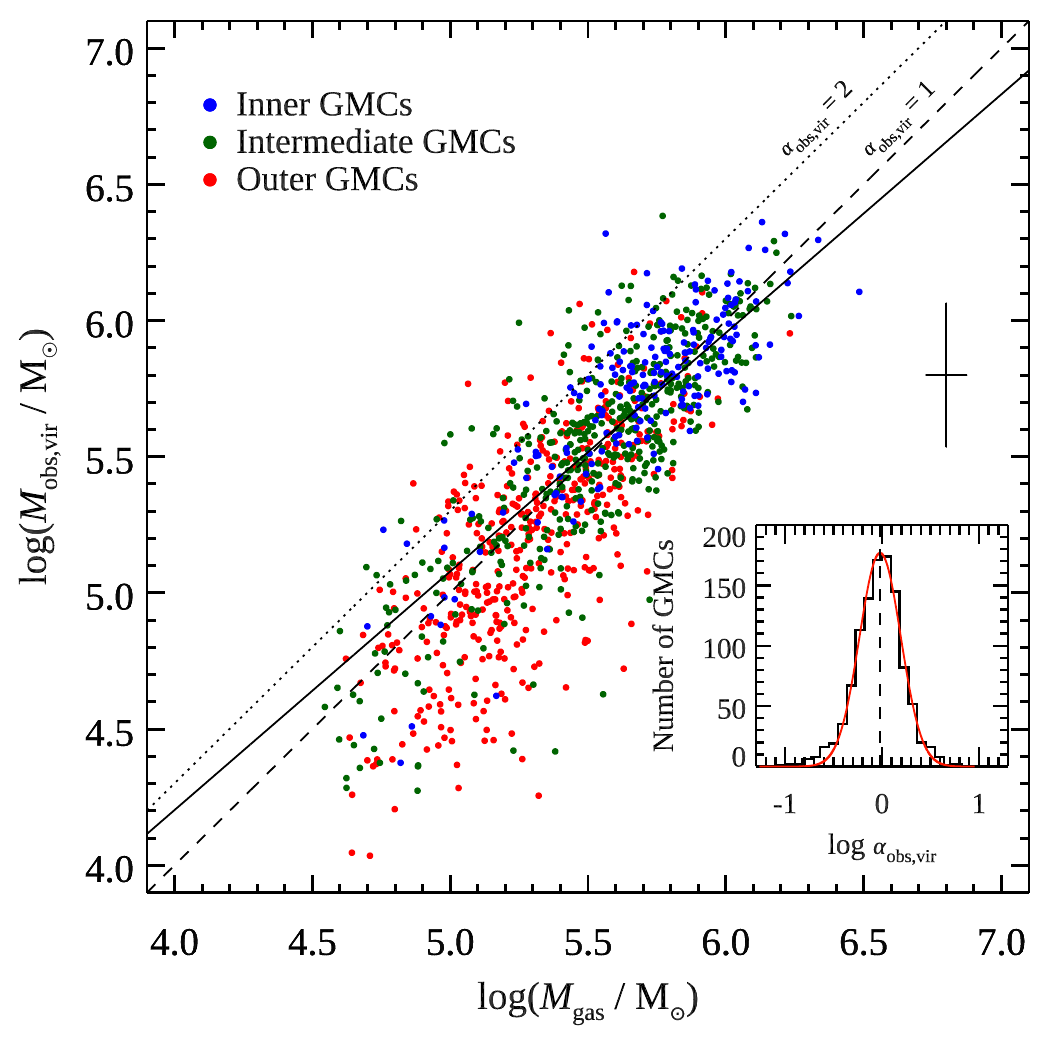}
    \caption{Correlation between the virial masses and the molecular gas masses of the resolved GMCs of NGC~1387.
    The GMCs of the different regions are shown in different colours.
    The solid line shows the best-fitting power-law relation, while the dashed and dotted diagonal lines show the $1:1$ ($\alpha_\mathrm{obs,vir}=1$) and $2:1$ ($\alpha_\mathrm{obs,vir}=2$) relations, respectively.
    The median error bar is shown in the centre-right of the panel.
    The distribution of $\log(\alpha_\mathrm{obs,vir})$ (black histogram) with a log-normal fit overlaid (red solid line) is shown in the inset.
    The dashed vertical line shows the mean of the log-normal fit. 
    }
    \label{fig:virial}
\end{figure}

\autoref{fig:virial} shows the comparison of the virial masses and the molecular gas masses of the resolved GMCs of NGC~1387.
GMCs from different regions of NGC~1387 are shown in different colours, as in previous figures.
Although the GMCs from different regions occupy slightly different mass ranges, they generally follow the same trend.
A power-law fit (carried out in logarithmic space) yields a best-fitting power-law index (i.e.\ a slope) of {$0.88$} (black solid line in \autoref{fig:virial}), suggesting that the estimated GMC virial masses are very similar to the molecular gas masses.
Indeed, the distribution of the virial parameters of the resolved GMCs of NGC~1387 is shown in the inset of \autoref{fig:virial}, and $80\%$ of the GMCs have $0.5<\alpha_\mathrm{obs,vir}<2$.
The resolved GMCs of NGC~1387 thus appear to be nearly perfectly virialised ($\alpha_\mathrm{obs,vir}=1$), with some scatter.
A Gaussian fit to the log-normal distribution of $\alpha_\mathrm{obs,vir}$ yields a mean virial parameter $\langle\alpha_\mathrm{obs,vir}\rangle=1.0$ with a standard deviation of $^{+0.6}_{\!-0.4}$.
This is very similar to that of MW disc \citep{2015ARA&A..53..583H} and Local Group galaxy \citep[e.g.][]{2003ApJ...599..258R, 2007ApJ...654..240R, 2011ApJ...737...40H} GMCs,
but is lower than that of the ETG NGC~4429 GMCs ($\langle\alpha_\mathrm{obs,vir}\rangle=4.0$; \citealt{2021MNRAS.505.4048L}).
{\citet{2015ApJ...803...16U} measured the virial parameters of the ETG NGC~4526 GMCs based on velocity gradient-subtracted velocity dispersions, which will be discussed in \autoref{sec:mw_compare}.}
Compared with the empirical threshold of $\alpha_\mathrm{obs,vir}=2$ for a gravitationally bound cloud \citep{Kauffmann:2013}, the $\alpha_\mathrm{obs,vir}$ distribution indicates that, on average, the resolved GMCs of NGC~1387 are both gravitationally bound and virialised.
When the GMCs are separated by region, there is a marginal trend of $\alpha_{\rm obs,vir}$ decreasing from the inner region to the outer region (see also \autoref{fig:properties}).
We note however that this interpretation only considers the internal gravitational force of each GMC.
Other factors such as magnetic fields, external pressure, and {the galactic gravitational potential} are not included in the calculation of $\alpha_\mathrm{obs,vir}$.

\subsection{{Uncertainty of $\alpha_{\rm CO(1-0)}$}}
\label{sec:alphaCO}
{
It is a long-standing challenge to calibrate the $\alpha_{\rm CO(1-0)}$ factor in different types of galaxy environments.
The value adopted in this work comes from the calibration for the MW disc environment, and the commonly adopted uncertainty of 30\% encompasses the variation across different calibration methods, such as CO isotopologues, dust extinction, dust emission, etc.\ \citep{2013ARA&A..51..207B}. 
Many works have shown deviations from this canonical $\alpha_{\rm CO(1-0)}$ at low metallicity \citep[e.g.][]{2011MNRAS.412..337G}, high stellar mass surface density \citep{1998ApJ...507..615D,2024ApJ...964...18C}, high CO line width \citep{2023ApJ...950..119T}, and in CMZs \citep[e.g.][]{2024PASJ...76..579K}. But some other studies suggest a small fraction of galaxies do not have such a deviation of $\alpha_{\rm CO(1-0)}$ in the centres \citep{2013ApJ...777....5S}.

Since no calibration has been done specifically for the environment of ETG centres and the calibrations for similar environments vary across galaxies, we do not correct for the potential deviation of $\alpha_{\rm CO(1-0)}$ from the MW value in this paper. 
But if the true value deviates from that, the following GMC measurements will be affected: the mass, the mass surface density, the mass spectrum, the virial parameter, and the Toomre parameter (see \autoref{sec:toomre}). Other measurements and analyses are not affected. 
Moreover, all the comparisons between the three ETGs (NGC~1387, NGC~4429, and NGC~4526), even for the affected quantities listed above, should remain valid because $\alpha_{\rm CO(1-0)}$ is likely to be very similar in the same ETG centre environment.
}

\section{Origin of velocity gradients in GMCs}
\label{sec:kine}

\subsection{Galactic rotation and turbulence}
\label{sec:kine_gal}

We now aim to constrain the origin of the internal kinematics of the GMCs of NGC~1387, in particular, whether it is inherited from the large-scale galactic rotation.
In the ETGs NGC~4526 and NGC~4429, the internal GMC rotation is aligned with the large-scale velocity field, suggesting that shear (i.e.\ galactic rotation) drives the observed internal GMC rotation (see e.g.\ figs.~7 and 8 of \citealt{2021MNRAS.505.4048L}).
Following the same approach, we measure the velocity gradient of each resolved GMC by fitting a plane to its first-moment map, weighted by intensity (i.e.\ the zeroth-moment map).
The fitting function is thus $v(x,y)=ax+by+c$, where $(x,y)$ are the positions on the sky and $a$, $b$ and $c$ are free parameters.
The projected angular velocity (i.e.\ the magnitude of the projected velocity gradient) is then $\omega_\mathrm{obs}=\sqrt{a^2+b^2}$, and the projected position angle (measured from north through east) of the rotation axis is $\phi_\mathrm{rot}=\tan^{-1}(b/a)$.

\begin{figure*}
    \includegraphics[width=0.8\textwidth]{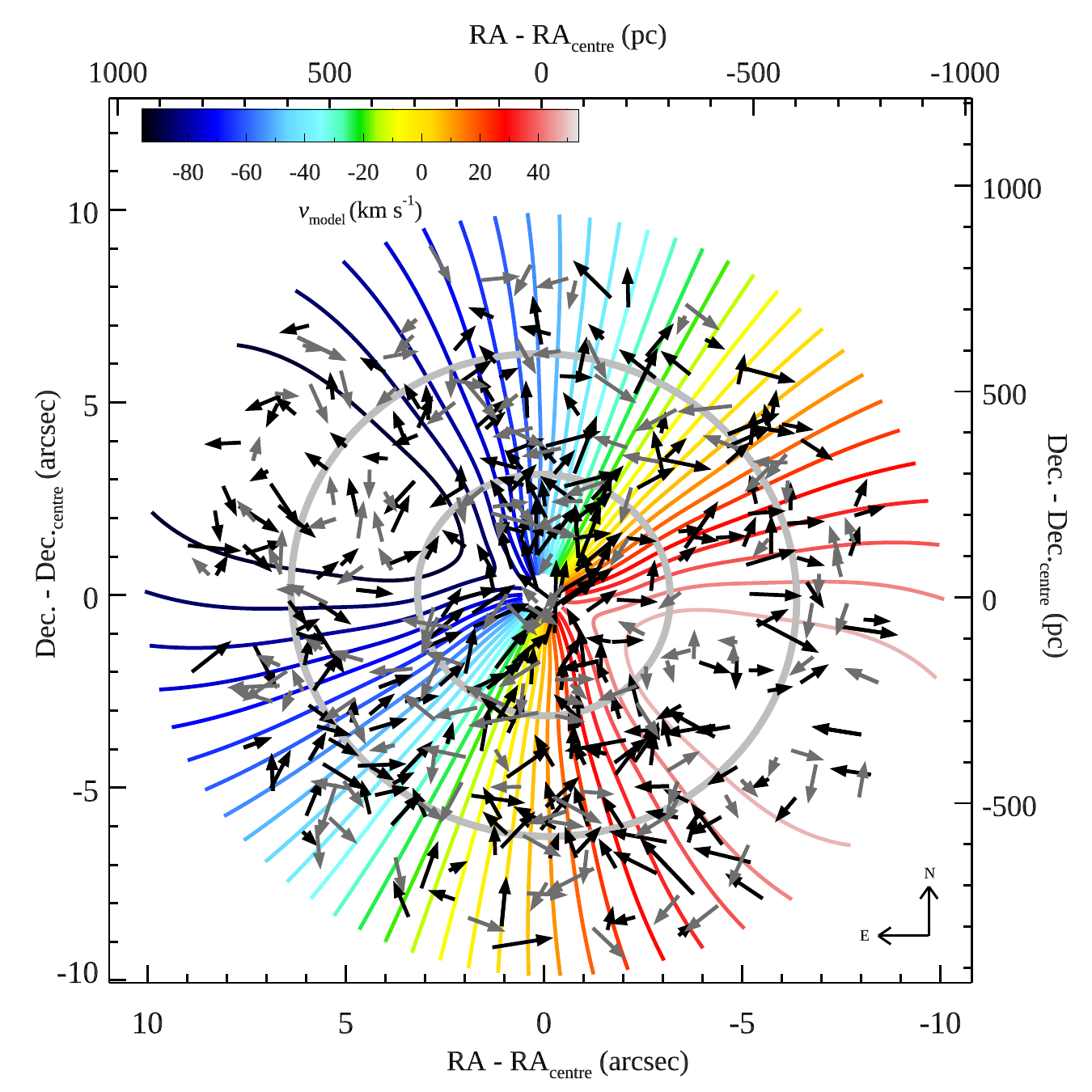}
    \caption{Projected circular velocity field of NGC~1387 ($v_\mathrm{model}$; coloured isovelocity contours).
    The projected angular momentum vectors of half of the resolved GMC of NGC~1387 (randomly selected, for clarity) are overplotted as arrows.     
    Black arrows indicate prograde GMCs, dark grey arrows retrograde GMCs, and the length of each arrow is proportional to $\omega_\mathrm{obs}$.
    The pale grey ellipses indicate the boundaries of the three regions defined in \autoref{sec:mom}.}
    \label{fig:v_field}
\end{figure*}

We show the projected angular velocity vectors of the resolved GMCs of NGC~1387 overlaid on the isovelocity contours (i.e.\ the velocity field) in \autoref{fig:v_field} (only half of the GMCs, randomly selected, are shown for clarity).
To calculate the isovelocity contours, we input the mass model described in \autoref{sec:hst} to the {\tt mge\_vcirc} routine of the {\tt Jean Anisotropic Modelling} ({\tt JAM}) package\footnote{\url{https://pypi.org/project/jampy/}} \citep{2020MNRAS.494.4819C}, and then used the {\tt kinms\_create\_velField\_oneSided} routine of the {\tt KinMS} package to create the LoS velocity map.

In \autoref{fig:v_field}, the projected angular velocity vectors do not align significantly with the large-scale galactic rotation, as traced by the isovelocity contours.
{We measure the angle between each cloud's projected angular velocity vector and the local isovelocity contour. We then divide}
GMCs into prograde ({angle smaller than $90\degr$}; black arrows in \autoref{fig:v_field}) and retrograde ({angle larger than $90\degr$}; dark grey arrows). 
The fraction of prograde clouds is larger than that of retrograde clouds in all three regions (and thus overall), but that fraction never approaches unity. In addition, there is a larger fraction of prograde GMCs in the inner region ($141/195=72\%$) than in the intermediate region ($278/448=62\%$) and outer region ($256/436=59\%$) of the galaxy.
This suggests that the inner-region GMCs couple best with the large-scale galactic rotation, and we elaborate on this in \autoref{sec:subsample}.
{On the other hand, the GMC prograde fractions of NGC~4429 \citep[figs. 7--8 of][]{2021MNRAS.505.4048L} and NGC~4526 \citep[figs. 7--8 of][]{2015ApJ...803...16U} are almost 100\%.}

\begin{figure*}
    \includegraphics[width=\textwidth]{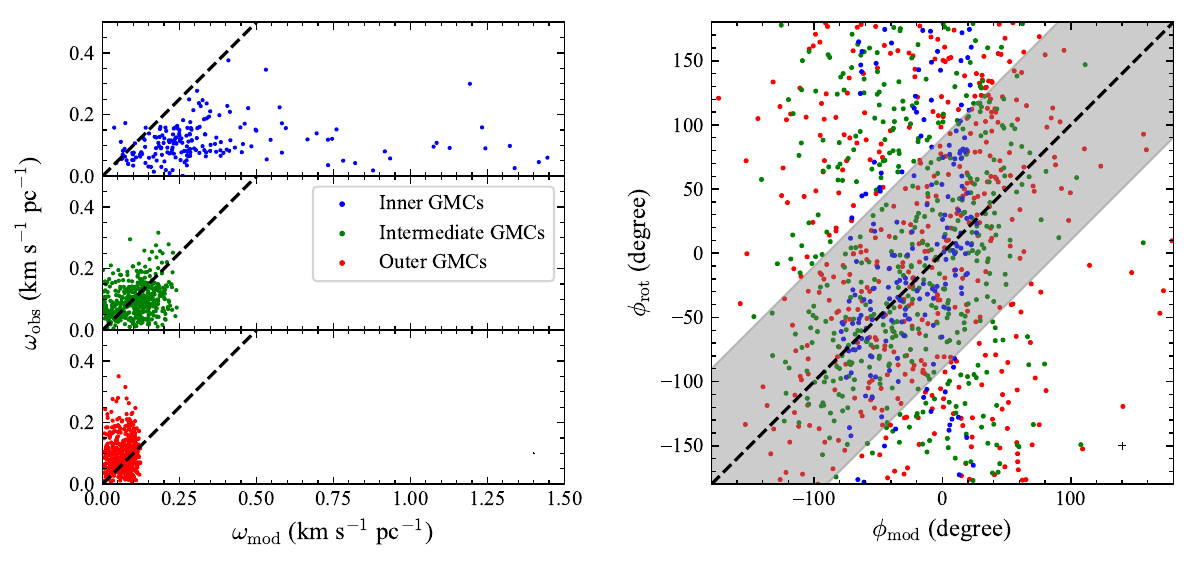}
    \caption{Comparison of the observed and modelled projected GMC angular velocity magnitudes ($\omega_\mathrm{obs}$ and  $\omega_\mathrm{mod}$, respectively; left panel) and position angles ($\phi_\mathrm{rot}$ and $\phi_\mathrm{mod}$, respectively; right panel) of the resolved GMCs of NGC~1387.
    The GMCs of the different regions are shown in different colours.
    The dashed lines show the $1:1$ relations.
    The grey shaded region in the right panel indicates an angular difference $\le90\degr$, which we adopt to define prograde clouds (conversely for retrograde clouds).
    The median uncertainty is shown as an error bar in the bottom-right corner of each panel (barely visible in the left panel).
    }
    \label{fig:shear}
\end{figure*}

In \autoref{fig:shear}, the measured $\omega_\mathrm{obs}$ and $\phi_\mathrm{rot}$ are compared with modelled quantities ($\omega_\mathrm{mod}$ and $\phi_\mathrm{mod}$), where the latter are measured in an identical manner but using a model data cube with circular velocity $v_\mathrm{circ}(R_\mathrm{gal})$.
As a result, $\phi_\mathrm{mod}$ is along the local isovelocity contour while $\omega_\mathrm{mod}$ is the magnitude of the expected velocity gradient.
As can be inferred from the left panel of \autoref{fig:shear} by comparing the three different regions, $\omega_\mathrm{mod}$ decreases with increasing galactocentric distance.
{For example, all GMCs with $\omega_\mathrm{mod}>0.55~{\rm km~s^{-1}}$ are within the central region defined by $R_{\rm gal}<100$~pc, and $80\%$ of GMCs at $R_{\rm gal}<100$~pc have $\omega_\mathrm{mod}>0.55~{\rm km~s^{-1}}$.}
This is due to a higher density of isovelocity contours in the central part of the galaxy, resulting from {both} a LoS projection effect {and the rapid change of the circular velocity near the centre}.
Most importantly, in both panels of \autoref{fig:shear}, there is no significant correlation between the observed and modelled quantities.
This is true for both the whole sample and the GMCs of individual regions.
The inner-region GMCs (blue data points) may have a weak positive correlation, but almost all inner-region GMCs have $\omega_\mathrm{obs}<\omega_\mathrm{mod}$.
We come back to this sub-population of GMCs in \autoref{sec:subsample}.

An alternative potential origin for the internal velocity gradients of GMCs is turbulence. Given the limited spatial resolution of our data, we can not currently measure the internal turbulence of the clouds.
The velocity gradients caused by turbulence can however be estimated in simulations, and \citet{2000ApJ...543..822B} report $(\omega_\mathrm{{turb}}/\mathrm{km~s^{-1}~pc^{-1}})=1.6\,(R_\mathrm{c}/0.1~\mathrm{pc})^{-1/2}$.
The expected $\omega_\mathrm{{turb}}$ caused by turbulence for GMCs of the sizes observed in NGC~1387 are thus $0.1$ -- $0.2$~km~s$^{-1}$~pc$^{-1}$, of the same magnitude as the measured $\omega_\mathrm{obs}$.
Turbulence might thus be at the origin of the observed GMC velocity gradients of NGC~1387, although further evidence is required to confirm this.

\subsection{Tidal radii}
\label{sec:tidal}

To further understand the potential impact of external (i.e.\ the galaxy's) gravity on the clouds, we estimate the tidal radii $R_\mathrm{t}$ of the resolved GMCs of NGC~1387. The $R_\mathrm{t}$ is defined as the distance from each GMC centre where the shear velocity due to differential galactic rotation is equal to the escape velocity of the GMC \citep{1991ApJ...378..565G, 2000ApJ...536..173T}.
We adopt a simplified form of equation~(52) of \citet{2021MNRAS.505.4048L}, which assumes a spherical galaxy mass distribution:
\begin{equation}
    R_\mathrm{t}=\left(\frac{G}{2A^2}\right)^{1/3}\,M_\mathrm{gas}^{1/3}\,\,,
\end{equation}
where $A$ is Oort's constant $A$, evaluated at the galactocentric distance of the GMC.

\begin{figure}
    \includegraphics[width=\columnwidth]{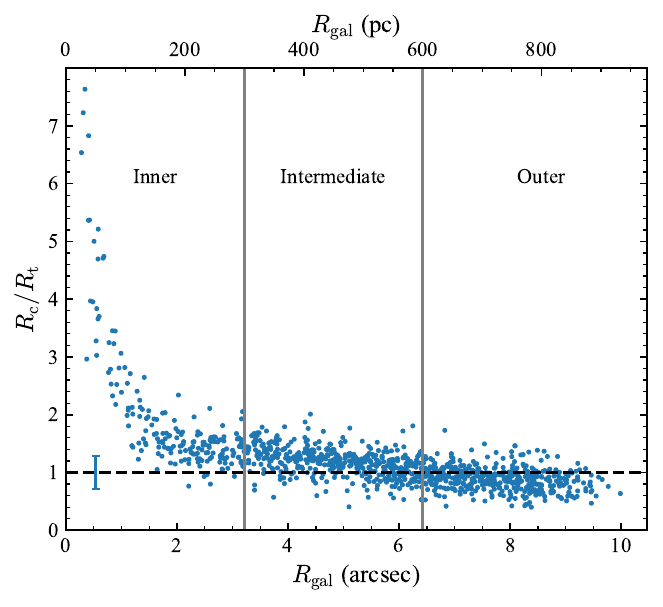}
    \caption{Measured GMC size to predicted tidal radius ratios ($R_\mathrm{c}/R_\mathrm{t}$) of the resolved GMCs of NGC~1387, as a function of GMC galactocentric distance $R_\mathrm{gal}$. The black horizontal dashed line indicates $R_\mathrm{c}=R_\mathrm{t}$ and also shows the median uncertainty of $R_\mathrm{c}/R_\mathrm{t}$ as a vertical error bar in the bottom-left corner. The pale grey vertical lines indicate the boundaries of the three regions defined in \autoref{sec:mom}.}
    \label{fig:tidal_radius}
\end{figure}

The measured GMC size to predicted tidal radius ratios $R_\mathrm{c}/R_\mathrm{t}$ of the resolved GMCs of NGC~1387 are shown as a function of GMC galactocentric distance $R_\mathrm{gal}$ in \autoref{fig:tidal_radius}.
Overall, most GMCs have sizes similar to their tidal radii, 
suggesting that most clouds are dominated by self-gravity rather than external gravity, although it could be that the cloud sizes have been limited by external gravity in the first place (thus driving $R_\mathrm{c}/R_\mathrm{t}$ to $1$).
These results are consistent with the virial parameters estimated in \autoref{sec:virial_parameter} and the kinematic analyses presented in \autoref{sec:kine_gal}, showing that most GMCs are decoupled from the large-scale galactic rotation.

\autoref{fig:tidal_radius} however also reveals a strong trend with galactocentric distance, whereby the inner-region GMCs essentially all have $R_\mathrm{c}>R_\mathrm{t}$, and their $R_\mathrm{c}/R_\mathrm{t}$ ratio increases rapidly with decreasing galactocentric distance. This primarily arises from rapidly increasing $A$ (i.e.\ shear) and thus rapidly decreasing $R_\mathrm{t}$ with decreasing galactocentric distance, as $R_\mathrm{c}$ varies little with $R_\mathrm{gal}$ (see \autoref{fig:properties}). 
This {predicts} that external gravity {should be} more important than self-gravity for the inner-region GMCs, and that its importance decreases with increasing galactocentric distance, as expected. In turn, this {further predicts} that the inner-region GMCs can not be gravitationally bound. 
{They can either be transient structures that should be sheared apart on a short timescale or} be bound by other forces.
While the virial parameters of the inner-region GMCs are indeed higher than those of the clouds farther out (see \autoref{fig:properties}), they remain overall modest, only a few inner-region clouds having $\alpha_\mathrm{obs,vir}>2$. The implications for the gravitational boundedness of the inner-region clouds arising out of the $R_\mathrm{c}/R_\mathrm{t}$ ratio analysis and the virial parameters thus appear contradictory. However, the concept of $R_\mathrm{t}$ is based on the assumption that the entirety of each GMC follows the galactic circular rotation. Throughout \autoref{sec:kine}, we demonstrate that this assumption does not hold for the {majority of} the GMCs of NGC~1387, particularly for the sub-sample of GMCs with $R_{\rm c}>R_{\rm t}$ (see \autoref{sec:subsample}). This may explain the limited predicting power of $R_\mathrm{t}$ regarding cloud boundedness indicated by $\alpha_\mathrm{obs,vir}$.

\subsection{Galactic rotation and GMC velocity gradients}
\label{sec:gs}

Although the observed internal velocity gradients of GMCs discussed in \autoref{sec:kine_gal} appear to arise primarily from internal turbulence, the fraction of prograde clouds is significantly greater than $50\%$ in all three regions, so the contribution of the large-scale galactic rotation to the gradients is potentially non-negligible.
We try to quantify this contribution in this sub-section, and follow \citet{2015ApJ...803...16U} and \citet{2021MNRAS.505.4048L} by estimating the `gradient-subtracted velocity dispersion' $\sigma_\mathrm{gs,los}$ of each cloud, arguably a better tracer of internal GMC turbulence than $\sigma_\mathrm{obs,los}$. 

The quantity $\sigma_\mathrm{gs,los}$ is estimated with the intention to remove the contribution of ordered internal bulk rotation from $\sigma_\mathrm{obs,los}$.
All the spectra of a given GMC along different lines of sight are thus shifted (according to their mean velocities) to a common systemic velocity before being stacked for the measurement of (the now gradient-subtracted) $\sigma_\mathrm{gs,los}$.
Under the assumption that internal GMC bulk motions arise purely from large-scale galactic circular rotation, the quadratic difference can be expressed as 
\begin{equation}
    \sigma_\mathrm{obs,los}^2-\sigma_\mathrm{gs,los}^2\approx b_\mathrm{e}R_\mathrm{c}^2\left(\Omega^2\sin^2\theta+(\Omega-2A)^2\cos^2\theta\right)\sin^2 i
\end{equation}
(see equations~(19), (20), and (B23) of \citealt{2021MNRAS.505.4048L}), where $i$ is the disc inclination, $\theta$ the azimuthal angle of the GMC in the galaxy plane and $\Omega$ the galactic circular angular velocity evaluated at the galactocentric distance of the GMC (i.e.\ $v_\mathrm{circ}(R_\mathrm{gal})/R_\mathrm{gal}$).
The geometric parameter $b_\mathrm{e}$ is again taken to be $1/5$, corresponding to a homogeneous mass volume density distribution (and allowing to compare with most previous works on GMCs).

\begin{figure*}
\centering
    \resizebox{\textwidth}{!}{
    \includegraphics[height=0.18\textheight]{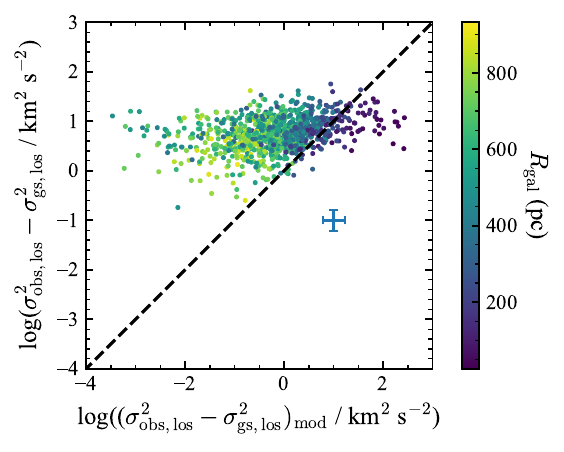} 
    \includegraphics[height=0.18\textheight]{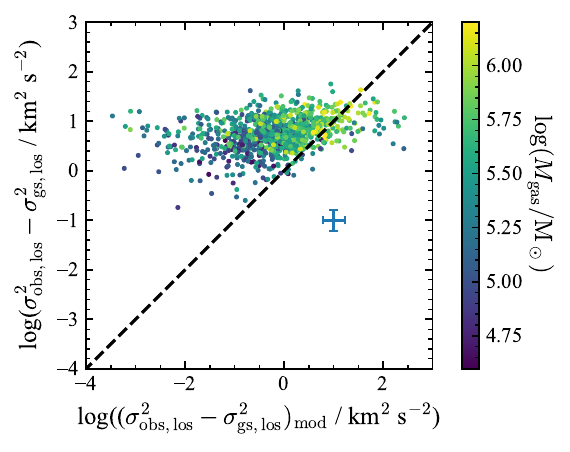} 
    \includegraphics[height=0.18\textheight]{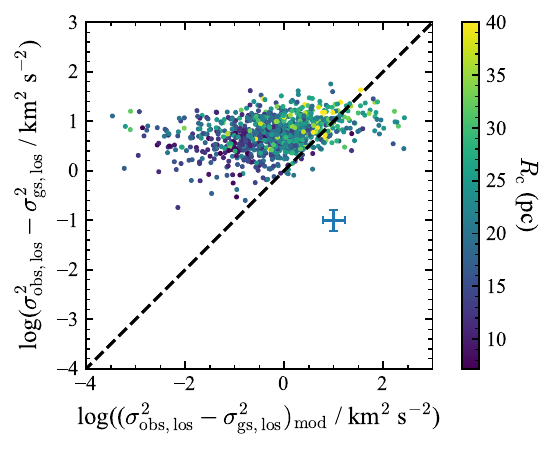} }
    \caption{Comparison of the observed and modelled differences $\sigma_\mathrm{obs,los}^2-\sigma_\mathrm{gs,los}^2$ for the resolved GMCs of NGC~1387.
    From left to right, different panels show the same measurements with different colour coding: $R_\mathrm{gal}$, $M_\mathrm{gas}$ and $R_\mathrm{c}$.
    The dashed lines show the $1:1$ relations.
    The median error bar is shown in the centre-right region of each panel.
    }
    \label{fig:sigma_gs_all}
\end{figure*}

By comparing the measured and modelled differences $\sigma_\mathrm{obs,los}^2-\sigma_\mathrm{gs,los}^2$, we can assess the contribution of the large-scale galactic circular rotation to the GMC velocity gradients.
\autoref{fig:sigma_gs_all} shows this comparison for all resolved GMCs of NGC~1387.
Different panels show the same measurements with different colour coding ($R_\mathrm{gal}$, $M_\mathrm{gas}$ and $R_\mathrm{c}$).
The overall distribution shows at best a very weak positive correlation {(the Pearson correlation coefficient $r_{\rm p} = 0.32$)} between the observed and modelled quantities, far from the one-to-one line expectation.
However, larger and more massive GMCs at small galactocentric distances tend to be closer to the one-to-one line.
This suggests that the GMCs of this particular sub-sample have their velocity gradients dominated by large-scale galaxy rotation, while most other GMCs are dominated by internal turbulence.

We note that the limited spectral resolution (channel width of $2$~km~s$^{-1}$) of the data could cause a bias to the measured {$\sigma_\mathrm{obs,los}$ and $\sigma_\mathrm{gs,los}$ (despite the spectral deconvolution), but it should not affect the modelled $(\sigma_\mathrm{obs,los}^2-\sigma_\mathrm{gs,los}^2)_{\rm mod}$. This bias is hard to quantify, but it should be more pronounced for smaller $\sigma_\mathrm{obs,los}$ and $\sigma_\mathrm{gs,los}$.}

\subsection{GMC sub-samples}
\label{sec:subsample}

\begin{figure*}
\centering
    \begin{subfigure}[c]{0.67\textwidth}
        \centering
        \includegraphics[width=\textwidth]{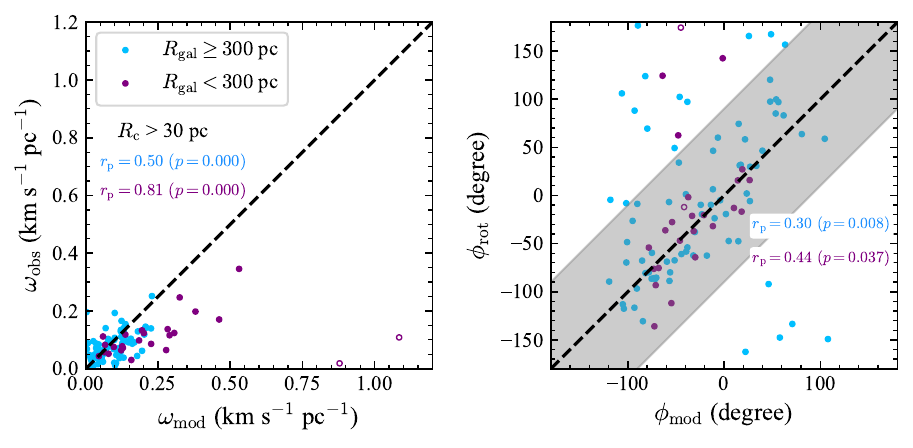} 
    \end{subfigure}
        \begin{subfigure}[c]{0.32\textwidth}
        \centering
        \includegraphics[width=\textwidth]{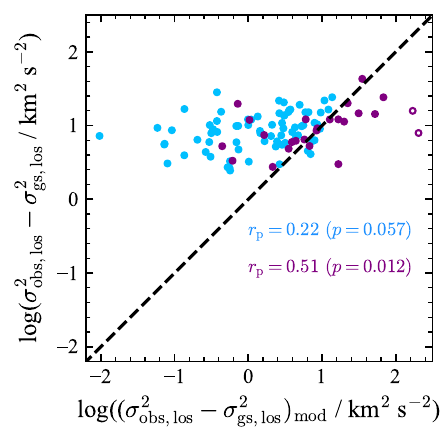} 
    \end{subfigure}

    \begin{subfigure}[c]{0.67\textwidth}
        \centering
        \includegraphics[width=\textwidth]{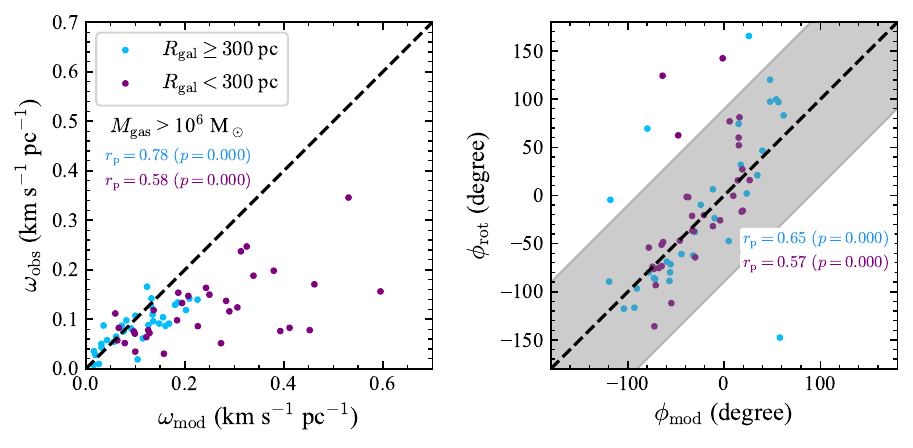} 
    \end{subfigure}
    \hfill
    \begin{subfigure}[c]{0.32\textwidth}
        \centering
        \includegraphics[width=\textwidth]{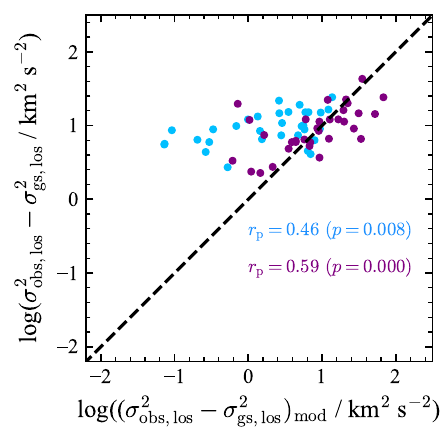} 
    \end{subfigure}
    
        \begin{subfigure}[c]{0.67\textwidth}
        \centering
        \includegraphics[width=\textwidth]{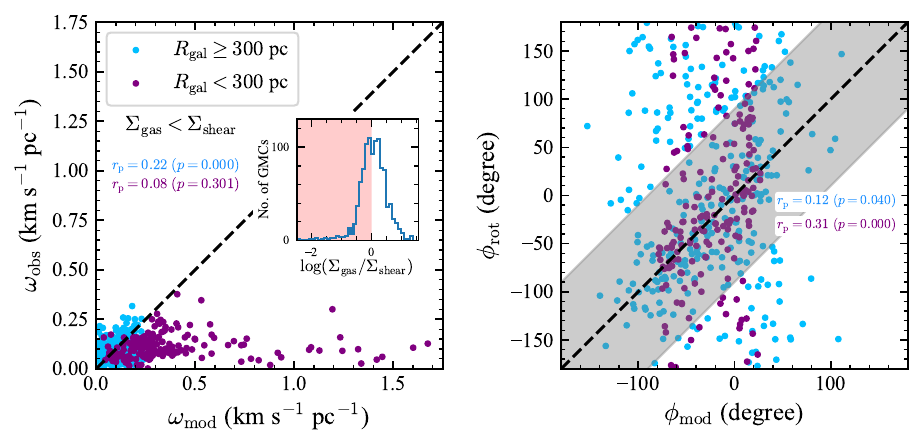} 
    \end{subfigure}
    \hfill
    \begin{subfigure}[c]{0.32\textwidth}
        \centering
        \includegraphics[width=\textwidth]{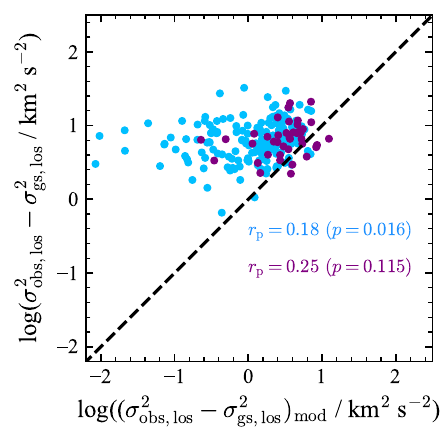}  
    \end{subfigure}

    \caption{Comparison of the observed and modelled GMC projected angular velocity magnitudes ($\omega_\mathrm{obs}$ and  $\omega_\mathrm{mod}$, respectively; left column), position angles ($\phi_\mathrm{rot}$ and $\phi_\mathrm{mod}$, respectively; middle column) and quadratic velocity dispersion differences ($\sigma_\mathrm{obs,los}^2-\sigma_\mathrm{gs,los}^2$ and $(\sigma_\mathrm{obs,los}^2-\sigma_\mathrm{gs,los}^2)_\mathrm{mod}$, respectively; right column), for three resolved GMC sub-samples of NGC~1387: $R_\mathrm{c}>30$~pc (top row), $M_\mathrm{gas}>10^6$~M$_\odot$ (middle row) and $\Sigma_\mathrm{gas}<\Sigma_\mathrm{shear}$ (bottom row).
    The inset in the bottom-left panel shows the distribution of the (logarithm of the) $\Sigma_\mathrm{gas}/\Sigma_\mathrm{shear}$ ratios of all the resolved GMCs of NGC~1387.
    Only the GMCs in the pink-shaded region ($\Sigma_\mathrm{gas}<\Sigma_\mathrm{shear}$) are plotted in the panels of this row.
    The data points are colour-coded according to the GMC galactocentric distance (purple, $R_\mathrm{gal}<300$~pc; blue,  $R_\mathrm{gal}\ge300$~pc).
    The black dashed lines show the $1:1$ relations.
    The grey shaded regions in the middle column indicate an angular difference $\le90\degr$, which we adopt to define prograde clouds (conversely for retrograde clouds).
    {The Pearson correlation coefficients ($r_{\rm p}$) and $p$-values of all correlations are stated, differentiating by the galactocentric distance following the same colour coding as the data points. In the first row, two outliers (hollow purple circles) are ignored from the calculation of $r_{\rm p}$; see the main text for details.}
    Note the different axis ranges of the panels in the left column.
    }
    \label{fig:kine_subsample}
\end{figure*}

Although the overall GMC population of NGC~1387 does not seem to be significantly affected by the large-scale galactic rotation, a specific sub-sample (primarily large and massive inner-region GMCs) does seem to be.
Using clues from the previous sub-sections, we thus construct three specific GMC sub-samples to test this possibility, requiring $R_\mathrm{c}>30$~pc, $M_\mathrm{gas}>10^6$~M$_\odot$ and $\Sigma_\mathrm{gas}<\Sigma_\mathrm{shear}$, respectively, where $\Sigma_\mathrm{shear}=4A^2R_\mathrm{c}/3 {\rm \pi G}$ is the minimum surface mass density required to remain bound against shear \citep{2021MNRAS.505.4048L}.
The last criterion selects low-surface density GMCs, which are more susceptible to galactic shear and thus more likely to align with the large-scale galactic rotation.
We also separate the GMCs by galactocentric distance in all the following comparisons, using $R_\mathrm{gal}=300$~pc (the boundary between the inner region and the intermediate region) as the threshold (all the thresholds were determined via trial and error, based on the tentative trends discussed in the previous sub-sections).
In \autoref{fig:kine_subsample}, we thus revisit the comparisons discussed in Sections~\ref{sec:kine_gal} and \ref{sec:gs} for these three sub-samples.

The first row of \autoref{fig:kine_subsample} shows the sub-sample of large resolved GMCs ($R_\mathrm{c}>30$~pc).
A positive correlation exists between the observed and modelled quantities of inner-region GMCs ($R_\mathrm{gal}<300$~pc, purple data points) in the left panel, although most data points are still lower than the $1:1$ line (the black dashed line).
{The two apparent outliers (shown as hollow purple circles) with $\omega_{\rm mod}>0.8~{\rm km~s^{-1}~pc^{-1}}$ have the smallest $R_{\rm gal}$ ($\lesssim 62~$pc) of this sub-sample (generally $>100~$pc). Their deviations from the main trend are probably due to the most extreme beam smearing effects and the most complex physical conditions close to the galaxy centre. 
If these two outliers are ignored, the Pearson correlation coefficient $r_{\rm p}=0.81$, indeed indicating a strong correlation.}
In the middle panel, most inner-region GMCs are close to the $1:1$ line, with only a few outliers. {With the same two GMCs excluded, $r_{\rm p}=0.44$ (nearly $1\sigma$ significance), indicating a weak correlation.}
In the right panel, the inner-region GMCs {(excluding again the same two GMCs) show a moderate} correlation {with $r_{\rm p}=0.51$ (more than $2\sigma$ significance) and moderate agreement with the $1:1$ line}.
The large GMCs of the intermediate region and the outer region ($R_\mathrm{gal}\geq 300$~pc) {have much lower $r_{\rm p}$ than those of the inner GMCs in each of the three panels. They also show much larger scatters and/or deviations from the $1:1$ line.} 

The second row of \autoref{fig:kine_subsample} shows the sub-sample of massive resolved GMCs ($M_\mathrm{gas}>10^6$~M$_\odot$).
There is a positive correlation {and/or substantial agreement with the $1:1$ line} in all three panels for all galactocentric distances. 
For this row (and the third row below), there is no apparent outlier to be excluded from the statistics. 
The $r_{\rm p}$ generally indicate strong or moderate correlations ($0.57$ -- $0.78$) and one weak correlation ($0.46$).
Most importantly, these correlations are much stronger than those of the entire GMC sample discussed in the previous sub-sections.

The third row of \autoref{fig:kine_subsample} shows the sub-sample of resolved GMCs with $\Sigma_\mathrm{gas}<\Sigma_\mathrm{shear}$.
There is no correlation {($r_{\rm p}=0.31$ at most) or $1:1$ agreement} between the observed and modelled quantities in any of the three panels at any galactocentric distance. 
 
We also tested resolved GMC sub-samples with $\Sigma_\mathrm{gas}<100$~$\mathrm{M_\odot~pc^{-2}}$, $\Sigma_\mathrm{gas}>500$~$\mathrm{M_\odot~pc^{-2}}$, small $\sigma_\mathrm{obs,los}$, large $\sigma_\mathrm{obs,los}$ and $R_\mathrm{c}>R_\mathrm{t}$ (not shown in \autoref{fig:kine_subsample}).
No correlation exists for any of these sub-samples.
In particular, {when a cloud has $R_\mathrm{c}>R_\mathrm{t}$, we expect galactic shear to strongly shape the cloud kinematics, as also discussed in \autoref{sec:tidal}.}
The absence of a correlation for the $R_\mathrm{c}>R_\mathrm{t}$ sub-sample, {along with the modest/low $\alpha_\mathrm{obs,vir}$ mentioned in Sections~\ref{sec:virial_parameter} and \ref{sec:tidal}}, suggests that although the discussion concerning the tidal radii in \autoref{sec:tidal} is physically motivated, further investigation of this aspect is required.

In summary, we see emerging correlations between GMC internal rotation and large-scale galactic circular rotation primarily for the most massive ($M_\mathrm{gas}>10^6$~M$_\odot$) resolved GMCs of NGC~1387, but to a lesser extent also for the largest ($R_\mathrm{c}>30$~pc) and/or innermost ($100<R_\mathrm{gal}<300$~pc) GMCs.
The requirement of small $R_\mathrm{gal}$ can be interpreted as the requirement of a steeply rising circular velocity curve and thus large shear, for which one expects stronger GMC coupling with galactic rotation.
For the other two requirements, more massive GMCs simply tend to be larger (and vice-versa), and larger GMCs are more affected by shear ($\Sigma_\mathrm{shear}\propto R_\mathrm{c}$).
As a GMC evolves, the (de-)coupling of its internal velocity gradients from galactic rotation is a competition between galactic shear and processes such as gas disc fragmentation (involving gravitational collapse; \citealt{2019A&A...621A.140J}), gas accretion \citep{2024MNRAS.528.2199R}, GMC mergers \citep{2023MNRAS.519.1887S} {and stellar feedback \citep{2024A&A...691A..70N}}.
Galactic shear imprints galactic differential rotation to the GMC, while all the others increase turbulence. To reveal the evolutionary path of a GMC, the relative strengths and timescales of these processes must be explored.

\section{Toomre parameter and cloud-cloud collision model}
\label{sec:toomre}

Since we have demonstrated in \autoref{sec:kine} that the majority of the resolved GMCs of NGC~1387 (except those with $R_\mathrm{gal}<300$~pc) are not significantly influenced by the large-scale galaxy rotation and shear, we now explore another mechanism potentially regulating GMC properties, i.e.\ cloud-cloud collisions.
\citet{2022MNRAS.517..632L} showed that where the Toomre instability parameter is below its critical threshold of unity, the disc is unstable and cloud-cloud collisions prevail in the galaxy NGC~404.

The Toomre parameter of a molecular gas disc is
\begin{equation}
    \label{eq:toomre}
    Q\equiv\frac{\kappa\sigma_\mathrm{disc}}{\pi {\rm G}\Sigma_\mathrm{mol}}
\end{equation}
\citep[e.g.][]{1964ApJ...139.1217T, 2008gady.book.....B}, where $\kappa$ is the epicyclic frequency (calculated from the circular velocity curve) and $\sigma_\mathrm{disc}$ is the azimuthally averaged molecular gas disc velocity dispersion (calculated from the second-moment map).
Even though the circular velocity curve is largely determined by the stellar contribution to the gravitational potential, this formulation only accounts for the mass surface density of the molecular gas, thus ignoring that of the stellar component.
As the stars in the inner parts of NGC~1387 are likely to be distributed in a rather thick spheroidal component, rather than in a thin disc coplanar with the molecular gas disc, adding their contribution to $Q$ would be non-trivial.
The $Q$ resulting from \autoref{eq:toomre} should thus be considered an upper limit (as any additional mass in the disc will only lower $Q$; e.g.\ \citealt{1994ApJ...427..759W}). 

\begin{figure}
    \includegraphics[width=\columnwidth]{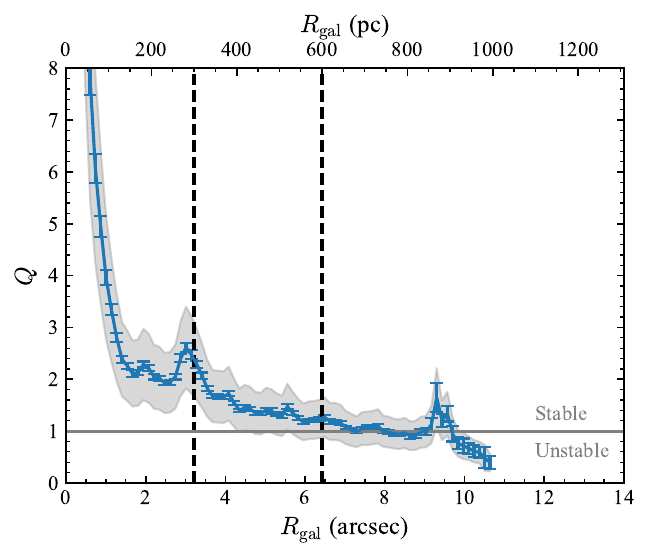} 
    \caption{
    Galactocentric radial profile of the Toomre instability parameter $Q$ (blue curve) of the molecular gas disc of NGC~1387.
    Error bars are $1$~$\sigma$ uncertainties propagated from the measurement uncertainties, while the grey shaded region includes an additional $30\%$ uncertainty to account for the CO-to-molecular gas conversion uncertainty.
    The black dashed vertical lines indicate the boundaries of the three regions defined in \autoref{sec:mom}.
    The grey horizontal solid line indicates $Q=1$.}
    \label{fig:toomre}
\end{figure}

\autoref{fig:toomre} shows the $Q$ galactocentric radial profile of the molecular gas disc of NGC~1387 (blue curve). 
$Q$ is high in the inner region (although at least partially because of beam smearing, leading to overestimated velocity dispersions; see \autoref{sec:mom}), but it generally decreases with galactocentric distance, primarily due to decreases of $\kappa$ and $\sigma_\mathrm{disc}$ rather than an increase of $\Sigma_\mathrm{mol}$.
At $R_\mathrm{gal}\gtrsim500$~pc, $Q$ remains below $\approx1.5$, close to the critical threshold of unity.
Even ignoring the stellar component, the molecular gas disc is thus close to being gravitationally unstable and having a tendency to fragment in that region. In turn, this means that cloud-cloud collisions could be important in the outer half of the molecular gas disc.
The bump in the $Q$ radial profile at $R_\mathrm{gal}\approx300$~pc is due to the narrow dip of $\Sigma_\mathrm{mol}$ at the same galactocentric distance (see the bottom-right panel of \autoref{fig:maps}), so we do not discuss this feature further.

Although $Q$ is not low enough anywhere to provide decisive support for the cloud-cloud collision model, one prediction of the model is that the slope of the size -- line width relation is $1/3$ (equation~(18) of \citealt{2022MNRAS.517..632L}).
The best-fitting slope of $0.29\pm0.06$ for all the resolved clouds of NGC~1387, 
as well as the slopes of the clouds in the intermediate and outer regions (see \autoref{tab:larson}), are in perfect agreement with that prediction. This is not the case for the clouds of the inner region, but $Q\gtrsim2$ there.

To probe other specific predictions of the cloud-cloud collision model, resolving the predicted characteristic cloud size is required, but this is predicted to always be $\lesssim3$~pc in the outer parts of the NGC~1387 molecular gas disc (equation~(36) of \citealt{2022MNRAS.517..632L}), significantly below the spatial resolution of our observations.
Higher resolution observations are thus required to test this model further.

\section{Radial gradients of GMC properties}
\label{sec:gradient}

\begin{figure*}
    \includegraphics[width=0.8\textwidth]{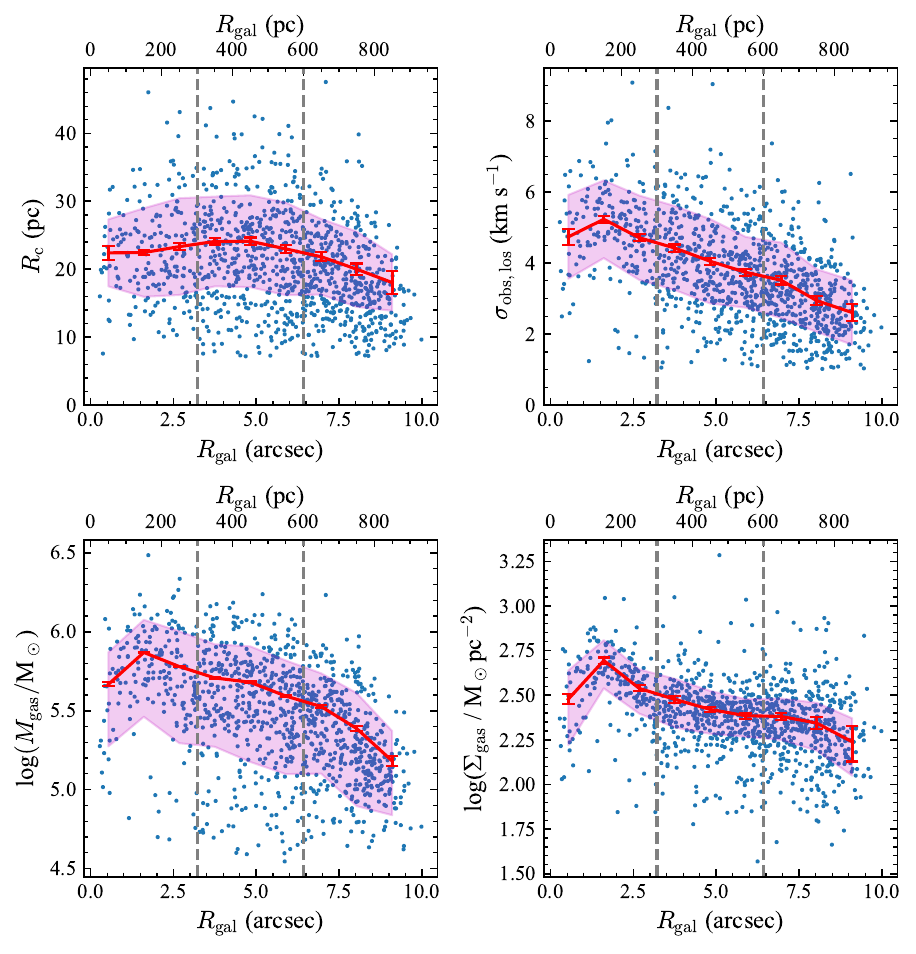}
    \caption{Properties of the resolved GMCs of NGC~1387 as a function of galactocentric distance:
    $R_\mathrm{c}$ (top-left), $\sigma_\mathrm{obs,los}$ (top-right), $M_\mathrm{gas}$ (bottom-left) and $\Sigma_\mathrm{gas}$ (bottom-right).
    Data points are shown in blue, azimuthal averages in red.
    For the averages, only measurements larger than their uncertainties are considered.
    Error bars indicate the uncertainty of the mean of each radial bin, while the magenta-shaded regions show the scatter (standard deviation) within each radial bin.
    The grey dashed vertical lines indicate the boundaries of the three regions defined in \autoref{sec:mom}.}
    \label{fig:radial}
\end{figure*}

As pointed out before (Sections~\ref{sec:GMC_properties} and \ref{sec:larson}), most properties of the resolved clouds of NGC~1387 reveal galactocentric distance (i.e.\ radial) gradients.
This is not the case for previously studied ETGs (e.g.\ NGC~4526, \citealt{2015ApJ...803...16U}; NGC~4429, \citealt{2021MNRAS.505.4048L}).
\autoref{fig:radial} shows the four fundamental properties $R_\mathrm{c}$, $\sigma_\mathrm{obs,los}$, $M_\mathrm{gas}$ and $\Sigma_\mathrm{gas}$ as a function of galactocentric distance in NGC~1387 (see also \autoref{fig:properties}).
The blue data points show individual resolved GMCs, while the red lines show azimuthal averages in galactocentric distance bins of width $100$~pc.
For the averages, only measurements larger than their uncertainties are considered.
All properties peak near the centre at $R_\mathrm{gal}\approx150$~pc ($R_\mathrm{c}$ at a slightly larger galactocentric distance), with a sharp drop in the innermost radial bin and a steady decrease with increasing galactocentric distance. 
Considering the $R_\mathrm{c}$ and $M_\mathrm{gas}$ trends, we interpret the decrease of $\Sigma_\mathrm{gas}$ from $\approx150$ to $550$~pc to be primarily driven by $M_\mathrm{gas}$, while the flattening of $\Sigma_\mathrm{gas}$ beyond that is driven by the simultaneous decrements of $R_\mathrm{c}$ and $M_\mathrm{gas}$. 

\section{Similarities to the MW disc}
\label{sec:physics_discussion}

Recent studies have shown that GMCs in extragalactic environments (especially in ETGs; \citealt{2015ApJ...803...16U, 2021MNRAS.505.4048L}) do not necessarily follow the MW disc or Local Group Larson relations, which were historically thought to be universal.
In light of this, the GMCs of NGC~1387 are surprisingly similar to those of the MW disc, suggesting a more varied and complex regulation of GMC properties {in ETGs}.

\subsection{{Deviation from other ETGs}}
\label{sec:etg_compare}

{We first summarise the key differences of GMC properties between NGC~1387 and the other ETGs. For the same reasons explained in \autoref{sec:GMC_properties}, we focus on the comparison between NGC~1387 against NGC~4526 and NGC~4429.
One major deviation of GMC properties is in the size -- line width parameter space. NGC~1387 shows a weak correlation ($r_{\rm s} = 0.47$) overlapping with the MW disc trend (see \autoref{sec:larson}). The GMCs of the other ETGs do not show such a correlation with $r_{\rm s} = 0.13$ for NGC~4429 \citep[fig.~13 and section~5.6 of][]{2021MNRAS.505.4048L} and $r_{\rm s} = -0.14$ for NGC~4526 \citep{2015ApJ...803...16U}. In addition, NGC~4429 and NGC~4526 GMCs all have higher line widths at a given size than MW disc GMCs, which are comparable to the MW centre trend. This may reflect a significant difference of turbulence levels in the GMCs in different ETGs. Similar discrepancies between NGC~1387 against NGC~4526 and NGC~4429 can also be found in the other two Larson scaling relations as well. 
Although NGC~4526 GMCs may be affected by the poorer spectral resolution (10~km~s$^{-1}$) to be biased high in line width, NGC~4429 data have the same spectral resolution as those of NGC~1387.

In addition to the dynamical state of GMCs, the kinematic orientation is the other major difference. The spins of NGC~1387 GMCs generally do not follow the galactic rotation (except for specific sub-samples), in drastic contrast with the other two ETGs (see details in \autoref{sec:kine_gal}). This is evidence for the deviation from the physical picture where shear drives GMC properties.
}

\subsection{{Origin of MW-disc-like GMC properties}}
\label{sec:mw_compare}

Apart from the galaxy morphological type (almost certainly a proxy for more relevant physical properties), we explore here other possible regulators {or tracers} of GMC properties.

Recent star formation may be one such {quantity. It reflects the recent dynamical state of GMCs (i.e. boundedness and tendency to collapse). Both NGC~4526 and NGC~4429 have little recent star formation activity \citep{2015ApJ...803...16U,2021MNRAS.505.4048L}.}
\citet{2019A&A...627A.136I} and \citet{2020MNRAS.496.2155Z} studied the SFRs and the star-formation efficiencies (SFEs), respectively, of a sample of galaxies including NGC~1387 as part of the Fornax3D project and the ALMA Fornax Cluster Survey.
The SFR of NGC~1387 is in the range $0.008$ -- $0.082$~M$_\odot$~yr$^{-1}$, depending on the correction of the H$\alpha$ flux for the {low ionisation nuclear emission line region (LINER)}.
{In light of the positive correlation between the total SFRs and the total stellar masses ($M_*$) of galaxies, i.e.\ the so-called star-forming main sequence, we compare the SFRs and $M_*$ of NGC~1387 and the MW.
The MW $M_*$ is $(6.08\pm1.14)\times10^{10}$~M$_\odot$ \citep{2015ApJ...806...96L}, 30\% higher than the $M_* = 4.7 \times 10^{10}$~M$_\odot$ of NGC~1387 \citep{2019A&A...623A...1I}. However, the MW SFR of $1.65\pm0.19$~M$_\odot$~yr$^{-1}$ \citep{2015ApJ...806...96L} is at least $20$ times higher than the NGC~1387 SFR. Indeed, NGC~1387 is $1$ -- $2$~dex below (i.e.\ at a lower SFR than) the star-formation main sequence of the local universe \citep{2014ApJS..214...15S}.
}
The SFE of NGC~1387 is in the range $-10.6 < \log(\mathrm{SFE}/\mathrm{yr^{-1}}) < -9.6$ (with our adopted $\alpha_{\rm CO(1-0)}$, same below), corresponding to a depletion time $\tau_\mathrm{dep}$ of $4$ -- $40$~Gyr, and the specific star-formation rate (sSFR) is in the range $-12.7 < \log(\mathrm{sSFR}/ \mathrm{yr^{-1}}) < -11.7$.
In the $\tau_\mathrm{dep}$ -- sSFR diagnostic diagram of \citet{Saintonge:2011}, NGC~1387 would be located in the region of high $\tau_\mathrm{dep}$ and small sSFR, along with other ETGs.

These parameters and comparisons indicate that the SFR of NGC~1387 is consistent with its early-type morphology.
Conversely, its star formation {does not echo its MW disc-like GMC properties. 
While these global properties (especially the SFE) provide an average indication of recent star formation (per unit cold gas amount), a spatially resolved study is desirable to constrain the environment in the vicinity of each GMC.
Such a study challengingly requires an accurate decomposition of the H$\alpha$ flux into star-formation and {LINER} contributions, or alternatively the use of another SFR indicator (free of {LINER} contamination).
\citet{2020MNRAS.496.2155Z} presented a spatially resolved analysis of the SFE of NGC~1387 (see their fig.~B4). They used line ratio diagnostics to separate star-forming dominated regions and composite regions (rather than decomposing the H$\alpha$ flux of each region). The majority of the star-forming regions have $\tau_\mathrm{dep}\approx20$~Gyr (with our adopted $\alpha_{\rm CO(1-0)}$), while the majority of composite regions have lower limits $\tau_\mathrm{dep}>3$~Gyr (again with our adopted $\alpha_{\rm CO(1-0)}$). Their results thus again indicate little star-formation activity per unit cold gas mass.
}
A more comprehensive study of the suppression of star formation {and stellar feedback} in NGC~1387 in the future would be desirable, to explore any potential causal link.

The gas-phase metallicity is another potential regulator of GMC properties, as metals serve as coolants of GMCs and affect processes such as the fragmentation of gas discs \citep[e.g.][]{2010A&A...510A.110H}.
With Very Large Telescope (VLT) Multi Unit Spectroscopic Explorer (MUSE) data from the Fornax3D project \citep[PI: M.\ Sarzi and E.\ Iodice,][]{2018A&A...616A.121S,2022A&A...660A.105L,2022Msngr.189....9S}, and using the \citet{Dopita:2016} calibration, we estimate the median gas-phase metallicity of the star-forming regions of NGC~1387 (identified through standard emission-line ratio diagnostic diagrams) to be $12+\log(\mathrm{O}/\mathrm{H})=9.07$, which is $0.4$~dex above solar metallicity.
Compared with the mass-metallicity relation shown in fig.~1 of \citet{2022A&A...660A.105L}, who used the same metallicity calibration, at a total stellar mass of $4.7\times10^{10}$~M$_\odot$ the gas-phase metallicity of NGC~1387 is $0.25$~dex larger than that of the general trend, and it is also higher than the MW metallicity \citep[e.g.][]{2011ApJ...738...27B}.

ETGs are generally located at the high-mass and high-metallicity end of the mass-metallicity relation, with a scatter of about $0.3$~dex (see fig.~9 of \citealt{2019MNRAS.484..562G} and references therein), suggesting that NGC~1387 is just a typical ETG in term of metallicity.
Furthermore, the gas-phase metallicity of galaxies is largely determined by the origin of the gas, i.e.\ either high-metallicity gas recycled from stellar mass loss \citep{2008ApJ...681.1215P,2010ApJ...714L.275M} {and/or accreted through hot halo cooling \citep{2015MNRAS.448.1271L}}, or low-metallicity gas accreted from {neighbouring} galaxies \citep{2019MNRAS.489L.108D,2019MNRAS.489.3739R,2022MNRAS.510.4485R,2023A&A...675A..59M}.
For NGC~1387, judging from its location near the centre of the Fornax Cluster, its deficit of atomic gas and the perfect kinematic alignment between cold gas and stars (see \autoref{sec:target}), the origin of its cold gas is {very likely} internal, in line with the analyses of cluster ETGs of \citet{2011MNRAS.417..882D}. Thus, the early-type classification, the metallicity and the likely origin of the molecular gas of NGC~1387 are all consistent with each other, but they do not provide an explanation of the GMC properties. A related but converse example is the ETG NGC~5128 (Centaurus~A), which harbours externally accreted cold molecular gas and exhibits LTG-like GMC properties \citep{2021MNRAS.504.6198M}.
{We are not aware of any gas-phase metallicity measurement for NGC~4429 or NGC~4526 in the literature.}

A third {physical quantity} we consider is the slope (power-law index $\beta$) of the spatial power spectrum of the CO zeroth-moment map, which has been used to characterise turbulent cascades and gas fragmentation \citep[e.g.][]{2004ARA&A..42..211E}.
\citet{2023MNRAS.526.5590G} measured the power-law index of the CO disc of NGC~1387 to be $2.09\pm0.14$, one of eight observed ETGs with $2.05<\beta<3.02$ and four observed LTGs with $2.45<\beta<2.81$. 
NGC~1387 is thus at the low end of the $\beta$ distribution of all observed galaxies in \citet{2023MNRAS.526.5590G}, {and it has a $\beta$ smaller than that of NGC~4429 ($\beta=2.34\pm0.22$)}. NGC~1387 is thus not typical of either the ETG or the LTG population.
However, compared with a sample of simulated galaxies from \citet{2023MNRAS.526.5590G}, the $\beta$ of NGC~1387 is intermediate between the typical $\beta$ of early-type-like galaxies ($\beta\approx2.5$) and late-type-like galaxies ($\beta\approx1.8$).
The number of galaxies with a robustly measured $\beta$ nevertheless remains small, preventing strong constraints on the general distribution of $\beta$ across the Hubble sequence, and it is of course possible that other physical drivers simultaneously determine both $\beta$ and the GMC properties. 

{Lastly, the near-face-on orientation of NGC~1387 may be the reason for its MW-like GMC properties.
As mentioned before, galactic rotation induces velocity gradients in GMCs and thus contributes to the observed velocity dispersions.
When such a contribution is analytically removed and only the turbulence-induced velocity dispersion remains, the average virial parameter of the NGC~4429 GMCs changes from $4.0$ (observed) to $1.3$ (after correction; \citealt{2021MNRAS.505.4048L}). Similarly, the average virial parameter of the NGC~4526 GMCs after velocity gradient subtraction is $0.99$ \citep{2015ApJ...803...16U}. The corrected virial parameters are thus very similar to those typical of LTGs. 
Given the small inclination of NGC~1387, and thus the small (projected) galaxy rotation velocities, it may be natural for the GMC velocity dispersions (and the resultant virial parameters) to be as small as those of LTGs.
In addition, turbulence-induced velocity dispersions may be anisotropic \citep[e.g.][]{2006ApJ...638..191K, 2011ApJ...738...88H,2022MNRAS.515.1663J}, so that the observed quantities would depend on the inclination angle. 

\citet{2022AJ....164...43S} also discussed such an inclination ($i$) dependence of molecular gas properties in LTGs at physical scales of $60$ -- $150$~pc ($4$ -- $10$ times larger than the spatial resolution of this work). The empirical velocity dispersion correction factor $(\cos i)^{0.5}$ they introduced would yield a negligible reduction ($1\%$) of our $\sigma_{\rm obs,los}$ measurements, but a much larger one ($39\%$) for NGC~4429. 
In turn, this would reduce the average observed virial parameter of the NGC~4429 GMCs from $4.0$ to $1.5$, the latter being similar to both the gradient-subtracted virial parameter of $1.3$ and the typical virial parameters of LTGs.

These results suggest that a high inclination angle, though not a physical driver of properties, can lead to some of the observed GMC property differences between ETG GMCs and LTG GMCs, by effectively revealing galactic shear (often stronger in ETGs than in LTGs) and/or turbulence anisotropy.
On the other hand, when observed at small inclinations, ETG GMCs and LTG GMCs should appear similar, as is the case here.
Unfortunately, resolved-cloud studies of ETGs are still too few to test for a systematic dependence on inclination. Further observations and simulations are thus required to explore this possibility.
}

\subsection{{Implication on star formation}}

{
From all the gas (cloud) properties discussed in this paper, NGC~1387 seems to have high potential for star formation. 
Specifically, it has moderately low Toomre parameter (indicating gravitational instability) in the majority of the molecular gas disc, low turbulence level in the GMCs, and fully virialised dynamical state of the GMCs. Yet, the star formation rate or efficiency of NGC~1387 is as low as a typical ETG, as discussed in \autoref{sec:mw_compare}.
This may pose challenges in star forming theories. 

We propose that the low inclination angle may be the cause of low line widths in \autoref{sec:mw_compare}, which underlie the claims of disc instability and GMC dynamical state. If this is indeed the case, line widths (and turbulence) in the other two dimensions, i.e.\ in the galactic plane, can play a major role in suppressing star formation. To test this, observations of a larger sample of ETGs at different inclination angles are needed.

Alternatively, a narrow internal density distribution of each GMC and a low Mach number may explain the low SFE in NGC~1387. 
The former can imply a low dense gas fraction. As stars mainly form in dense clumps and cores within a GMC, this could further lead to low SFE \citep[e.g.][]{2004RvMP...76..125M,2004ApJ...606..271G}.
The Mach number can be used as a tracer of the prevalence of shock-induced compression in GMCs \citep{2012ApJ...761..156F,2021NatAs...5..365F}. In this picture, a low Mach number leads to less such compression and therefore low SFE, although the opposite effect of a low Mach number (i.e.\ less stabilisation against gravity) leading to high SFE has also been proposed \citep[e.g.][]{2024A&A...690A..43H}.
To constrain the internal density distribution and the Mach number of GMCs, radiative transfer modelling is required with observations of multi-$J$ transitions and/or more molecule species such as the optically thinner $^{13}$CO. 

Apart from pursuing a physical origin for the low SFE, the accuracy of the H$\alpha$ emission based SFR is worth re-visiting as well. In general, H$\alpha$ suffers from dust extinction, contamination from other ionising mechanisms, and, like in the case of this paper, poorer angular resolution than ALMA gas observations. A summary of alternative SFR indicators is beyond the scope of this paper, but one candidate is the 33-GHz radio continuum dominated by free-free emission from HII regions \citep{2011ApJ...737...67M}. Progress towards better SFR indicators in the future will contribute to the understanding of star formation suppression in ETGs. 
}

\section{Summary and conclusions}
\label{sec:sum}

We obtained high-resolution ($\approx0\farcs17\times0\farcs14$ or $16\times13$~pc$^2$ spatially, $2$~km~s$^{-1}$ spectrally) ALMA observations of the central region of the lenticular galaxy NGC~1387, with a sensitivity of $1.11$~mJy~beam$^{-1}$ ($1.12$~K).
We identified $1285$ GMCs, $1079$ of which are spatially and spectrally resolved, by applying a modified version of the {\tt CPROPStoo} package.
We measured the basic properties of the GMCs (radius $R_\mathrm{c}$, LoS velocity dispersion $\sigma_\mathrm{obs,los}$, molecular gas mass $M_\mathrm{gas}$ and molecular gas mass surface density $\Sigma_\mathrm{gas}$), probed the inferred GMC mass spectra, Larson relations and virial parameters, and investigated the origin of the GMCs' internal rotation (including large-scale galactic rotation, gravitational instabilities, and cloud-cloud collisions).
All of these properties were compared to those of GMCs in other galaxies.
Our key findings are summarised below.

\begin{itemize}
    \item The means (standard deviations) of the basic properties of the resolved GMCs of NGC~1387 are: $\langle R_\mathrm{c}\rangle=20$~pc ($7$~pc), $\langle\sigma_\mathrm{obs,los}\rangle=3.5$~km~s$^{-1}$ ($1.5$~km~s$^{-1}$), $\langle\log(M_\mathrm{gas}/\mathrm{M_\odot})\rangle=5.5$ ($0.4$) and $\langle\log(\Sigma_\mathrm{gas}/\mathrm{M_{\odot}~pc^{-2}})\rangle=2.4$ ($0.2$). 

	\item The mass spectrum (i.e.\ the cumulative probability distribution function of GMC mass) of NGC~1387 has a slope of $-1.8$ (best-fitting truncated power law), comparable to that of MW disc GMCs.
    The cut-off mass of $1.5\times10^6$~M$_\odot$ indicates a lack of high-mass GMCs.
 
    \item The Larson relations and virial parameters of the NGC~1387 GMCs are very similar to those of MW disc GMCs, despite the galaxy's early-type morphology.
    This shows that not all ETG GMCs deviate from the MW scaling relations.
    
    \item Overall, the internal rotation of the GMCs does not seem to arise from the large-scale circular rotation of the galaxy (e.g.\ through shear and tides), {again in contrast to previously studied ETGs}. 
    The prograde fraction of all resolved GMCs is nevertheless high ($63\%$), with a slight decrease from the inner region ($72\%$) to the outer region ($59\%$). In addition, massive and/or large GMCs at small galactocentric distances do appear to be better coupled to the large-scale galactic rotation.
    
    \item The Toomre instability parameter $Q\lesssim1.5$ in the outer half of the molecular gas disc, indicating that cloud-cloud collisions could be effective in that region.
    The shallow slope of the size -- line width relation ($0.29\pm0.06$) is as expected from cloud-cloud collisions.
    \end{itemize}

{Overall, we have revealed a broader variety of GMC properties in the ETG environment than previously reported. Specifically, the GMCs of NGC~1387 exhibit similarity to MW disc GMCs. We have discussed the potential origin of such variety of GMC properties, and identified the viewing-angle dependency as a plausible one. The implication of such GMC properties on star formation theories is also intriguing. A larger sample of ETGs with multi-molecule multi-transition observations of the cold molecular gas will further advance the understanding on these fronts.}

\section*{Acknowledgements}

{We thank the anonymous referee for the helpful comments.}
FHL thanks Jiayi Sun for helpful discussions and acknowledges support from the ESO Studentship Programme, the Scatcherd European Scholarship of the University of Oxford, and the European Research Council’s starting grant ERC StG-101077573 (`ISM-METALS').
MB was supported by STFC consolidated grant `Astrophysics at Oxford' ST/H002456/1 and ST/K00106X/1.
JG gratefully acknowledges funding via STFC grant ST/Y001133/1.
ST acknowledges the support of the Scientific and Technological Research Council of Türkiye (TUBITAK), 2219 International Postdoctoral Research Fellowship Program for Turkish Citizens.
TGW acknowledges funding from the European Research Council (ERC) under the European Union’s Horizon 2020 research and innovation programme (grant agreement No.\ 694343).

This paper makes use of the following ALMA data: ADS/JAO.ALMA\#2016.1.00437.S and ADS/JAO.ALMA\#2016.2.00053.S. ALMA is a partnership of ESO (representing its member states), NSF (USA) and NINS (Japan), together with NRC (Canada), MOST and ASIAA (Taiwan), and KASI (Republic of Korea), in cooperation with the Republic of Chile. The Joint ALMA Observatory is operated by ESO, AUI/NRAO and NAOJ. This paper is also based on observations collected at the European Southern Observatory under ESO programme 296.B-5054(A).

This research used observations made with the NASA/ESA Hubble Space Telescope, which is a collaboration between the Space Telescope Science Institute (STScI/NASA), the Space Telescope European Coordinating Facility (ST-ECF/ESA), and the Canadian Astronomy Data Centre (CADC/NRC/CSA); the NASA/IPAC Extragalactic Database (NED), which is operated by the Jet Propulsion Laboratory, California Institute of Technology, under contract with the National Aeronautics and Space Administration; the Mikulski Archive for Space Telescopes (MAST), for which STScI is operated by the Association of Universities for Research in Astronomy, Inc., under NASA contract NAS5-26555; NASA's Astrophysics Data System Bibliographic Services; Cube Analysis and Rendering Tool for Astronomy ({\tt CARTA})\footnote{\url{https://cartavis.org}} for data visualisation and measurements \citep{2021ascl.soft03031C}; and {\tt adstex}\footnote{\url{https://github.com/yymao/adstex}} for literature referencing.

\section*{Data Availability}
\label{sec:data_available}

The raw data underlying this article are available from the ALMA archive at \url{https://almascience.eso.org/aq} using the project codes listed above. 
The {\it HST} images are available at the MAST archive: \url{https://mast.stsci.edu/}.
The WISDOM data products are made available from \url{https://www.wisdom-project.org}.
The analysis scripts are made available at \url{https://github.com/ericliang45/NGC1387-GMC}.




\bibliographystyle{mnras}
\bibliography{NGC1387} 





\appendix

\section{Continuum image of NGC~1387}
\label{sec:app_continuum}

\begin{figure}
    \centering
    \includegraphics[width=\columnwidth]{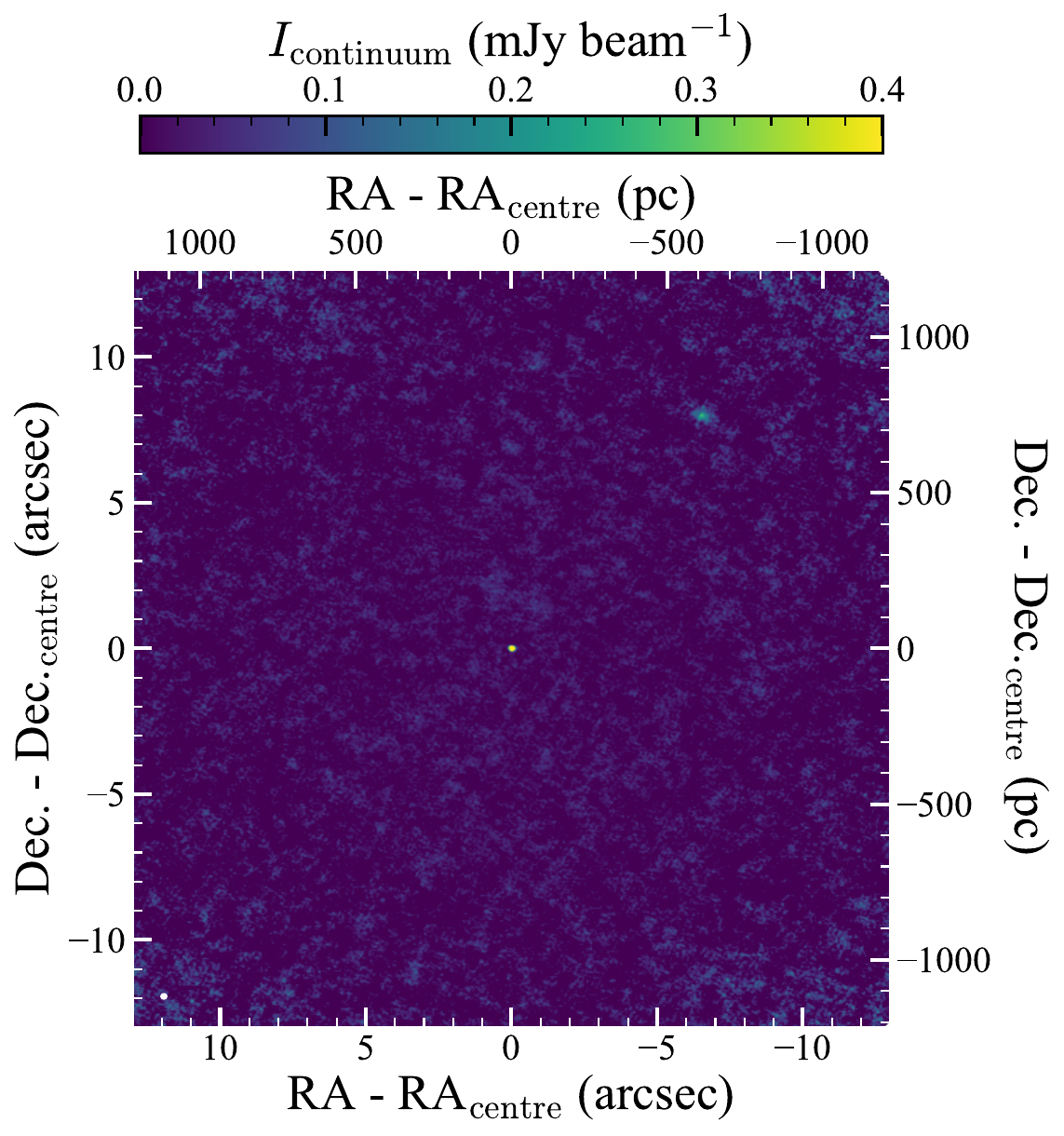}
    \caption{{PB-corrected continuum image of NGC~1387.
    The centre of the galaxy is well detected, but there is an additional faint diffuse source $10\farcs3$ ($970$~pc) to the north-west.
    The ALMA beam is shown as a white ellipse in the bottom-left corner.
    The maximum of the colour bar is much lower than the peak intensity of the central point source, to better visualise the northwestern source and background noise.}}
    \label{fig:continuum}
\end{figure}

{
Following the data reduction steps detailed in \autoref{sec:co}, the PB-corrected $230$-GHz continuum emission of NGC~1387 was imaged and the full field of view is shown in \autoref{fig:continuum}. 
The central point source is discussed in the main text (\autoref{sec:co}) and a zoom-in image of it is presented by \citet{2024MNRAS_dominiak}.
However, a second faint diffuse source is present $10\farcs3$ ($970$~pc) north-west of the galaxy centre.

We use the PB-uncorrected continuum image to measure the noise ($0.02$~mJy~beam$^{-1}$), manually estimate its size (for standard error propagation), and spatially integrate its flux, to gauge the statistical significance of this detection ($S/N=14$).
The integrated (PB-corrected) flux of the source is $2.2\pm0.2$~mJy.
As there is no apparent counterpart to the source in CO or at other wavelengths, one possibility is that the source is a background galaxy.
}

\section{Result robustness as a function of {\tt CPROPStoo} parameters}
\label{sec:robustness}

The identification of GMCs from our adopted data cube has some arbitrariness in the choices of the {\tt CPROPStoo} parameters.
Indeed, as described in \autoref{sec:cprops}, several steps require manual input based on trial and error.
Here, we quantify the changes to the main derived GMC properties when the two most important parameters are varied, and show that our results are robust against these choices.

\begin{figure*}
    \includegraphics[width=\textwidth]{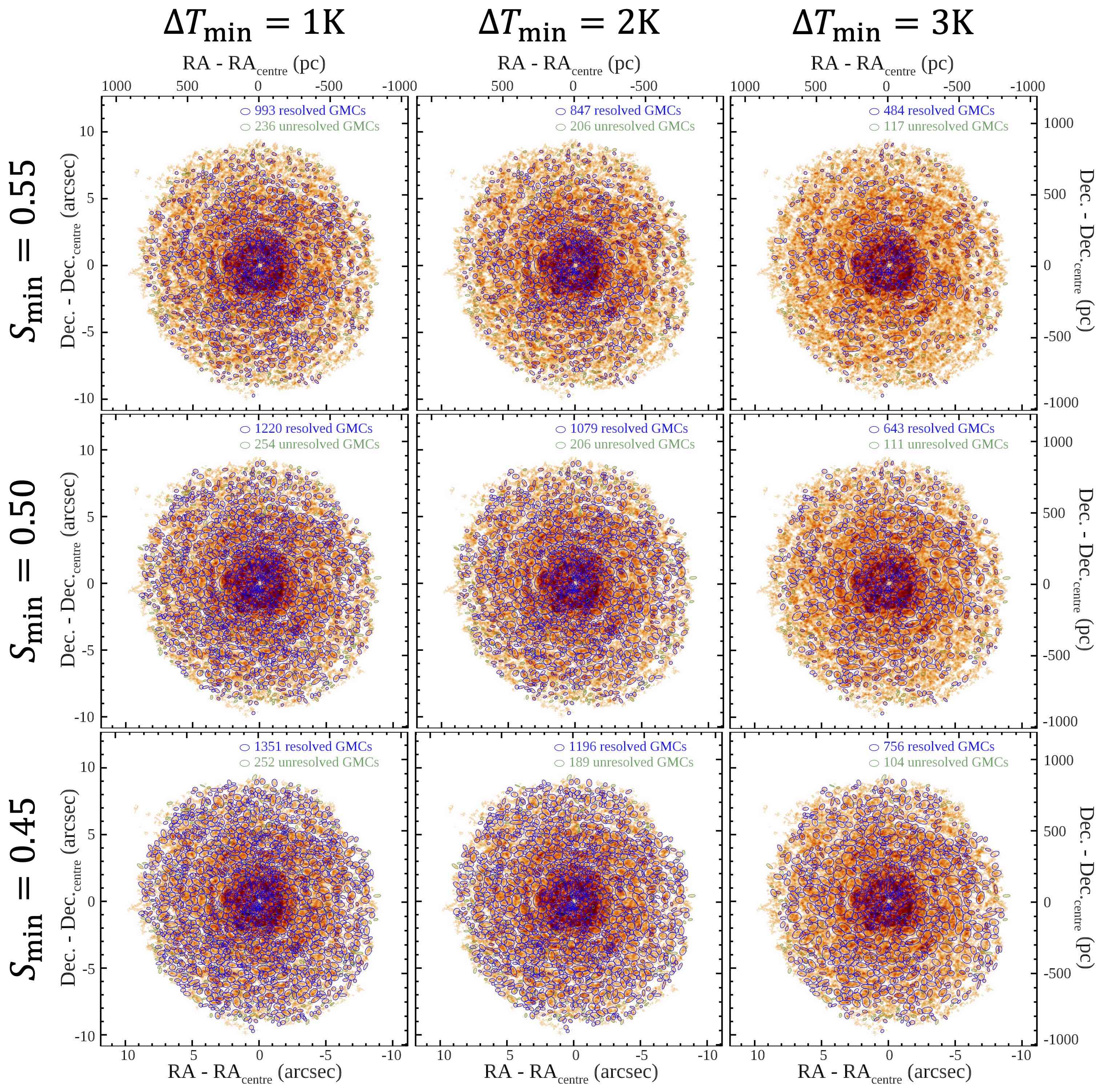}
    \caption{GMC identification as a function of the {\tt CPROPStoo} parameters $\Delta T_\mathrm{min}$ (from left to right: $1$, $2$ and $3$~K) and $S_\mathrm{min}$ (from bottom to top: $0.45$, $0.50$ and $0.55$).
    All panels are as \autoref{fig:cprops_identification}, with a few annotations removed for clarity.
    The central panel shows the results for the set of parameters adopted in the main body of this paper.}
    \label{fig:variation_gmc}
\end{figure*}

\begin{figure*}
    \includegraphics[width=\textwidth]{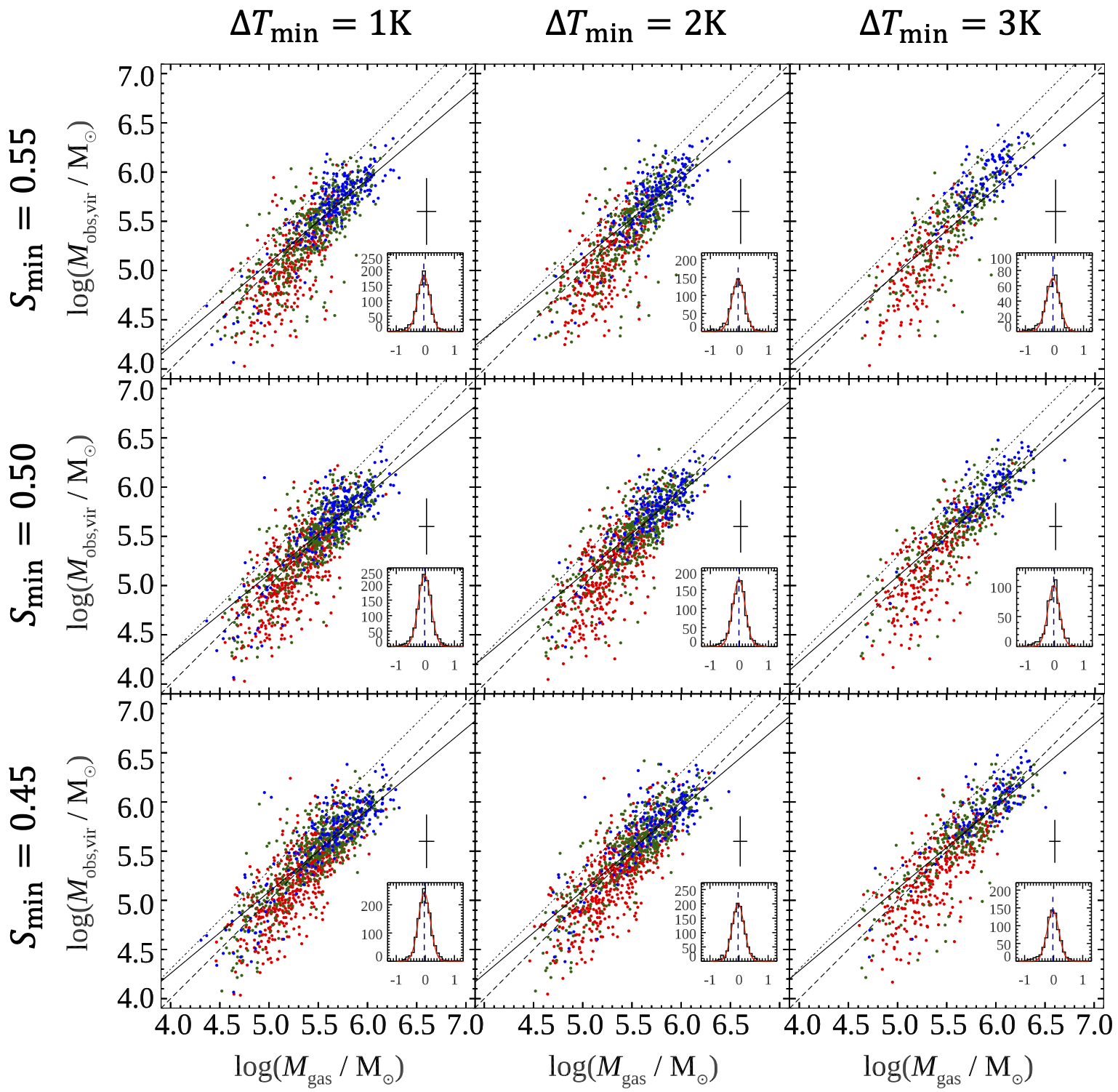}
    \caption{As \autoref{fig:variation_gmc}, but for the comparison of the GMC virial masses ($M_\mathrm{obs,vir}$) and molecular gas masses ($M_\mathrm{gas}$), as well as the distributions of the virial parameters ($\alpha_\mathrm{obs,vir}$).
    All panels are as \autoref{fig:virial}, with a few annotations removed for clarity.}
    \label{fig:variation_alpha}
\end{figure*}

We perform the same GMC identification and analyses for nine different sets of parameters.
The minimum contrast ($\Delta T_\mathrm{min}=1$, $2$ and $3$~K) and minimum convexity ($S_\mathrm{min}=0.45$, $0.50$ and $0.55$) are varied while all other parameters are kept constant.
The GMC identification results are shown in \autoref{fig:variation_gmc} and the GMC virial mass ($M_\mathrm{obs,vir}$) -- molecular gas mass ($M_\mathrm{gas}$) comparisons are shown in \autoref{fig:variation_alpha} (along with the distributions of the resulting virial parameters, $\alpha_\mathrm{obs,vir}$).
The central panels of both figures show the results for the set of parameters adopted in the main body of this paper.
We note that the GMC identification is carried out in 3D (i.e.\ using the adopted data cube), so the visualisation of the GMC catalogues using the 2D zeroth-moment maps shown in \autoref{fig:variation_gmc} does not reflect the process perfectly.

As expected, \autoref{fig:variation_gmc} shows that the GMC catalogue using the strictest combination of parameters (top-right panel of each figure; $\Delta T_\mathrm{min}=3$~K and $S_\mathrm{min}=0.55$) has the fewest GMCs ($484$ resolved and $117$ unresolved GMCs), while the opposite combination (bottom-left panel of each figure; $\Delta T_\mathrm{min}=1$~K and $S_\mathrm{min}=0.45$) has the most GMCs ($1351$ resolved and $252$ unresolved GMCs).
A higher $\Delta T_\mathrm{min}$ leads to identify only the largest overdensities and fails to identify GMCs across the relatively smooth parts of the molecular gas disc.
A lower $S_\mathrm{min}$ leads to the inclusion of slightly larger structures with more abundant sub-structures.
A similar conclusion is reached by inspecting the moment maps of individual GMCs (not shown).
The two parameters $\Delta T_\mathrm{min}$ and $S_\mathrm{min}$ are thus impacting the GMC identification in the manner expected.

\autoref{fig:variation_alpha} shows that the two masses $M_\mathrm{obs,vir}$ and $M_\mathrm{gas}$ correlate very well with each other for all sets of parameters, and the $\alpha_\mathrm{obs,vir}$ distributions are also all very similar, being centred around $\alpha_\mathrm{obs,vir}=1$ with very similar scatters.

We also probed the robustness of the Larson relations for all nine sets of parameters.
Considering the scatters, they all roughly agree with the MW disc GMC relations, although the best-fitting slopes and zero-points do vary a little.
For example, the slope of the size -- line width relation ranges from $0.18$ to $0.44$, all nevertheless flatter than that of the MW disc GMCs ($0.5$).

Based on the tests in this appendix, we therefore conclude that, within reasonable ranges, the choice of the parameters used for the GMC identification does not affect our conclusions.


\bsp	
\label{lastpage}
\end{document}